\documentclass[dvipsnames, 12pt]{article}
\usepackage[margin=30mm]{geometry}
\usepackage{amsmath}\allowdisplaybreaks
\usepackage{amsfonts}
\usepackage{amssymb}
\usepackage{verbatim}
\usepackage{setspace}
\usepackage{rotating}
\usepackage{booktabs}
\usepackage{longtable}
\usepackage{tabularx}
\usepackage{etoolbox}\AtBeginEnvironment{thebibliography}{\linespread{1}\selectfont}
\newcommand{\zerodisplayskips}{%
  \setlength{\abovedisplayskip}{-5pt}%
  \setlength{\belowdisplayskip}{5pt}%
  \setlength{\abovedisplayshortskip}{-5pt}%
  \setlength{\belowdisplayshortskip}{5pt}}
\appto{\normalsize}{\zerodisplayskips}
\appto{\small}{\zerodisplayskips}
\appto{\footnotesize}{\zerodisplayskips}
\usepackage{tikz}
\usetikzlibrary{fit,positioning}
\usepackage{amsthm}
\usepackage{bm}
\usepackage{dsfont}
\usepackage{lscape}
\usepackage{geometry}
\usepackage{float}
\usepackage{verbatim}
\usepackage{subcaption}
\usepackage[toc,page,header]{appendix}
\usepackage{minitoc}
\usepackage{algorithm}
\usepackage{algpseudocode}

\usepackage{hyperref}
\hypersetup{
    colorlinks=true,
    linkcolor=cyan,
    filecolor=cyan,      
    urlcolor=cyan,
    citecolor=cyan,
}

\usepackage{natbib}
\usepackage{authblk}

\usepackage{listings}
\begin{document}
\doparttoc 
\faketableofcontents 

\title{Explaining Recruitment to  Extremism:\\ A Bayesian Case-Control Approach}

\author[1]{\small Roberto Cerina\thanks{Author contributions: NK and CB conceived of the study. RC developed the models. RC, CB, and NK contributed to the analysis. RC and CB developed the code for the R package and the documentation. CB, NK and AZ contributed to the data collection. CB, NK and AZ developed the literature review. RC, CB, NK, and AZ contributed to the writing. We received helpful feedback and advice from Sir David Cox, Thomas Hegghammer, Bjørn Høyland, Bent Nielsen, Jacob Aasland Ravndal, and Frank Windmeijer. \cite{hertog_ISIS_2021} reached out to us as we were finalizing our manuscript. Their analysis also uses leaked ISIS recruitment data to analyze the socio-economic correlates of joining the movement. To implement the methods described in this paper, see the associated R package, as well as documentation and vignettes: \url{http://extremeR.info}}}
\author[2]{Christopher Barrie}
\author[3]{Neil Ketchley}
\author[4]{Aaron Y. Zelin}
\affil[1]{\small Maastricht University}
\affil[2]{University of Edinburgh}
\affil[3]{University of Oxford}
\affil[4]{Brandeis University}

\date{\today}
\maketitle
\thispagestyle{empty}

\begin{abstract}
\noindent Who joins extremist movements? Answering this question poses considerable methodological challenges. Survey techniques are infeasible and selective samples provide no counterfactual. Recruits can be assigned to contextual units, but this is vulnerable to problems of ecological inference. In this article, we take inspiration from epidemiology and elaborate a technique that combines survey and ecological approaches. The multilevel Bayesian case-control design that we propose allows us to identify individual-level and contextual factors patterning the incidence of recruitment, while accounting for rare events, contamination, and spatial autocorrelation. We validate our approach by matching a sample of Islamic State (ISIS) fighters from nine MENA countries with representative population surveys enumerated shortly before recruits joined the movement. High status individuals in their early twenties with university education were more likely to join ISIS. There is more mixed evidence for relative deprivation. We provide software for applied researchers to implement our method.\end{abstract}

\newpage
\pagenumbering{arabic} 
\doublespacing
\section{Introduction}

Identifying who is more or less likely to join extremist movements is a pressing issue for both political science and public policy.\footnote{We understand extremists to be movements or individuals that reject or seek to subvert democracy, often through violent means \citep[see e.g.][]{muddePoliticalExtremismVols2004}.} However, empirical research on this topic is beset by methodological challenges. Population surveys offer little insight into the phenomenon as recruits to extremism are tiny minorities in any society, and so are tiny minorities in survey samples. This is before obvious problems related to eliciting truthful responses to questions probing illicit actions. Recent innovations in survey and online digital trace methodologies have allowed researchers to obtain more accurate measures of support for extremism \citep{Bail2018a, Blair2013a, corstange_sensitive_2009, lyall_explaining_2013, Mitts2019}. However, these approaches capture attitudes, rather than behaviour. For researchers interested in why some individuals join extremist movements and not others, the most common strategy is to collect a convenience sample of recruits after which one of two approaches are commonly preferred. A first sees researchers simply report sample proportions of a given characteristic, e.g. the percentage of recruits who have university education. While often descriptively useful, this fails to account for population baselines and other confounding factors affecting the incidence of recruitment. An alternative strategy is to assign recruits to meaningful contexts and use the characteristics of those places to explain variation in the recruitment rate. While this approach provides a counterfactual and allows for multivariate analysis, it suffers from familiar problems of ecological inference \citep{robinsonEcologicalCorrelationsBehavior1950}. 

The method we propose in this paper allows researchers to leverage both survey \textit{and} contextual data to make robust inferences about the individual and ecological correlates of recruitment to extremism. To do so, we take inspiration from the case-control design used in epidemiology and show how it can be adapted to combine a convenience sample of cases (extremists) with controls (respondents from a representative survey). In this, we build on the recent introduction of case-control methods to political science found in the study of protest participation \citep{rosenfeld2017reevaluating}. Several statistical challenges arising from the nature of extremism remain, however. Popular approaches for modelling rare events [e.g. \citealt{king2001logistic}] do not account for spatial autocorrelation in the incidence of recruitment or the possibility of contamination between cases and controls. Further, conventional rare events approaches face potential separation problems in the absence of regularized priors for coefficients of interest. Finally, while existing approaches incorporate informative priors on the true prevalence of the outcome of interest (here: recruitment to extremism), this information is often not known.

Our approach offers a complete solution to these statistical problems and can be described as a multilevel, Bayesian case-control design that is robust to rare events, contamination, and spatial autocorrelation patterning the incidence of recruitment \citep{rota2013re}. The Bayesian approach is preferable for a number of reasons. First, Bayesian probabilistic programming software provides unique flexibility in the modeling of the complex hierarchical structures characterizing recruitment into extremism. Second, it permits the use of informative priors to account for the true prevalence of the event of recruitment, as well as to regularize coefficient estimates to account for separation bias and instability when carrying out regressions \citep{heinze2002solution}. Finally, in the absence of prior knowledge of the overall propensity of being a recruit in a given context, the model can estimate the propensity from the data \citep{rota2013re}. 

A great strength of our method is that applied extremism researchers can choose those parameters most relevant for their case. When sampling from national populations, the risk of contamination between cases and controls may be sufficiently low such that it does not pose a threat to inference. On the other hand, recruitment may not qualify as a rare event when comparing recruits to certain sub-populations. So too, spatial autocorrelation in recruitment may not apply if sampling from a small area or closed context. As we show, our modelling strategy if flexible to the inclusion or exclusion of these parameters, depending on the case at hand. In support of our approach, and to help guide the modelling decisions of future practitioners, we provide an extensive simulation study that compares our model to alternative frameworks, and show its robustness and superiority in predicting the true underlying probability of recruitment under various bias-inducing scenarios. 

To display some of the key properties of our modeling strategy, we analyze recruitment of Sunni Muslim males in nine MENA countries to the Islamic State in Iraq and Syria (ISIS). We focus our analysis on an individual's level of education and social status --- two key variables associated with recruitment to extremism found in the literature on violent Islamist movements. We show how our approach can be used to perform two types of analyses. In the first, we leverage a multilevel regression model trained on a cross-national sample of ISIS recruits and non-recruits. This provides a robust descriptive analysis about the individual-level characteristics of recruits across countries and sub-national administrative units. A second analysis focuses on two countries for which we have rich contextual information: Egypt and Tunisia. This analysis adds value by adjusting for local heterogeneity with the addition of relevant ecological covariates, allowing us to ascertain the potential sensitivity of individual-level findings to unobserved contextual confounding. 

For the purposes of illustration, we implement the complete solution described above, accounting for the possibility of contamination, spatial autocorrelation, and separation in our regression coefficients. Overall, we find that high-status males with university education in their early twenties were more likely to join ISIS. We also find that relatively deprived males in Egypt were more likely to join ISIS, but not in Tunisia. This heterogeneity in the individual and contextual correlates of violent extremism demonstrates the importance of accounting for both individual and context specific factors. 

While our analysis focuses on recruitment to ISIS, our hope is that this paper inspires political scientists to apply case-control methods to other instances of extremism where data on recruits and population surveys are available. Examples include participation in the 2021 attack on the Capitol Building in Washington D.C. \citep{pape_opinion_2021}, recruitment to far-right movements and white supremacist groups \citep{klandermansExtremeRightActivists2006, simiAddictedHateIdentity2017}, as well as other examples of violent extremism \citep{dellaportaClandestinePoliticalViolence2013}. In this spirit, the dedicated R package accompanying this paper allows applied researchers working on such cases to implement our proposed technique.\footnote{See: \url{http://extremeR.info}.}

\section{Explaining recruitment to extremism}
The principle strategy available to researchers interested in the correlates of recruitment to extremism is to sample on the dependent variable, obtaining relevant demographic information on individual extremists or members of extremist movements. In the ideal scenario, researchers are able to obtain movement membership lists, which can reveal information on tens of thousands of individuals \citep[e.g.][]{biggs_explaining_2012}, although in practice such complete data is rare. Absent such lists, a well-established strategy is to leverage data from arrests or killings to generate samples of participants \citep[e.g.][]{berrebiEvidenceLinkEducation2007, ibrahimAnatomyEgyptMilitant1980, Ketchley2017a, Krueger2003}. Alternatively, researchers can look to collect demographic information on extremists by either interviewing former recruits \citep[e.g.][]{berubeConvergingPatternsPathways2019, dellaportaClandestinePoliticalViolence2013}, or by reconstructing the biographical profiles of prominent extremists from open source information \citep[e.g.][]{Gambetta2016, jensenRadicalizationViolencePathway2020}. Of course, a principle limitation of these samples is that they do not provide information on individuals outside of the subpopulation of interest, meaning that it is not possible to compare recruits to the population from which they are drawn. To remedy this, researchers either confine attention to variation amongst recruits \citep[e.g.][]{morrisWhoWantsBe2020}, or else assign individuals to meaningful contexts, e.g. universities, cities, or countries, and then use the characteristics of those units to explain cross-sectional variation in the recruitment rate \citep[e.g.][]{barrie_is_protest_2018, Ketchley2017a, pape_opinion_2021}. While this latter approach is undoubtedly superior to simply analyzing sample proportions, it inevitably relies on ecological inference.\footnote{An alternative approach adopted in the extremism literature is the ``life history" method, whereby researchers draw on small convenience samples of extremists in an effort to identify events and individual-level characteristics that made them more likely to participate in extremism \citep[e.g.][]{simiAddictedHateIdentity2017}. As with simply reporting sample proportions, this method lacks an obvious counterfactual, meaning that we cannot rule out the possibility that factors identified as leading individuals to pursue extremist behavior are also present in individuals who did not become extremists \citep{ashworthDesignInferenceStrategic2008}.} 

\subsection{A multilevel Bayesian case-control design}

In what follows, we suggest two new methods for analyzing recruitment to extremism. The first leverages a cross-national, multilevel regression model trained on a complete sample of recruits and survey respondents. This  provides a robust descriptive analysis about the individual-level factors which characterize recruits across countries and subnational units. The model uses random effects to control for unobservable subnational heterogeneity; these are preferable to fixed effects due to potentially heavily imbalanced area-level sample sizes \citep{gelman2006data,clark2015should}. The model further uses a conditionally auto-regressive prior \citep{besag1991bayesian,morris2019bayesian} to account for spatial smoothing. The second analysis focuses on single country studies where rich contextual information is available. The added value of this analysis lies in controlling for local heterogeneity in order to ascertain the robustness of any individual-level findings to contextual confounding. Taken together, our proposed setup thus plots a way forward for researchers to combine survey and ecological information for the robust analysis of recruitment to extremism.

\subsection{Simple case-control set-up}

We begin by describing the backbone of our model, which is a logistic regression accounting for case-control sampling protocol via an offset. Borrowing from Rota et al. \citeyearpar{rota2013re}, we define $r_i=\{0,1\}$ as the set of states that observation $i$ in our sample of size $n = n_0 + n_1$ can obtain, where $r_i = 1$ implies the observation is a `case', $r_i = 0$ defines a control, $n_1 = \sum_i^n \mathds{1}(r_i = 1)$ and $n_0 = \sum_i^n \mathds{1}(r_i = 0)$. In our application, a `case' would refer to a known extremist; a `control' to a survey respondent. Recall that cases are selected entirely on the dependent variable while controls come from the population that cases are drawn from. Take $N_1$ to represent the number of cases in the population of interest, and $N_0$ the number of controls. The probability of being included in the sample ($s_i = 1$) conditional on the true state of any individual can hence be understood as $P_1 = \mbox{Pr}(s_i = 1 \mid r_i = 1) = \frac{n_1}{N_1}$, while that of being sampled as a control is $P_0 = \mbox{Pr}(s_i = 1 \mid r_i = 0) = \frac{n_0}{N_0}$. The log-ratio of these sampling probabilities can then be used as an `offset' in a logistic regression, to account for the sampling protocol. The hierarchical specification of the model follows, with regression coefficients being assigned a very weakly informative prior;\footnote{See \url{https://github.com/stan-dev/stan/wiki/Prior-Choice-Recommendations} for prior-choice advice when using \texttt{Stan}. Note further that the normal distribution in our model (and in \texttt{Stan}) is parametrized by mean and standard deviation.}

\begin{align}
    r_i & \sim \mbox{Bernoulli}(\rho_i);\\
    \mbox{logit}(\rho_i) & = \mbox{log}\left(\frac{P_1}{P_0}\right) + \sum_k x_{i,k} \beta_{k};\\
    \beta_k & \sim N(0,10).
\end{align}
The above hierarchical model thus contains three layers: layer (1) is a model of the true state of an observation, conditional on their latent propensity $\rho$; layer (2) describes this latent propensity, by accounting for systematic variation due to heterogeneity in covariates; layer (3) models the effects of each covariate by assigning a prior probabilistic model.

\subsection{Contaminated controls}

Recall that that the case-control setup as described above takes known recruits and combines them with `controls' taken from survey respondents. While we know that our cases are correctly labeled, we do not know whether this is true of our controls. That is, our controls may be `contaminated' as survey respondents may have become recruits \citep{lancaster1996case}. This is especially concerning when researchers have access to biographical information on tens of thousands of extremists  \citep[e.g.][]{biggs_explaining_2012} or are comparing recruits to small sub-populations \citep[e.g.][]{Ketchley2017a, ketchleyWhoSupportedEarly2021}. \cite{rota2013re} outline a `latent variable' formulation of their contamination model. Below we present our version of that same model as a mixture, which we find more intuitive.

The `label' of an observation, $y_i=\{0,1\}$, is observed for all observations, while the true `state' of an observation, $r_i=\{0,1\}$, is only observed for cases. The implied probability distribution of labels conditional on being a control is:

\begin{align*}
    \mbox{Pr}(y_i = 1 &\mid r_i =0 , s_i = 1 ) = 0 = \theta_{0};\\  
     \mbox{Pr}(y_i = 0 &\mid r_i =0 , s_i = 1 ) = 1 = (1-\theta_{0} );
\end{align*}

Due to contamination, it is possible that observations characterized by $y_i=0$ are actually in state $r_i = 1$; hence we need a probability distribution for $y \mid r_i = 1$. Let $\pi = \frac{N_1}{N_1 + N_0}$ be the prevalence of recruits in the population of interest, and let $n_u = \sum_i^n \mathds{1}(y_i = 0)$ be the number of unlabeled observations. We expect there to be $\pi n_u$ cases amongst the unlabeled observations. We can then characterize the probability distribution of labels, conditional on being a case, as:

\begin{align*}
      \mbox{Pr}(y = 1 \mid r =1 , s = 1) = \frac{n_1}{n_1 + \pi n_u} = \theta_{1};\\
      \mbox{Pr}(y = 0 \mid r =1 , s = 1) = \frac{\pi n_u}{n_1 + \pi n_u} = (1-\theta_{1}).
\end{align*}

Finally, our model for the latent state $r_i$ must reflect the possibility of contamination. We do this by re-defining the relative-risk of being sampled as: 

\begin{align*}
\frac{P_1}{P_0} = \frac{\frac{n_1 + \pi n_u}{N_1}}{\frac{ (1-\pi)n_u}{N_0}} = \frac{n_1}{\pi n_u} + 1.
\end{align*}

The updated, hierarchical specification for the case-control model accounting for contaminated controls is then: 

\begin{align}
    y_i & \sim \mbox{Bernoulli}(\theta_{r_i});\\
    r_i & \sim \mbox{Bernoulli}(\rho_i);\\
    \mbox{logit}(\rho_i) & = \mbox{log}\left(\frac{n_1}{\pi n_u} + 1\right) + \sum_k x_{i,k} \beta_{k};\\
    \beta_k & \sim N(0,10).
\end{align}

In summary, we derive our labels via two distinct data generating processes, identified by a latent state $r_i = \{1,0\}$. In the event that the latent state of a given record is that of a true control, $r_i = 0$, it is then impossible for this record to be labeled $y_i = 1$; conversely, if the latent state is that of a true case, $r_i = 1$, then it is still possible for a record to be labeled $y_i = 0$, with probability $(1-\theta_1)$. This latter model describes the issue of contamination. Note that in our application, $\theta$ is always observed, and fed to the model as data. 

\subsection{Area-level random effects}\label{rand_eff}

Survey data and information on recruits often contain information on the origin or location of residence of individuals. And we can understand individuals as nested within geographical units of increasing sizes. Generalizing, we can exploit variance at three levels: the individual, some small-area, and some large-area. 

These area effects could be incorporated in the model via fixed-effects, by expanding the design matrix to include relevant dummy-variables for each area of interest. We consider this strategy unwise when trying to explain recruitment to extremism and prefer a random-effects approach. In the case of rare forms of political behaviour, our geographical units at all levels of analysis will have relatively few observations \citep{gelman2006data}. Additionally, for many units, we will have no cases. Finally, we know that lists of recruits are unlikely to be exhaustive; that is, we will not have data for every recruit hailing from every subnational unit or country. Here, a sample of recruitment data or similar can be treated as a non-probability sample --- it is unlikely that we can have complete confidence the sample constitutes a complete or random sample of the population of interest. Given these concerns, a random effects approach is preferable as it means: 1) we are able to borrow strength across areas, which also increases efficiency, to produce more realistic estimates for the area-level coefficients \citep{clark2015should,baio2012bayesian}; 2) in the absence of more detailed knowledge about the data-generating process, the shrinkage effect obtained by partial pooling is more likely to shield our estimates from any systematic sampling bias among our cases \citep{gelman2006data}.

We can also relax some of the theoretical bias associated with the shrinkage induced by random effects via incorporating observable area-level heterogeneity in the design-matrix as fixed effects \citep{gelman2006data}. This is what we elect to do in single country analyses. Finally, it is worth highlighting that our goal is not to make inferences about area-level effects. Rather, we seek to strip our individual-level effects estimates of contested variance that may be associated with the provenance of the recruit. The resulting hierarchical model is as follows:

\begin{align}
    y_i & \sim \mbox{Bernoulli}(\theta_{r_i});\\
    r_i & \sim \mbox{Bernoulli}(\rho_i);\\
    \mbox{logit}(\rho_i) & = \mbox{log}\left(\frac{n_1}{\pi n_u} + 1\right) + \sum_k x_{i,k} \beta_{k} + \phi_{l[i]} + \eta_{j[i]} ;\\
    \beta_k & \sim N(0,10);\\
    \phi_l & \sim N(0,\sigma_\phi);\\
    \sigma_\phi = \frac{1}{\sqrt{\tau}_\phi}, \mbox{ } \tau_\phi &\sim \mbox{Gamma}(\epsilon,\epsilon);\\
    \eta_j & \sim N(0,\sigma_\eta);\\
    \sigma_\eta = \frac{1}{\sqrt{\tau}_\eta}, \mbox{ } \tau_\eta &\sim \mbox{Gamma}(\epsilon,\epsilon);
\end{align}

\noindent where $\epsilon$ stands for some arbitrary number, chosen as a compromise to minimize the prior information and maximise the Markov chain Monte Carlo (MCMC) convergence speed and stability.

\subsection{Spatial autocorrelation}

The network ties connecting actors across space play an important role in recruitment to high-risk activism and extremism \citep{centola_complex_2007, mcadam_specifying_1993}. Sometimes the ties connecting recruits will be available; more commonly this information will not be recoverable.

In the absence of detailed network information, we propose controlling for potential network effects at levels of varying scale. We work on the assumption that network ties are more likely to form between individuals who are geographically proximate. Depending on the richness of the data on recruits, we may generate distance matrices between geographical units of varying size.

To account for area-level spatial autocorrelation, we propose incorporating a version of the conditional auto-regressive (CAR) model \citep{besag1991bayesian}. This approach has been used in individual-level models of behaviour, enabling local smoothing of predictions according to behaviour observed in neighbouring areas \citep{selb2011estimating}. The key ingredients of a CAR model are $\bm{\omega}$, a distance-weight matrix; $\alpha$, a parameter governing the degree of autocorrelation, where $\alpha =0$ implies spatial independence, and $\alpha =1$ implies an intrinsic conditional auto-regressive (ICAR) model \citep{besag1995conditional}; and $\sigma_{\psi}$, the standard deviation of the subnational unit effects. The resulting model for spatial random effect $\psi_l \mbox{ } \forall \mbox{ }l = \{1,...,L\}$   is then:

\begin{equation*}
\psi_l \mid \psi_{l^\prime}   \sim N\left(\alpha\sum_{l^{\prime} \neq l} \omega_{ll^{\prime}} \psi_{l^\prime},\sigma_{\psi} \right).\\
\end{equation*}

In practice, we implement the ICAR specification of the model, with $\alpha = 1$, and take $\bm{\omega}$ to be the neighbourhood matrix. The neighbourhood matrix has diagonals zero (a unit cannot neighbour itself) and off-diagonal zero or one depending on whether the given units are neighbours. We choose this specification of the distance matrix because of the efficiency gains it affords in a Bayesian context  \citep{morris2019bayesian}. This leads to:

\begin{equation*}
\psi_l \mid \psi_{l^\prime}  \sim N\left(\frac{\sum_{l^{\prime} \neq l} \psi_{l^\prime}}{d_{l,l}},\frac{\sigma_{\psi}}{\sqrt{d_{l,l}}} \right),
\end{equation*}

\noindent where $d_{l,l}$ is an entry of the diagonal matrix $D$ of size $L \times L$, whose diagonal is defined as a vector of the number of neighbours of each area. The joint distribution of this model is simply a multivariate normal distribution $\bm{\phi} \sim N(0,[\tau_\psi(D - W)]^{-1})$, $\tau_\psi = \frac{1}{\sigma^2_\psi}$, which is conveniently proportional to the squared pairwise difference of neighbouring effects. Note that the sum-to-zero constraint is needed for identifiability purposes, as in its absence any constant added to the $\psi$s would cancel out in the difference.\footnote{Generally, this model has the disadvantage of being an improper-prior, as its density does not integrate to unity and is non-generative (i.e., it can only be used as a prior and not as a likelihood) though it serves our purposes within the context of a hierarchical model. Finally, we should note that this prior encodes an intrinsic dependence between subnational units. It can no longer detect the degree of spatial autocorrelation supported by the data but instead assumes that areas are explicitly dependent, and estimates coefficients accordingly.} Following \cite{morris2019bayesian}, setting the precision to $1$ and centering the model such that $\sum^L_l \psi_l = 0$, we arrive at:

\begin{equation*}
\mbox{log } p(\bm{\psi} ) \propto \mbox{exp} \left\{ -\frac{1}{2} \sum_{l^{\prime} \neq l}(\psi_l - \psi_{l^\prime})^2 \right\};
\end{equation*}

The hierarchical model we implement to incorporate the spatial component is within the Besag-York-Mollié (BYM) family \citep{besag1991bayesian}. For a given level of analysis, say the city or province in a cross-country analysis, BYM models are characterized by two random effects which explain unobserved heterogeneity: $\phi_l$ defines a non-spatial component while $\psi_l$ defines systematic variance due to spatial dependency. The typical challenge with BYM is that the two areal effects cannot be identified without imposing some structure since they are mutually dependent, meaning either component is capable of accounting for contested variance at the area-level. This leads to inefficient posterior exploration of any MCMC sample, and subsequent lack of convergence \citep{riebler2016intuitive}. To overcome this, we implement a state-of-the-art solution leveraging penalized-complexity priors \citep{simpson2017penalising}, which proposes modelling the two effects as a scaled mixture such that:

\begin{align*}
\gamma_l = \sigma \left( \phi_l\sqrt{(1-\lambda)} + \psi_l\sqrt{(\lambda/s)}  \right);
\end{align*}

\noindent where $\phi$ and $\psi$ are random effects scaled to have unitary variance and $\lambda \in [0,1]$ is a mixing parameter, defining the proportion of residual variation attributable to spatial dependency. In order for the spatial and unstructured effects to share $\sigma$, they must be on the same scale \citep{riebler2016intuitive}. We must therefore scale the ICAR-distributed effects, as their original scale is defined by the local neighbourhood. A proposed scaling factor is chosen such that the geometric mean of the variance parameters over the areal units is $1$, $\mbox{Var}(\psi_l)= 1$. Note that this scaling factor, $s$ in the equation above, can be calculated directly from the adjacency matrix, and hence it is not to be estimated but passed to the model as data. 

The resulting hierarchical specification of our model follows:
\begin{align}
    y_i & \sim \mbox{Bernoulli}(\theta_{r_i});\\
    r_i & \sim \mbox{Bernoulli}(\rho_i);\\
    \mbox{logit}(\rho_i) & = \mbox{log}\left(\frac{n_1}{\pi n_u} + 1\right) + \sum_k x_{i,k} \beta_{k} + \gamma_{l[i]} + \eta_{j[i]} ;\\
    \beta_k & \sim N(0,10);\\
    \gamma_l & = \sigma \left( \phi_l\sqrt{(1-\lambda)} + \psi_l\sqrt{(\lambda/s)}  \right);\\
    \lambda & \sim \mbox{Beta}(0.5,0.5);\\
    \phi_l & \sim N(0,1);\\
    \psi_l \mid \psi_{l^\prime}  & \sim N\left(\frac{\sum_{l^{\prime} \neq l} \psi_{l^\prime}}{d_{l,l}},\frac{1}{\sqrt{d_{l,l}}} \right)\\    
    \sigma & \sim  \frac{1}{2}N(0,1);\\
    \eta_j & \sim N(0,\sigma_\eta);\\
    \sigma_\eta = \frac{1}{\sqrt{\tau}_\eta}, \mbox{ } \tau_\eta &\sim \mbox{Gamma}(\epsilon,\epsilon);
\end{align}

\noindent where $\frac{1}{2}N$ denotes a half-normal distribution, which is the recommended prior for the variance of BYM effects \citep{morris2019bayesian}. 

\subsection{Regularizing prior coefficients}

Multiple contributions have highlighted problems with logistic regression coefficient estimates under rare-events \citep{king2001logistic}. The intuition behind these challenges is typically described as some variation on the standard separation problem where any given covariate or simple combination thereof perfectly separates cases from controls. This leads to biased and unstable point-estimates with large associated uncertainty \citep{heinze2017logistic}. A number of regularization techniques have been proposed to reduce bias and stabilize the coefficient estimates. Our preferred regularization method is that proposed by \cite{gelman2008weakly} and \cite{ghosh2018use}. The approach assumes it should be unlikely to observe unit-changes in the (standardized) covariates that would lead to outcome changes as large as $5$ points on the logit scale. Using a slight variation on this approach to ensure sufficient regularisation, we use a Cauchy prior with scale-parameter set to $1$ for the regression coefficients, and a `looser' scale of $10$ logit points on the intercept to accomodate for the rarity of the event in the sample. The advantages of the Cauchy prior lie in its fat tails, which avoid over-shrinkage of large coefficients \citep{ghosh2018use}. We apply this prior to our fixed effects exclusively, as the likelihood of our random effects is already structured and penalized. Our final model specification is then as follows:

\begin{align}
    y_i & \sim \mbox{Bernoulli}(\theta_{r_i});\\
    r_i & \sim \mbox{Bernoulli}(\rho_i);\\
    \mbox{logit}(\rho_i) & = \mbox{log}\left(\frac{n_1}{\pi n_u} + 1\right) + \sum_k x_{i,k} \beta_{k} + \gamma_{l[i]} + \eta_{j[i]} ;\\
    \beta_1 & \sim \mbox{Cauchy}(0,10);\\
    \beta_k \mid k>1 & \sim \mbox{Cauchy}(0,1);\\
    \gamma_l & = \sigma \left( \phi_l\sqrt{(1-\lambda)} + \psi_l\sqrt{(\lambda/s)}  \right);\\
    \lambda & \sim \mbox{Beta}(0.5,0.5);\\
    \phi_l & \sim N(0,1);\\
    \psi_l \mid \psi_{l^\prime}  & \sim N\left(\frac{\sum_{l^{\prime} \neq l} \psi_{l^\prime}}{d_{l,l}},\frac{1}{\sqrt{d_{l,l}}} \right)\\    \sigma  &\sim \frac{1}{2}N(0,1);\\
    \eta_j & \sim N(0,\sigma_\eta);\\
    \sigma_\eta = \frac{1}{\sqrt{\tau}_\eta}, \mbox{ } \tau_\eta &\sim \mbox{Gamma}(\epsilon,\epsilon);
\end{align}

\subsection{Bayesian modeling in \texttt{Stan}}\label{stan}

The model outlined above is amenable to Bayesian estimation via Monte Carlo Markov Chain (MCMC) methods. Previous contributions to the case-control literature [e.g., \citealt{rota2013re, rosenfeld2017reevaluating}] have used \texttt{WinBUGS} \citep{lunn2000winbugs} or \texttt{JAGS} \citep{plummer2003jags} as software to implement some variations on a simple Gibbs sampler. Due to the heavy computational burden imposed by the spatial prior, we propose instead to innovate by estimating this model in \texttt{Stan} \citep{carpenter2017stan}. \texttt{Stan} leverages a version of Hamiltonian Monte-Carlo (HMC) called the `No U-Turn Sampler' (NUTS) \citep{hoffman2014no}, which dramatically improves the efficiency and speed of convergence of our Markov-Chains. A challenge we face is that \texttt{Stan} cannot handle the sampling of latent discrete parameters ($r_i$ in our hierarchical model above), posing a problem for the estimation of mixture models. The state-of-the-art solution is to marginalize the latent parameter out. In practice this means replacing our model for the observed labels $y$ with the following mixture of Bernoulli distributions:

\begin{equation}
f ( y_i \mid  \rho_i ) =  \rho_i  \mbox{ Bernoulli} ( y_i \mid \theta_{1 } )  + ( 1 - \rho_i ) \mbox{ Bernoulli} (y_i \mid \theta_{0} ).
\end{equation}

Beyond allowing for model parameters to be informed by $y_i$ according to the mixed structure above, marginalization provides significant advantages for posterior exploration and MCMC efficiency as it leverages expectations rather than sampling of discrete parameters. Listing \ref{lst:stan_dist} in the Appendix presents the \texttt{Stan} code for our final model. Note that fixed-effects covariates are standardized.\footnote{We standardize both dichotomous and continuous variables as this aids convergence.} Estimates of regression coefficients on the original, unstandardised scale are computed and available in the Appendix.

\subsection{Simulation study}\label{simulation}
The model that we propose extremism researchers should adopt is significantly more complex than the standard case-control design using rare-events logistic regression and requires a substantial understanding of Bayesian methods to be fully appreciated. Moreover, the model's estimation becomes roughly exponentially more computationally challenging as the sample size increases. To provide evidence that our approach is nevertheless preferable to a more straightforward case-control design, in Appendix Section \ref{appendix:simulation_study} we report the results of a comprehensive simulation study that compares the performance of our model against the King and Zeng model \citeyearpar{king2001logistic}, as well as a simple fixed-effects logistic regression. We score these models according to their ability to accurately predict the underlying latent propensity of recruitment, $\mu_i = \mbox{logit}({\rho_i})$. We further investigate these models' performance in accurately estimating the intercept, regression coefficients, and residual area effects. 

Simulations show that our model is robust and general. The results suggest that in a rare-event scenario, our model outperforms King and Zeng's rare-events logistic regression thanks to its ability to account for spatial auto-correlation, whilst also remaining largely unbiased to discrepancies in sample and population prevalence. As prevalence increases, our model retains a degree of robustness that neither a simple fixed-effects logistic regression, nor the `rare events logit', can offer -- largely thanks to the contamination layer. This robustness extends not just to the ability to correctly estimate latent propensity $\bm{\mu}^\star$, but actively reduces bias and RMSE in the estimation of coefficients. 

\section{Correlates of Islamist extremism} 

Determining the demographic profile of recruits to violent extremism is a central concern in studies of terrorism and political violence. Among the most commonly studied characteristics is education \citep{Krueger2003, Krueger2017, morrisWhoWantsBe2020, Ostby2019, Mesquita2005}. In their review of the literature, \cite{Ostby2019} summarize three mechanisms linking education to reduced political violence. First, the availability of educational infrastructure might lower societal grievances and lead to a consequent decline in support for extremist groups. Second, the opportunity costs of engaging in violence are greater for more highly educated individuals. Third, education helps spread norms of tolerance and civic engagement, lowering support for forms of political extremism.

An alternative body of literature argues that greater access to higher education may increase recruitment to violent extremism, in part as a result of relative deprivation \citep{Davies1962, Gurr1970}. This argument has been popular among scholars studying political violence in the MENA where it is argued that mass educational expansion created a pool of highly educated unemployed individuals who were attracted to high risk activism \citep{Campante2012, Lia2005, Malik2013}. A parallel literature focuses on the educational profiles of Islamist and jihadist groups. The key finding of this scholarship is that recruits are drawn disproportionately from the highly educated \citep{Berrebi2007, Krueger2003, Gambetta2016}. Again, relative deprivation arising from unrealized social status is often cited here as an explanatory factor with MENA countries having expanded higher education out of step with labor market demand, creating a ``lumpen intelligentsia" vulnerable to recruitment \citep{Roy1994}. An alternative interpretation is that the highly educated often have greater feelings of efficacy and a sense that they can understand and change politics \citep{Ketchley2017a}.

An adjacent body of research focuses on recruitment to transnational terrorist groups such as ISIS and Al-Qaeda. These contributions, which tend to be ecological and operate at the country level, find associations between human development indices and horizational inequalities in educational attainment as key factors patterning the flow of recruits to groups \citep{Verwimp2016, benmelech2018, Gouda2019}. Using a sample of leaked border documents similar to those we exploit below, \cite{worldbank2016} find that recruits tend to be highly educated and have high social status. Other research stresses the local-level determinants of ISIS recruitment, comparing the demographic characteristics of fighters with local-level averages for the general population. While not multivariate, these studies conclude that fighters tend to hail from marginalized regions and areas with prehistories of contention, but that there remains significant across-context variation in the correlates of recruitment \citep{Rosenblatt2016, Sterman2018}.

In summary, there are somewhat competing predictions for why some individuals may be more likely to be recruited to violent extremism. University education, relative deprivation, and high social status are commonly advanced to explain an individual's propensity to join violent Islamist groups. Scholarship also points to important subnational, ecological factors in the incidence of recruitment. Taken together, the literature would benefit from a method that combines information at the individual \emph{and} ecological levels to appraise those factors that make recruitment more or less likely.

\section{Data}

\begin{figure}[htp]
    \centering
    \includegraphics[scale = 0.585]{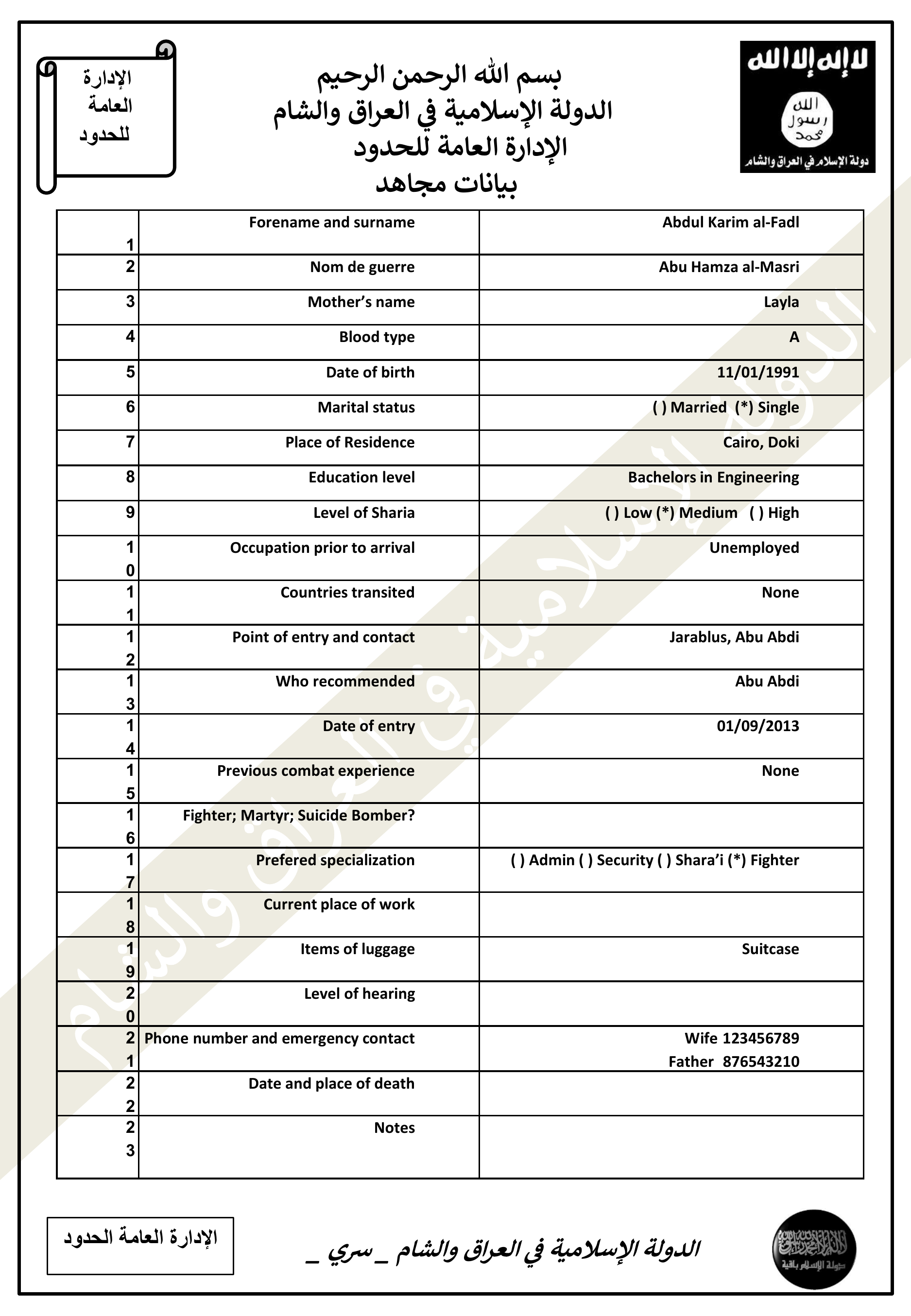}
    \caption{Example of ISIS border document (details changed)}
    \label{fig:sampledoc}
\end{figure}

Our key data source to study ISIS is a set of leaked border documents. This leak was widely covered in international news media and has been used to provide descriptive statistics on the geographical distribution and demographic characteristics of ISIS fighters from multiple MENA countries \citep{worldbank2016, zelin2018, Sterman2018}.\footnote{Dodwell et all \citeyearpar{dodwell_caliphates_2016} suggest that around 98\% of these individuals can be matched against records for ISIS recruits held by the U.S. Department of Defense.} When an individual entered Islamic State territory from Turkey they passed though an ISIS-controlled border control point. The leaked documents derive from these border control offices. They contain detailed information on the home residence of each recruit, age, education, marital status, previous employment, employment status, previous combat experience, and date of entry into ISIS-controlled territory. Figure \ref{fig:sampledoc} provides an imitation of one of the border documents. We use data for nine countries in the MENA that were included in the leak. These are: Algeria, Egypt, Jordan, Kuwait, Lebanon, Libya, Morocco, Tunisia, and Yemen. In total, we have complete records for 1,051 recruits. 

For the case-control design, we combine individual-level ISIS recruitment data with a nationally representative sample of individuals from Wave III of the Arab Barometer \citeyearpar{arabarom2014} survey. The fieldwork for these surveys was completed \textit{before} most recruits recorded in our border documents entered ISIS-held territory, and so may be vulnerable to contamination.\footnote{The surveys were in the field at different times for each country: December, 2012-January, 2013 for Jordan; February, 2013 for Tunisia; March-April, 2013 for Egypt and Algeria; April-June, 2013 in Morocco; July 2013 in Lebanon; November-December 2013 in Yemen; and February-March 2014 in Kuwait \citep{arabarom2014}.} Our choice of covariates to use from this survey is naturally constrained by the information included in the border documents. We elect to include covariates for age, marital status, university education, and student status. We combine two variables for unemployed and employment in agricultural or manual labor to create a composite variable designed to measure ``low status" activity. An interaction between this variable and our university education variable is designed to capture the relative deprivation hypothesis; that is, whether highly educated individuals engaged in low status economic activity are more likely to become recruits.

As per our models above, we present a ``bird's eye" and ``worm's eye" view of ISIS recruits. The first approach uses just individual-level information for recruits across our nine countries. Here, we leverage a multilevel regression model trained on the complete sample of $1,051$ recruits and $5,093$ unlabeled records.\footnote{We lose $248$ cases and $225$ unlabeled records due to incompleteness. We considered incorporating a multiple imputation model in our Bayesian analysis, but decided against it: a) in the interest of clarity, as a number of layers of complexity are already present; b) in the interest of estimation stability, as incorporating further uncertainty could destabilize the estimates of coefficients; c) we have no a-priori reason to believe missingness is correlated with the outcome, and hence we do not expect that listwise deletion will bias the estimates.} The second incorporates contextual information for Egypt ($n_1 = 66, n_0 = 551$ complete records\footnote{We lose $15$ controls and $95$ cases to listwise deletion of incomplete records in the Egypt dataset.}) and Tunisia ($n_1 = 426$, $n_0 = 589$ complete records\footnote{We lose $5$ controls and $20$ cases to listwise deletion of incomplete records in the Tunisia dataset.}) at the district level. We focus on these two countries due to the availability of contextual information at the district level that is not easily accessible for the other countries in our sample. Again, the added value of this analysis lies in controlling for observable district-level heterogeneity in order to ascertain the robustness of any individual-level findings to contextual confounding. As such, for each country we include variables to capture subnational differences in demographic and labor-market composition, employment opportunities, as well as more context-specific variables designed to capture support for Islamist political organizations and prehistories of contentious politics. Our choice of contextual variables is based on existing research finding that lack of employment opportunities, prehistories of mobilization and repression, as well as support for political Islam, are predictive of ISIS recruitment \citep{worldbank2016, Rosenblatt2016, grewal_counting_2020, barrie_is_protest_2018}. As such, the inclusion of these variables provides insight into the potential magnitude and origin of contextual confounding when analyzing recruitment to extremism. We describe the sources and operationalization of these ecological variables in Appendix \ref{appendix:independent_variables}.

\section{Results}
In this section we present the results of our two analytic approaches: the `bird's eye' view, referring to our cross-country analysis, and the `worm's eye' view, referring to our two country-specific analyses. For each, we present: i) convergence diagnostics; ii) the posterior density of fixed and random effects according to our models; 3) the posterior predictive distribution across potential recruitment profiles.

\subsection{MCMC Convergence}

Convergence diagnostics provide a first measure of the reliability of our parameter estimates for both the Bird's Eye and Worm's Eye models. Here, we follow Vehtari et al \citeyearpar{vehtari2021rank} and implement multiple state-of-the-art tests. 

Per Appendix Figures \ref{fig:bird_rhat_all}, \ref{fig:egypt_rhat_all}, and \ref{fig:tunisia_rhat_all}, we examine four versions of the Gelman-Rubin statistic ($\hat{R}$) to verify convergence is obtained broadly, as well as when we encounter heteroskedasticity across chains, or when these are heavy-tailed. There exist various convergence-thresholds in the literature -- the most stringent requires $\hat{R}<1.01$, a medium-stringency threshold suggests $\hat{R}<1.05$ (especially if we are estimating a large number of parameters), whilst the historical recommendation was $\hat{R}<1.1$ \citep{gelman1992inference}. Recent work demonstrates that this latter threshold prematurely diagnoses convergence in most cases \citep{vats2021revisiting}. The parameters of all of our models are broadly convergent under the harshest $1.01$ threshold for all of the measures of $\hat{R}$, with the exception of a very small number of spatial effects which are convergent under a slightly more laxed threshold, though still well below the `premature convergence' threshold $\hat{R}<1.1$. \footnote{Note that in the Bird's Eye model, this struggle is slightly exacerbated by the inclusion of governorates from Israel and Saud Arabia, for which we have no observations, and whose effects are fully interpolated via the spatial process.}

Appendix Figures \ref{fig:bird_ESS_all}, \ref{fig:egypt_ESS_all}, and \ref{fig:tunisia_ESS_all} present five measures of Effective Sample Size (ESS), which tell us about the true number of independent draws from the joint posterior distribution after accounting for auto-correlation within chains. The measures check that the independent sample is `large enough' to ensure stability of summaries of the distribution at various moments (e.g. overall, at the median, at the tails, etc.). Mirroring the performance of the $\hat{R}$, the posterior samples for most of our estimates parameters are well above the recommended threshold ($ESS>400$) for ensuring stability of the central and tail estimates. 

We further explore convergence at different quantiles of the posterior distribution of our least-convergent parameters -- those with the lowest bulk and tail ESS (Figure \ref{fig:bird_min.ESS_all} in the Appendix presents these measures for the Bird's Eye model -- the Worm's Eye analysis is available upon request).  
The inference is that if these relatively low-ESS parameters showcase satisfactory ESS at every quantile, we can be reassured that the whole model has converged. These plots suggest broad reliability of estimates at every section of the distribution. Finally, we explore the mixing properties of our chains for these least-convergent parameters (Figure \ref{fig:bird_mix.ESS_all} in the Appendix).\footnote{Here, we choose to include mixing diagnostics for the Bird's Eye model. The Worm's Eye analysis is available upon request.} 
These plots broadly suggest good mixing properties of our model, even for these relatively inefficient posterior samples. 

\subsection{Fixed and Random Effects}
\begin{figure}[t!]
    \centering
    \includegraphics[scale = 0.85]{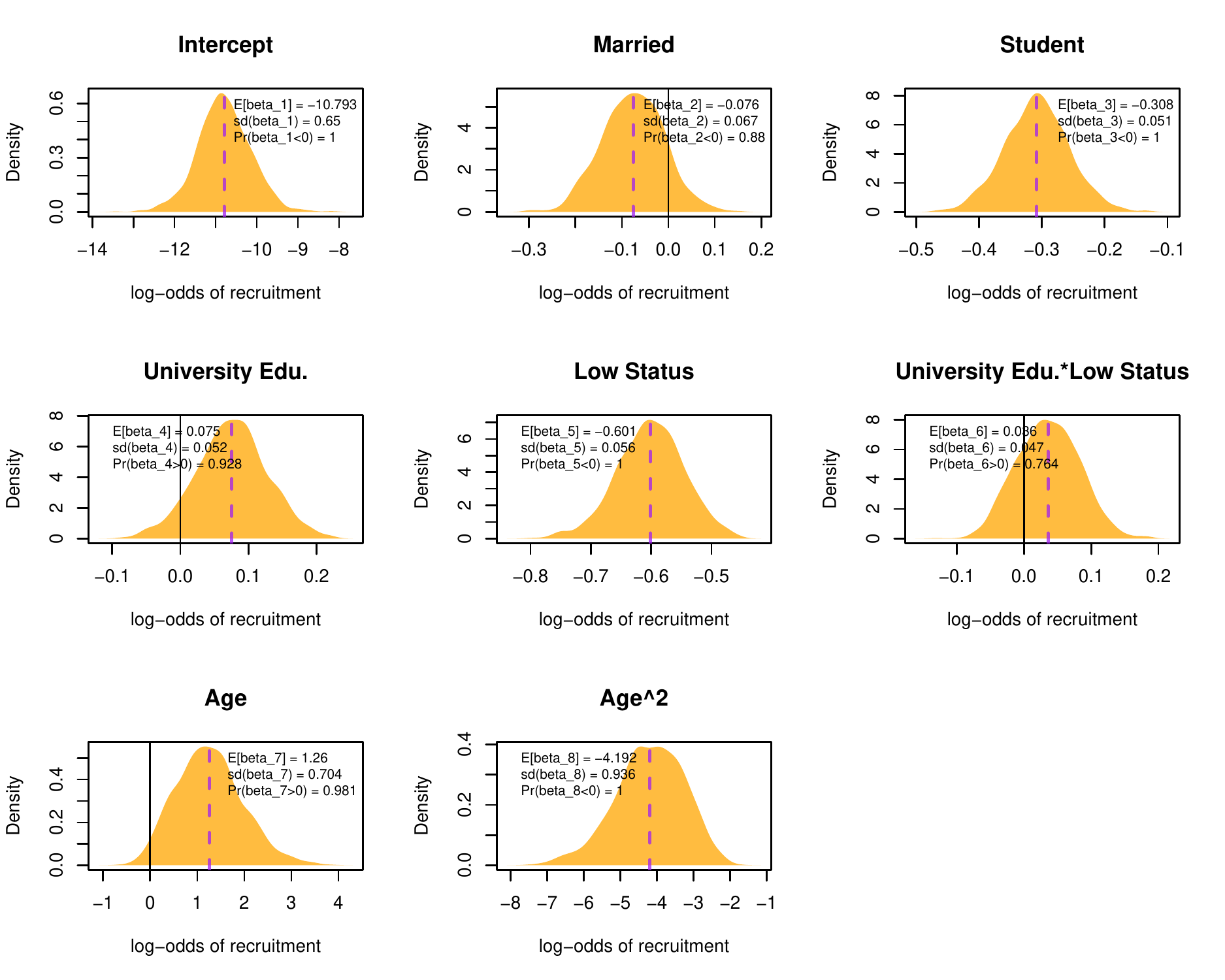}
    \caption{Posterior density of fixed-effect coefficients for the \emph{Bird's Eye} model.}
    \label{fig:bird_fixedeff_all}
\end{figure}

\begin{figure}[htbp]
\centering
\begin{subfigure}{0.8\textwidth}
    \includegraphics[width=1\linewidth]{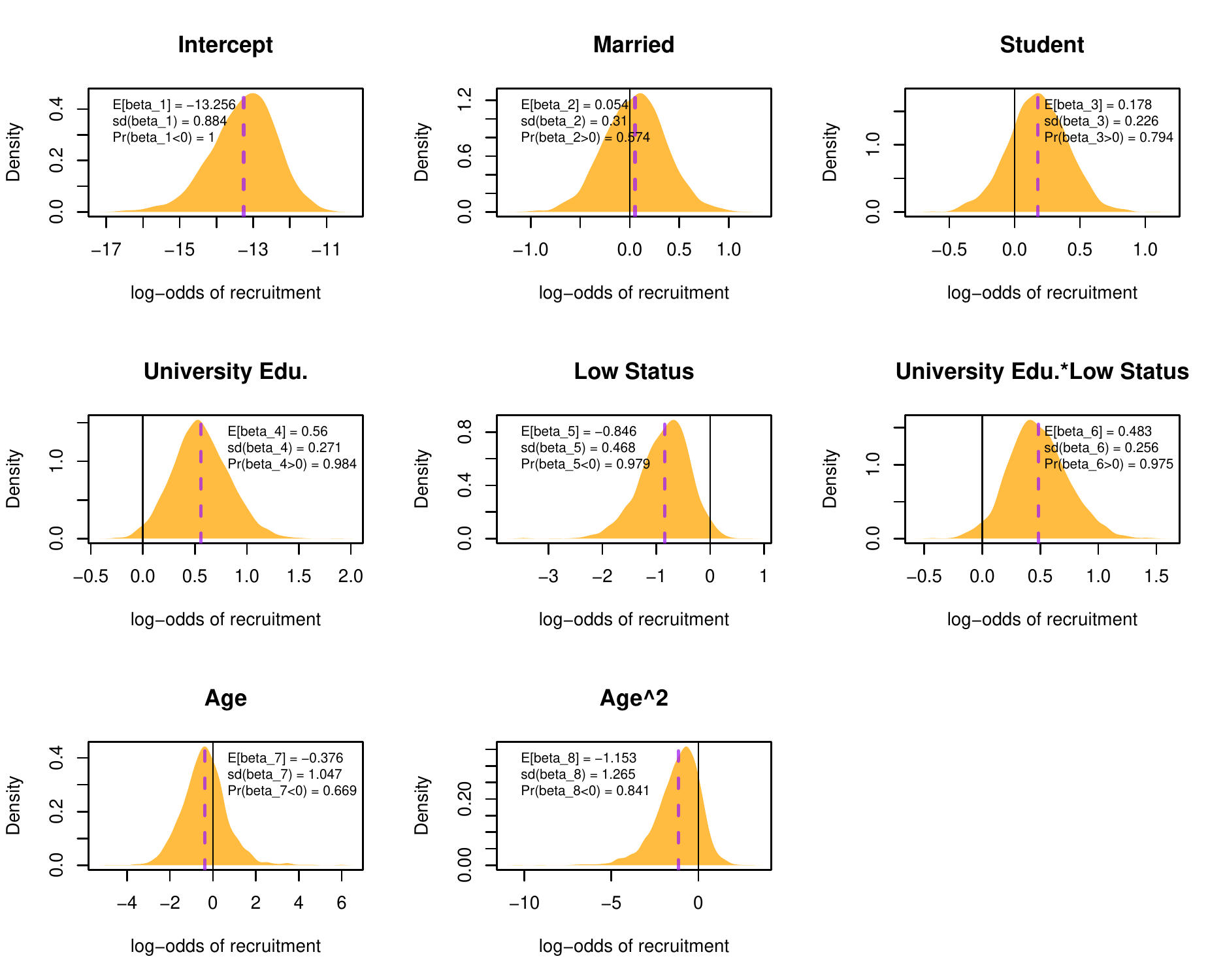}
    \caption{Egypt}
    \label{fig:worm_fixedeff_egypt_ind}
    \end{subfigure}

\begin{subfigure}{0.8\textwidth}
    \includegraphics[width=1\linewidth]{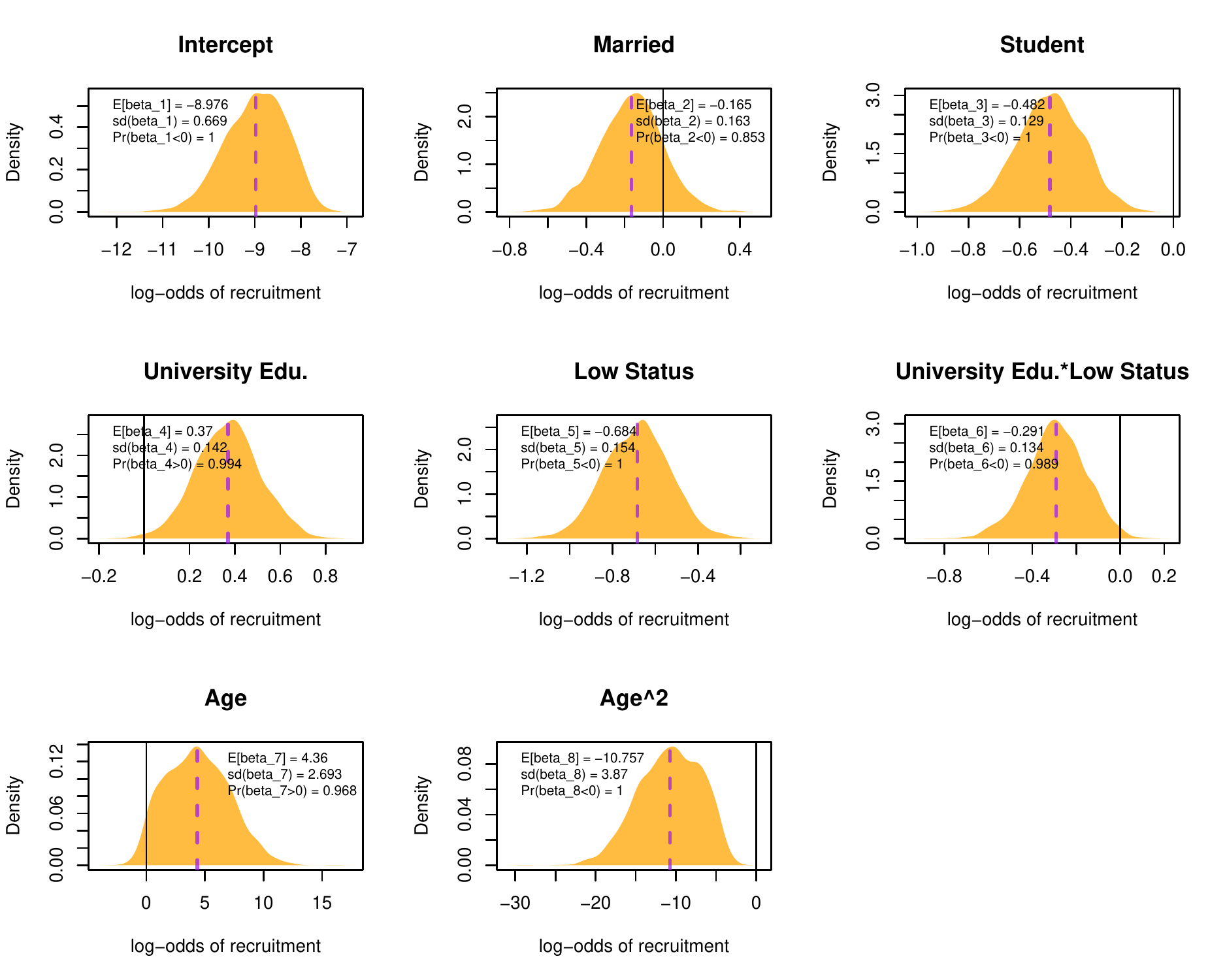}
    \caption{Tunisia}
    \label{fig:worm_fixedeff_tunisia_ind}
    \end{subfigure}
    \caption{Posterior density of fixed-effect coefficients for the \emph{Worm's Eye} models.}
\end{figure}

Figure \ref{fig:bird_fixedeff_all} presents the posterior density of the individual-level fixed effects in the Bird's Eye model; Figures \ref{fig:worm_fixedeff_egypt_ind} and \ref{fig:worm_fixedeff_tunisia_ind} present the Worm's Eye equivalent. These plots contain the main results of our models. Note that all the covariates, including dummies, are centered and scaled, hence the coefficients are to be interpreted in terms of standard deviations from the mean of each covariate (Figures \ref{fig:bird_fixedeff_original}, \ref{fig:worm_fixedeff_egypt_ind_original} and \ref{fig:worm_fixedeff_tunisia_ind_original} in the Appendix are the individual-level posterior densities on the original, non-standardized scale). Since we are principally interested in the robust estimation of individual-level predictors, we display only the posterior density of individual fixed effects for all of our models.\footnote{Figures \ref{fig:worm_fixedeff_egypt} and \ref{fig:worm_fixedeff_tunisia} display the standardized district-level posterior densities, whilst \ref{fig:worm_fixedeff_egyptl_original} and \ref{fig:worm_fixedeff_tunisia_original} present district-level coefficients on the original, non-standardized scale.} To aid with interpretation, the mean and standard deviation of each covariate are reported in the legend of each plot.

The estimated intercepts for the three models are extremely low. For the Bird's Eye model, the log-odds are in the order of negative $11$. For the Egypt Worm's Eye model, it is just over negative $13$; in Tunisia it is negative $9$. The size of the intercept is primarily driven by the size of the offset, which is in turn determined by the overall prevalence of recruitment. It is therefore not surprising that Egypt's intercept is so dramatically low, given the close-to-zero prevalence of recruitment when compared to population size ($\pi = \frac{4}{100,000}$) versus Tunisia where this prevalence is higher ($\pi = \frac{2}{1,000}$). For the Bird's Eye model, a different offset is provided for observations coming from different countries, to account for country-specific prevalence. The large and negative intercept underscores an important challenge in the explanation of why individuals join movements like ISIS: a linear combination of features capable of pushing an individual to become a recruit has to be extremely large, on the log-odds scale, to meaningfully affect the otherwise extremely low probability of recruitment. 

\begin{figure}[t!]
    \centering
    \includegraphics[trim=0 0 0 55, clip, width = 1\textwidth]{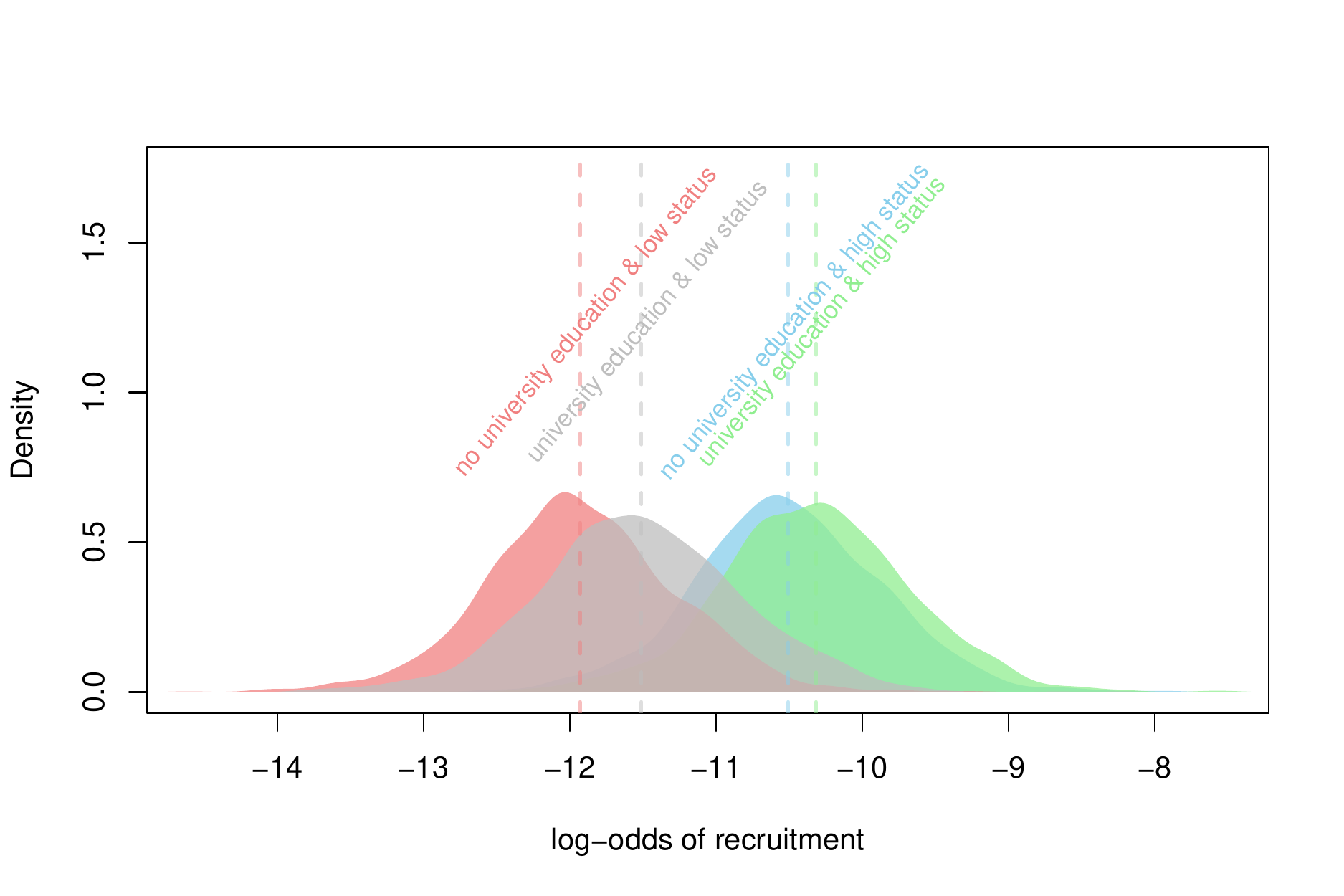}
    \caption{Predicted propensity of recruitment for relative-deprivation profiles according the \emph{Bird's Eye} model.}
    \label{figure::relative_deprivation}
\end{figure}

\begin{figure}[htbp]
\centering
\begin{subfigure}{.75\textwidth}
    \includegraphics[width=1\linewidth]{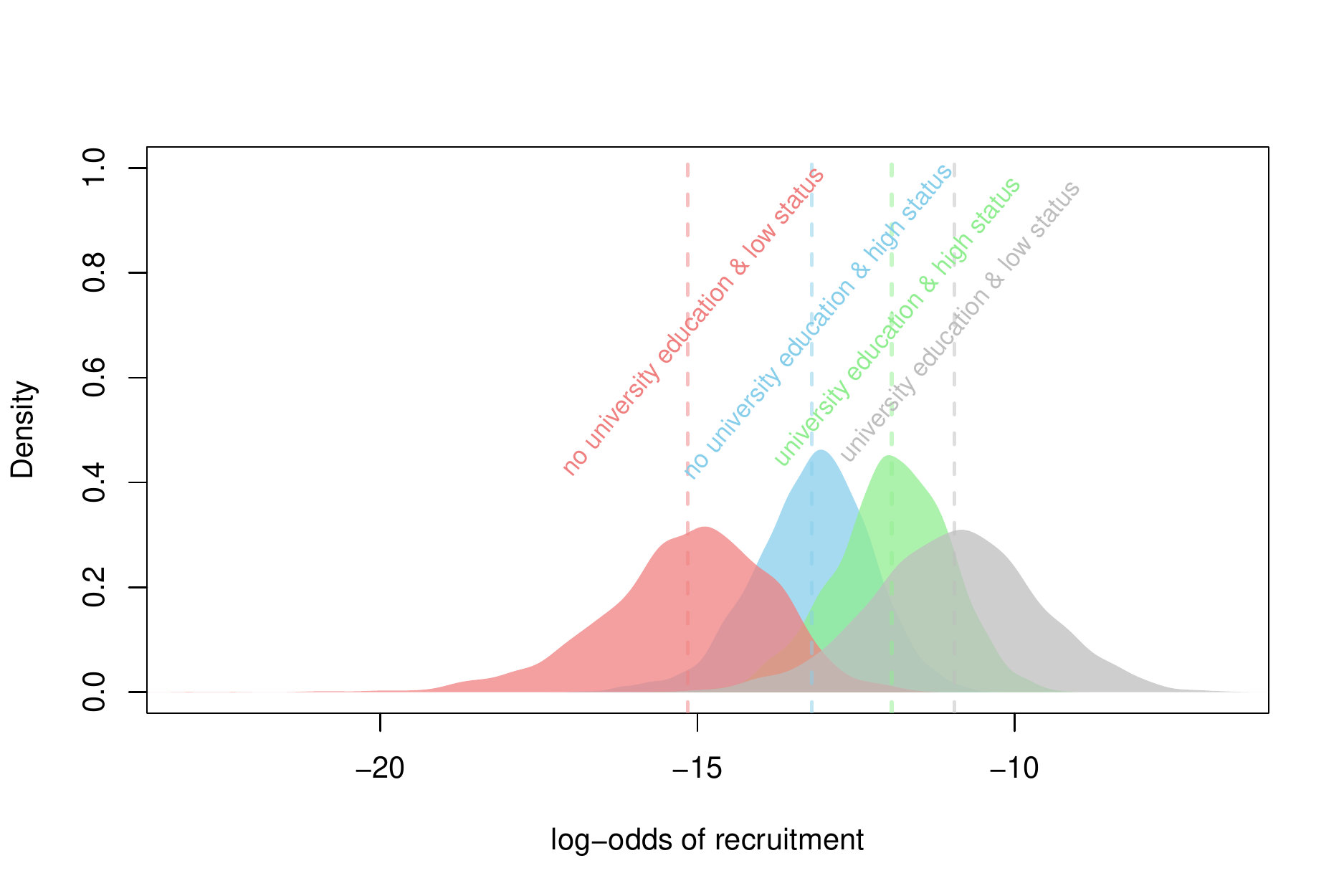}
    \caption{Egypt}
    \label{figure::relative_deprivation_egypt}
    \end{subfigure}
\begin{subfigure}{.75\textwidth}
    \includegraphics[width=1\linewidth]{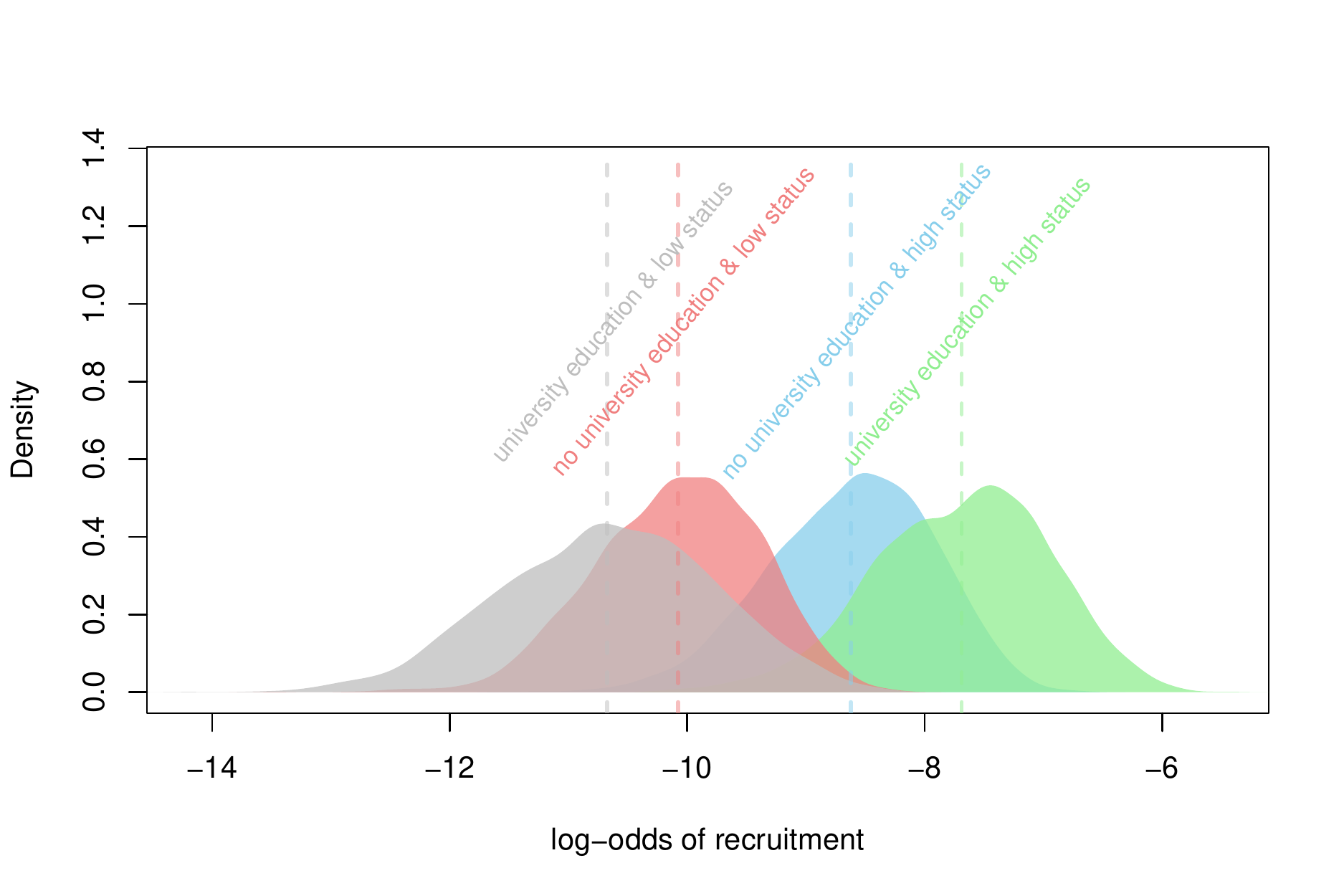}
    \caption{Tunisia}
    \label{figure::relative_deprivation_tunisia}
    \end{subfigure}
    \caption{Predicted propensity of recruitment for relative-deprivation profiles according to the \emph{Worm's Eye} models.}
\end{figure}

For the purposes of illustration, we focus primarily on testing the role of education and social status in an individual's decision to join ISIS. An individual who has university education and low-status is assumed to be relatively deprived. We compare predicted log-odds, as opposed to predicted probabilities, as these are scarcely comparable due to the powerful effect of the intercept, which drags probabilities of most profiles close to zero (though see Figures \ref{figure::bird_relative_deprivation_relprob},
\ref{figure::egypt_relative_deprivation_relprob}, and \ref{figure::tunisia_relative_deprivation_relprob} for predicted probabilities of recruitment relative to the `average' profile, and Figures \ref{figure::bird_relative_deprivation_n}, \ref{figure::egypt_relative_deprivation_n}, and \ref{figure::tunisia_relative_deprivation_n} for expected counts under different relative-deprivation scenarios). The total logit effects on probability of recruitment for different relative-deprivation profiles are shown in Figure \ref{figure::relative_deprivation} for the Bird's Eye model, and in Figures \ref{figure::relative_deprivation_egypt} and \ref{figure::relative_deprivation_tunisia} for the Worm's Eye. The relative deprivation hypothesis finds mixed support: at the Bird's eye level, it seems clear that being high-status plays a key role in increasing propensity of being recruited, while having university education plays a more minor role. A similar pattern is evident in Tunisia, though the effect of being high-status and having university education is starker, meaningfully increasing the propensity to join ISIS by around $3$ points on the log-odds scale compared to relatively deprived individuals. In Egypt the effects are more consistent with the traditional relative-deprivation hypothesis, though note the large prediction intervals around the total effects of relatively-deprived individuals. There is also substantial overlap between the distributions in all plots. This is largely due to the uncertainty around the intercept, which plays a role in marginalising these effects. Note further that varying prediction intervals on the effects reflect the highly unbalanced prevalence of the groups in our study. All in all, the evidence from these analyses suggest that high-status individuals were more likely to be recruited by ISIS, and that having university education on-top of a high social status further increases the likelihood of recruitment. The large prediction intervals as a result of uncertainty around the intercept underscores that much remains unknown about the underlying systematic determinants of recruitment. 

\begin{figure}[htbp]
\begin{subfigure}{0.65\textwidth}

\includegraphics[width=1\linewidth,trim={2cm, 4.25cm 0cm 3cm},clip]{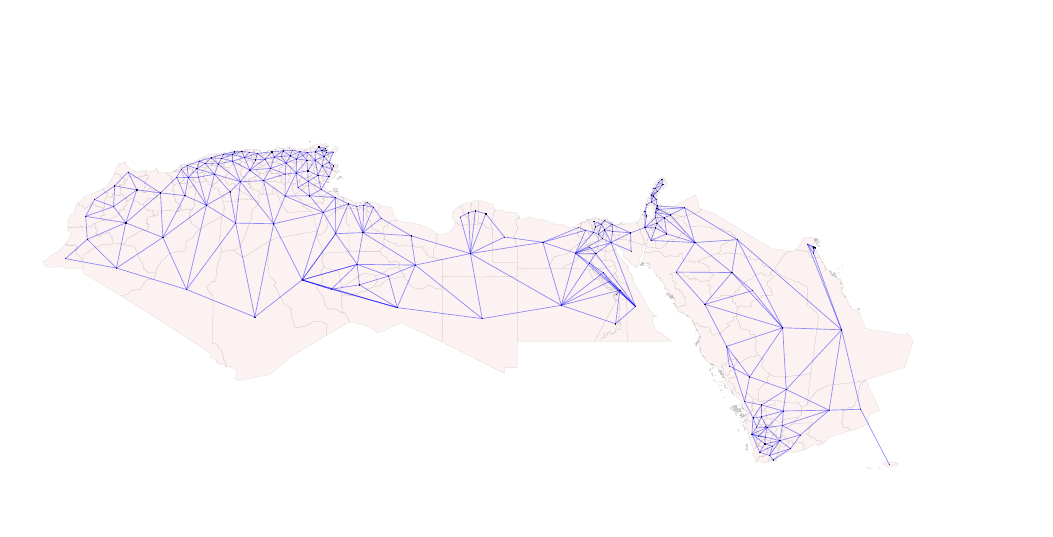}
\caption{}
\label{fig:bird_fullyconnected_all}
\end{subfigure}
\begin{subfigure}{0.33\textwidth}
\includegraphics[width=1\linewidth]{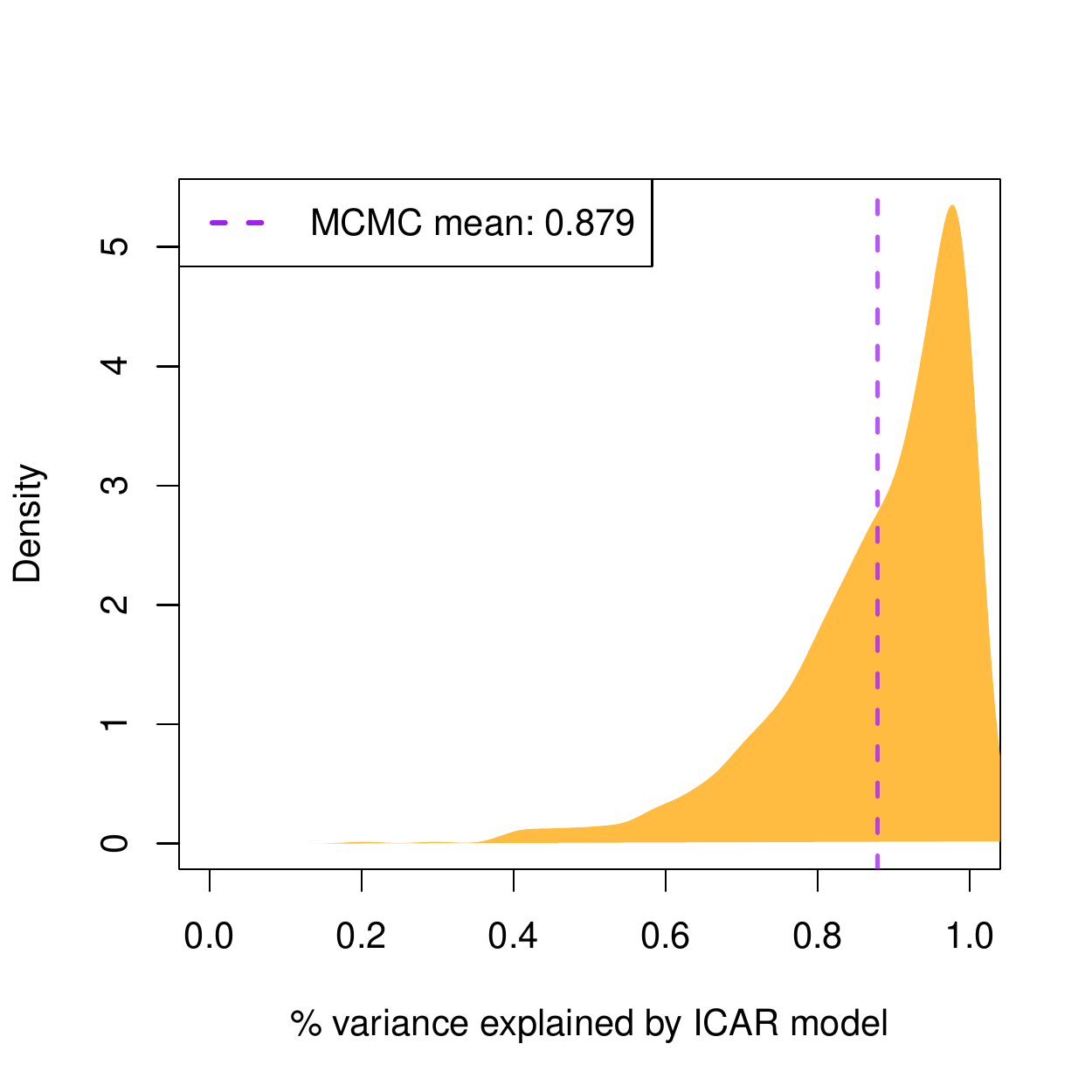}
\caption{} 
\label{fig:bird_lambda}
\end{subfigure}
\caption{Fully-connected graph for the Bird’s Eye model (a) and Governorate-level variance mixing parameter - $\lambda$ (b).} \label{fig::bird_lambda}
\end{figure}

 To fit the ICAR model we implemented the fully-connected graph shown in Figure \ref{fig:bird_fullyconnected_all}.\footnote{In the graph, edges connect nodes identified by the centroids of governorates for each country. Minor adjustments were performed to ensure the absence of islands or sub-graphs, which would have made the analysis needlessly complicated. Note also that Israel and Saudi Arabia are included for the purpose of obtaining this fully-connected graph, but no observations were available for either of these countries in terms of recruits or Arab Barometer observations, and hence the direction of the estimates for their governorates is entirely driven by the spatial component. Figure \ref{fig::bird_spatial_outcome_residual} displays the observed number of recruits per area, along with the residual for each governorate.} The spatially autocorrelated component dominates the governorate-level variance, as shown by the posterior of mixing parameter $\lambda$ in Figure \ref{fig:bird_lambda}, estimated via Monte-Carlo mean at close to $0.9$, suggesting around $90\%$ of the variance at the governorate level can be explained by the ICAR model.\footnote{The spatial distribution of point estimates for governorate and country-level random effects are presented in Figure \ref{fig:bird_spatial_effects}.}

\begin{figure}[t!]
\begin{subfigure}{0.5\textwidth}
\includegraphics[width=\linewidth]{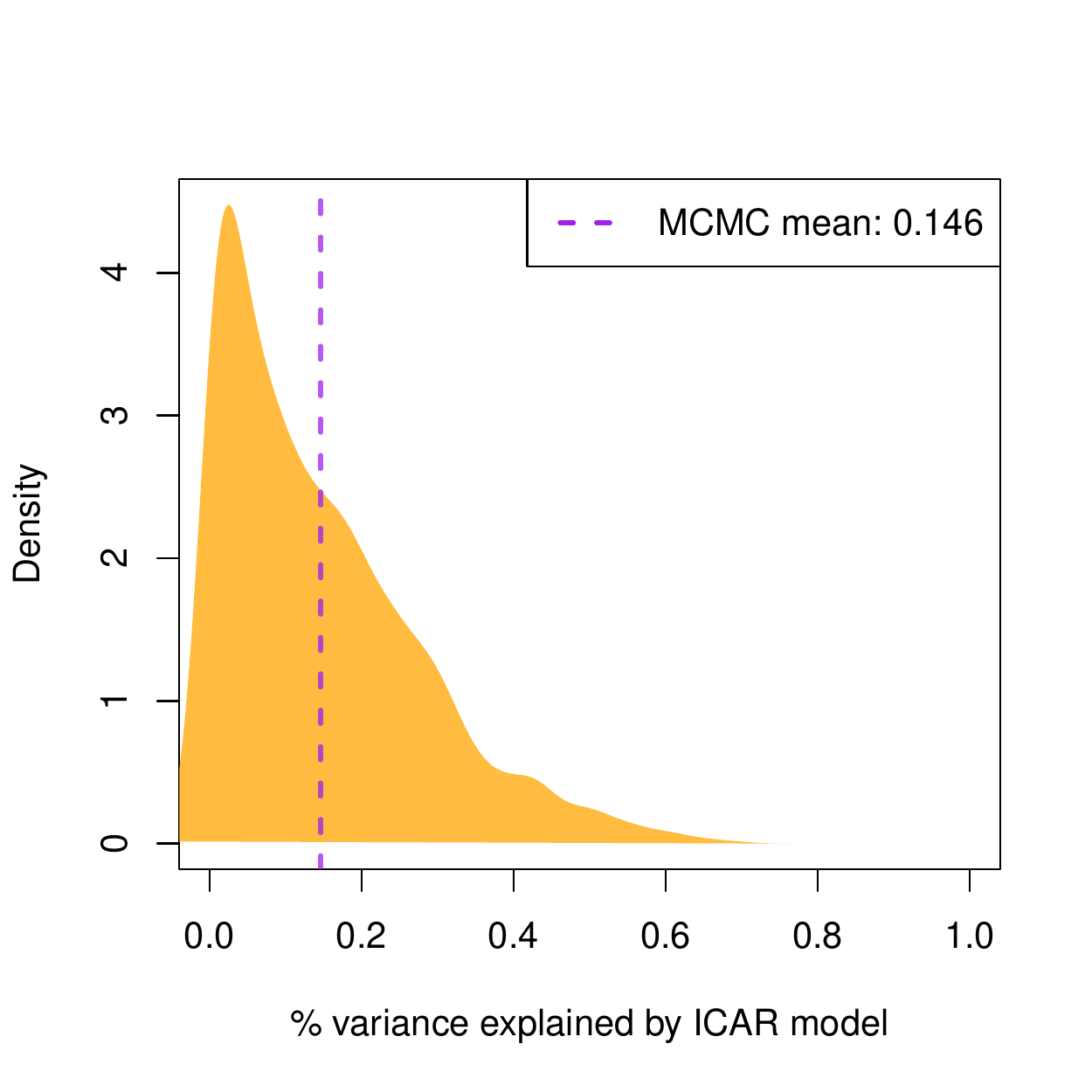}
\caption{} 
\end{subfigure}\hspace*{\fill}
\begin{subfigure}{0.5\textwidth}
\includegraphics[width=\linewidth]{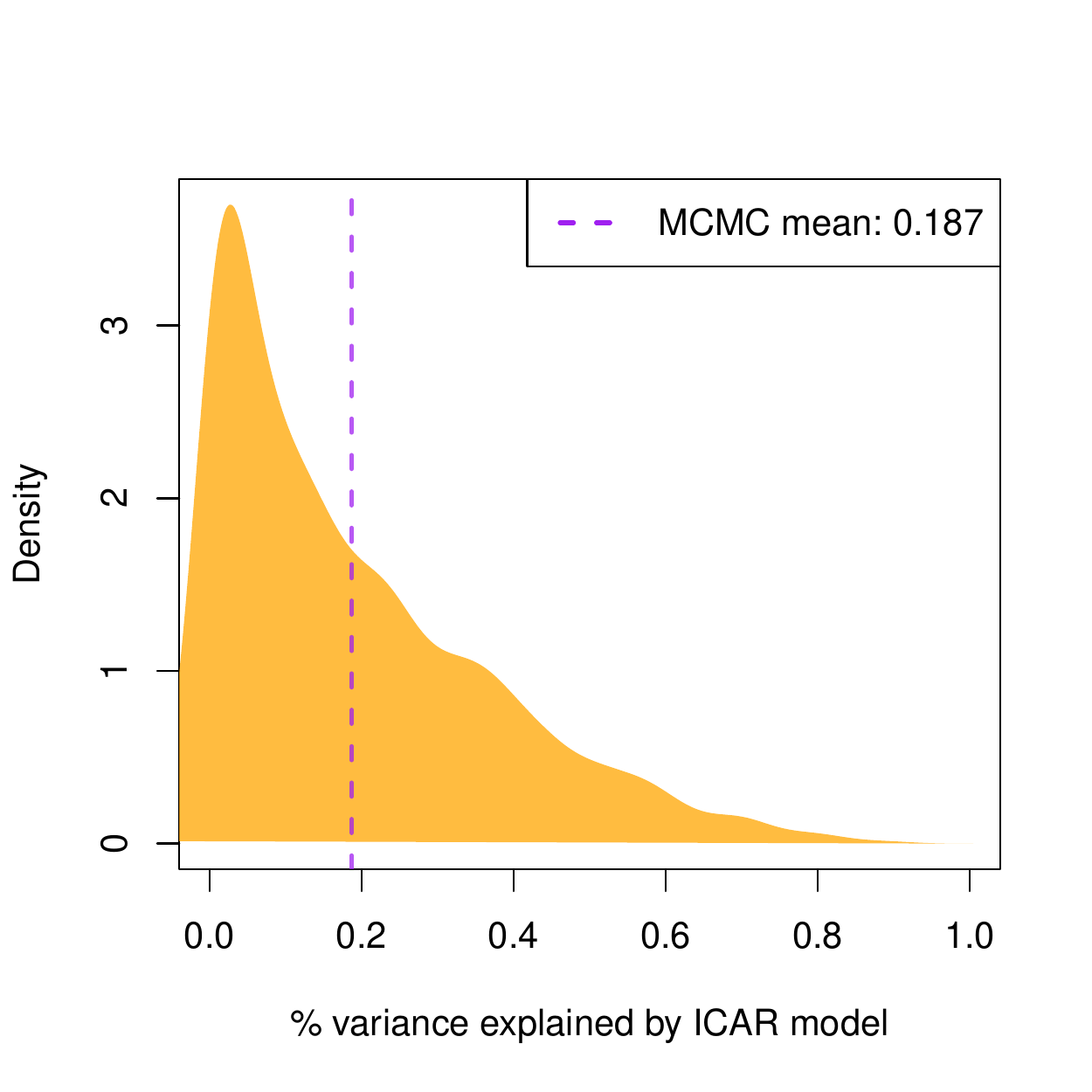}
\caption{} 
\end{subfigure}
\caption{District-level variance mixing parameter - $\lambda$ - for Egypt (a) and Tunisia (b).} \label{fig::worm_lambda}
\end{figure}

We repeat these analyses for Egypt and Tunisia. Figure \ref{fig::worm_lambda} shows similar mixing among spatial and non-spatial components for the two countries, with around $15\%$ of the district-level variance in Egypt being explained by spatial patterns, and $19\%$ in Tunisia. It is noteworthy that very few of our contextual variables have explanatory power for predicting recruitment. Coupled with the low percentage of variance being explained by the spatial components, our analysis suggests that, in spite of our best efforts to account for observable heterogeneity, there exist a vast array of unobserved, non-spatial district-level effects, which accounts for over $80\%$ of the unexplained district-level variance in both Egypt and Tunisia. Hence this contextual variance, while properly accounted for, remains unexplained. In the Appendix we also describe Moran's I statistics for the Worm's Eye analysis as well as point estimates for the district and governorate effects in Egypt and Tunisia. 

\subsection{Predicted propensity of recruitment by profile}
To conclude our analysis, we present inferences derived from the posterior predictive distribution of the out-of-sample probability of recruitment, focusing on individual-level characteristics. We produce descriptive plots on the log-odds scale, to aid interpretation and avoid the intercept-trap which would collapse all propensities to zero; plots on the relative probability and absolute probability are also presented in the Appendix, enabling the reader to appreciate the sharp non-linear `jump' which characterises the recruitment profiles of different socio-demographc groups. Note further that these predicted probabilities differ from the within-sample propensities, in that they are calculated after removing the offset, hence generalized to apply to any sampling protocol, as opposed to the specific case-control setting of our application.

\begin{figure}[htp]
    \centering
    \includegraphics[width = 1\textwidth]{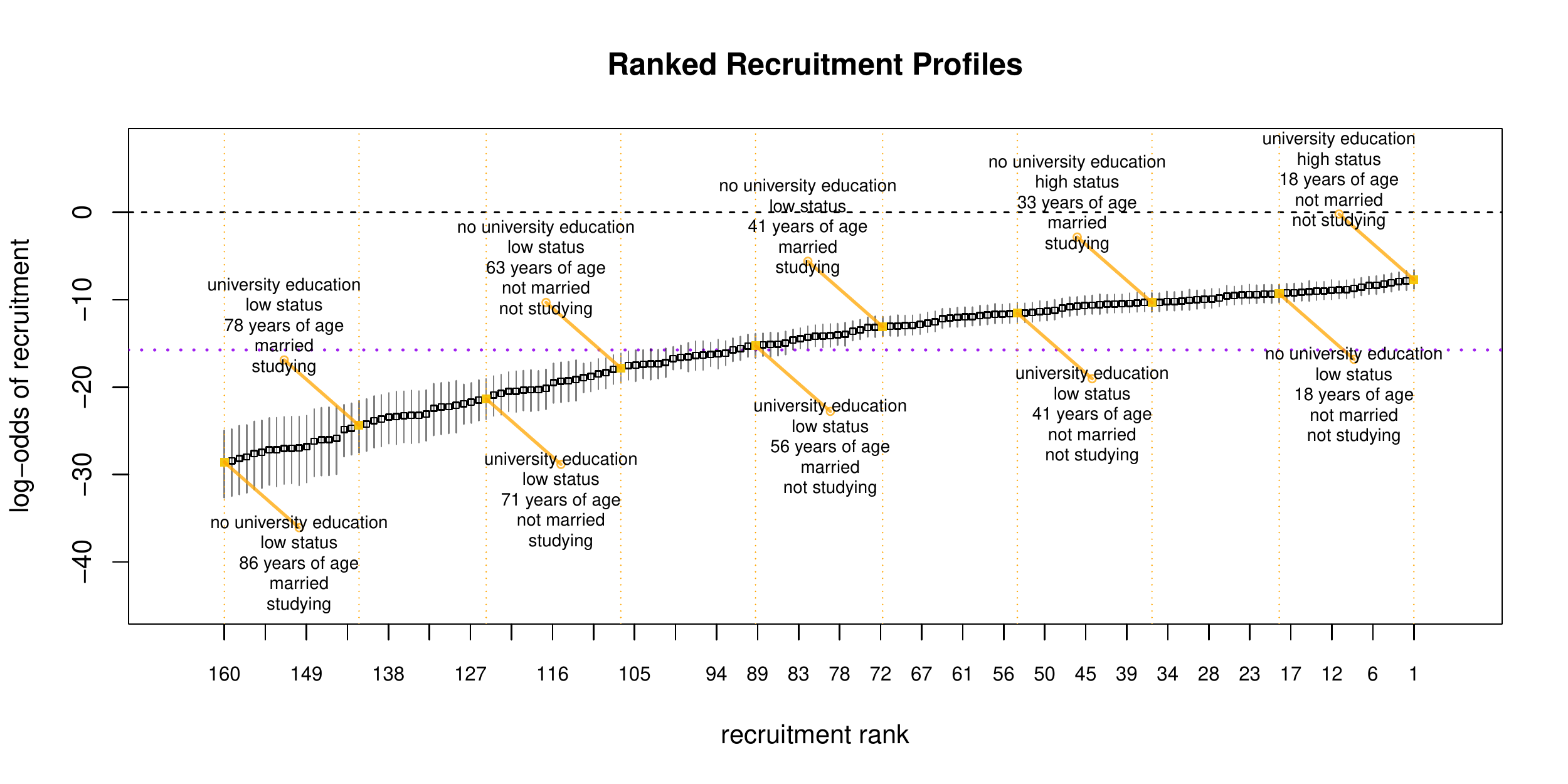}
    \caption{Distribution of the predicted probabilities prediction intervals, presented on the log-odds scale to aid cross-profile comparisons. The black dotted line highlights the the zero-log-odds point, whilst the purple dotted line notes a central estimate for the median recruitment propensity across profiles.}
    \label{figure::predicted_probabilities_all}
\end{figure}

What is the profile of individuals `at risk' of recruitment to ISIS according to our models? We attempt to answer this question by analyzing the predicted probabilities of all possible theoretical profiles, defined by the individual-level characteristics available in our data. Every profile is assumed to come from a hypothetical `average district'. Figure \ref{figure::predicted_probabilities_all} presents point estimates and prediction intervals for the log-odds of recruitment, over $160$ possible profiles in the Bird's eye model. Similar plots displaying the absolute and relative probabilities of recruitment are available in Figures \ref{figure::predicted_probabilities_bird_prob} and \ref{figure::predicted_probabilities_bird_odds} in the Appendix. Table \ref{table::predicted_probabilities_all_1pct} presents the profiles of the top $10$ most likely profiles to be recruited, providing four useful metrics to interpret the results: predicted probability; predicted rate per $10,000$ people; predicted odds relative to the average profile; and log-odds.
\vspace{10pt}

\begin{spacing}{1}
\begin{table}[htp]
\centering
\scalebox{0.62}{
\begin{tabular}{c|ccccc|cccc}
  \hline
  \hline
 Rank & Married & Student & University Edu & Low-Status & Age & $\hat{\mbox{P}}(r = 1 \mid X = x)$ & Predicted Rate & $\frac{\hat{\mbox{P}}(r = 1 \mid X = x)}{\hat{\mbox{P}}(r = 1 \mid X = \bar{X})}$ & $\mbox{logit}\left(\hat{\mbox{P}}(r = 1 \mid X = x)\right)$ \\ 
  \hline
1 &       0 &       0 &       1 &       0 &      18 & 0.000460 & 5/10000 & 22.561 & -7.684 \\ 
  2 &       1 &       0 &       1 &       0 &      18 & 0.000398 & 4/10000 & 19.273 & -7.828 \\ 
  3 &       0 &       0 &       0 &       0 &      18 & 0.000376 & 4/10000 & 18.632 & -7.885 \\ 
  4 &       1 &       0 &       0 &       0 &      18 & 0.000324 & 3/10000 & 16.013 & -8.036 \\ 
  5 &       0 &       0 &       1 &       0 &      26 & 0.000281 & 3/10000 & 13.745 & -8.177 \\ 
  6 &       1 &       0 &       1 &       0 &      26 & 0.000241 & 2/10000 & 11.763 & -8.331 \\ 
  7 &       0 &       0 &       0 &       0 &      26 & 0.000230 & 2/10000 & 11.287 & -8.379 \\ 
  8 &       1 &       0 &       0 &       0 &      26 & 0.000197 & 2/10000 & 9.695 & -8.530 \\ 
  9 &       0 &       1 &       1 &       0 &      18 & 0.000168 & 2/10000 & 8.295 & -8.690 \\ 
  10 &       1 &       1 &       1 &       0 &      18 & 0.000147 & 1/10000 & 7.211 & -8.828 \\ 
   \hline
   \hline
\end{tabular}
}
\caption{Top 10 recruitable theoretical profiles according to the Bird's eye model. Profiles are ordered by predicted probability of recruitment net of sampling protocol. $10$ ages are evaluated, starting at $18$ (to avoid non-existent profiles) and ending at the largest observed age ($86$). The last four columns represent respectively: i. the predicted probability of recruitment; ii. the predicted rate of recruitment per $10,000$ people; iii. the predicted odds of recruitment, relative to the `average' profile; iv. the log-odds of recruitment.}
\label{table::predicted_probabilities_all_1pct}
\end{table}
\end{spacing}{}

From the Bird's Eye prediction intervals we notice that the predicted probability of recruitment is centered around $-15$ on the log-odds scale, again underscoring how rare the event in question is, and how few are the profiles of people who are susceptible to recruitment. A select number of profiles approach a predicted probability around $-7$, and translate to meaningful rates of recruitment; these are highlighted in the predicted-probabilities table, which show the $10$ most recruitable profiles. Looking at Table \ref{table::predicted_probabilities_all_1pct}, we can say that the most likely recruitable profile (loosely characterised as a young, high-status, Sunni male with some university education who is unmarried and not currently studying), is around $23$ times as likely to be recruited as an average Sunni male from an average area in the MENA. Moreover, for every $10,000$ members of the most recruitable profile across the region, we expect $5$ to have joined ISIS. 
It is worthwhile to note that, in agreement with Figure \ref{figure::relative_deprivation}, all the most recruitable profiles are high-status individuals, and a majority of them has had some university education. Unsurprisingly, all of these profiles are under-25, and not currently studying. 

\begin{figure}[H]
    \centering
    \includegraphics[width = 1\textwidth]{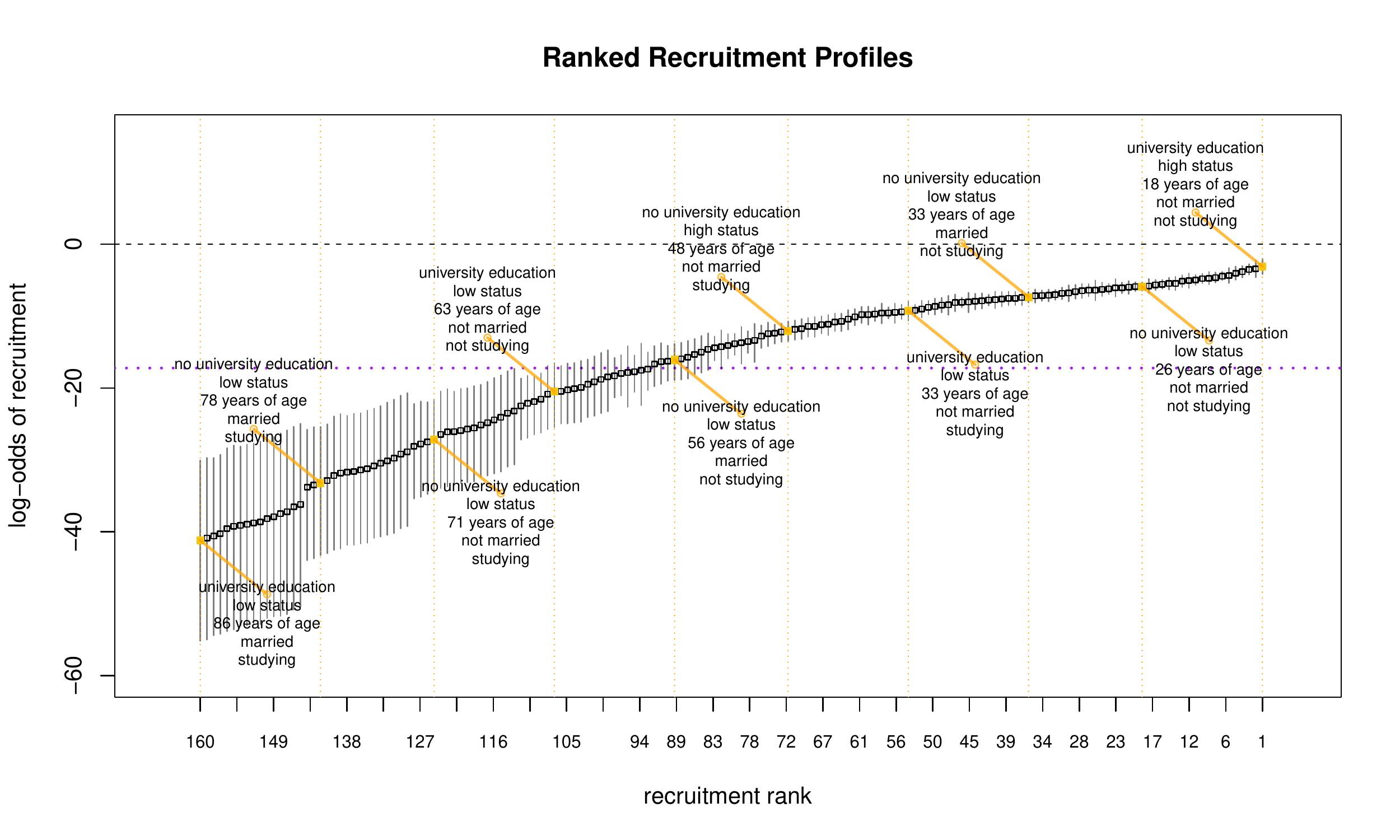}
    \caption{Worm's Eye (Tunisia) distribution of the predicted probabilities prediction intervals, presented on the log-odds scale to aid cross-profile comparisons. The black dotted line highlights the the zero-log-odds point, whilst the purple dotted line notes a central estimate for the median recruitment propensity across profiles.}
    \label{figure::predicted_probabilities_tunisia_logit}
\end{figure}

\begin{figure}[H]
    \centering
    \includegraphics[width = 1\textwidth]{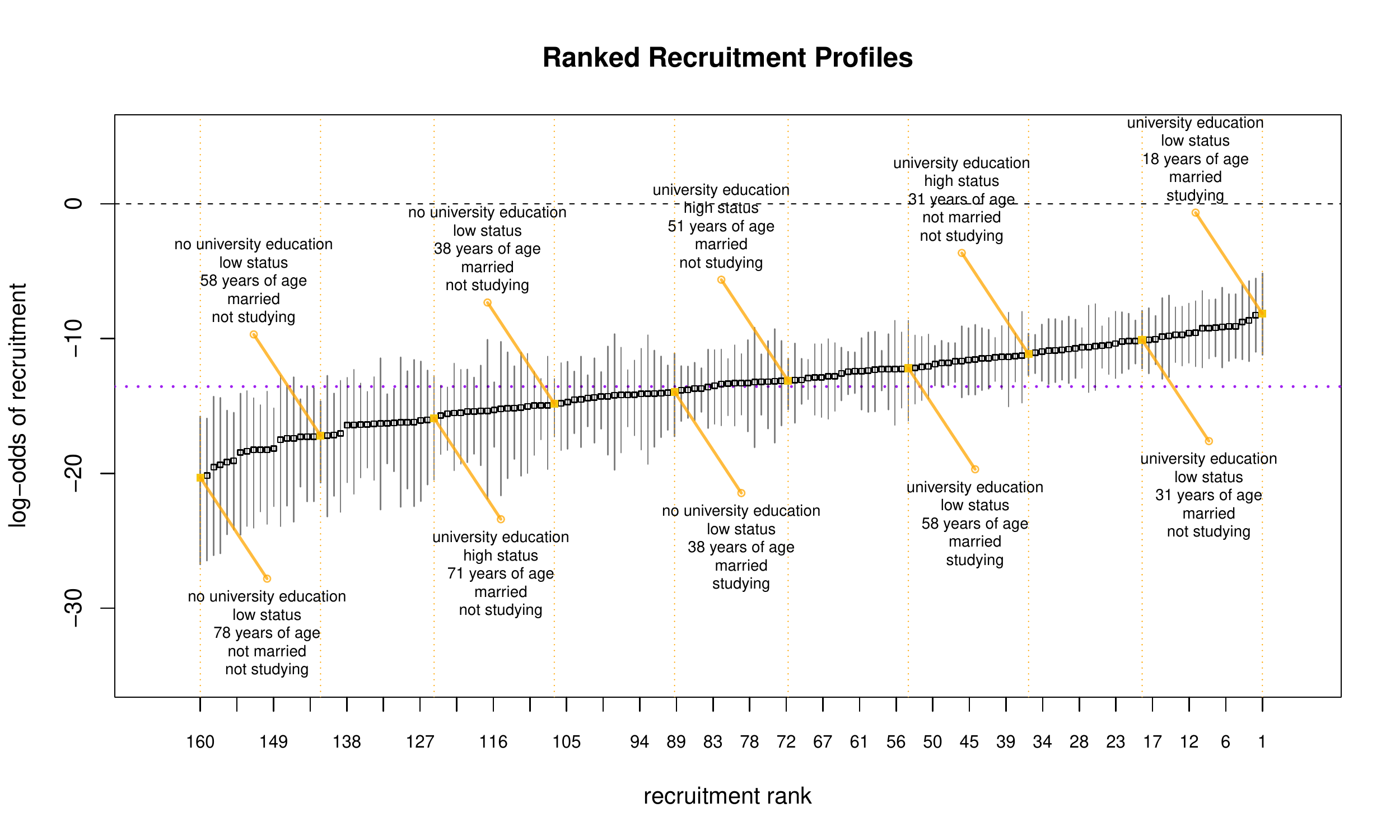}
    \caption{Worm's Eye (Egypt) distribution of the predicted probabilities prediction intervals, presented on the log-odds scale to aid cross-profile comparisons. The black dotted line highlights the the zero-log-odds point, whilst the purple dotted line notes a central estimate for the median recruitment propensity across profiles.}
    \label{figure::predicted_probabilities_egypt_logit}
\end{figure}

The individual-level profiles highlighted at the Bird's eye level are comparable to those highlighted within Tunisia in the Worm's eye level (Figure \ref{figure::predicted_probabilities_tunisia_logit} and Table \ref{table::predicted_probabilities_tunisia_1pct}), whilst the Egypt analysis reveals more evidence for the relative-deprivation hypothesis, with a majority of the highly-recruitable profiles being relatively-deprived (Figure \ref{figure::predicted_probabilities_egypt_logit} and Table \ref{table::predicted_probabilities_egypt_1pct}.\footnote{For absolute and relative probabilities of recruitment from the Worm's eye models, see Appendix Figures \ref{figure::predicted_probabilities_egypt_prob}, \ref{figure::predicted_probabilities_egypt_odds}, \ref{figure::predicted_probabilities_tunisia_prob}, and \ref{figure::predicted_probabilities_tunisia_odds}.} The relative recruitability of the most susceptible profiles in Egypt and Tunisia is also greater. In Egypt, the most recruitable profile (loosely characterised as a young, low-status, Sunni male with some university education who is married and is currently studying) is around $157$ times as likely to be recruited as the average Egyptian Sunni male -- though the expected rate of recruitment for this group is lower than for the Bird's Eye models' most recruitable profile, clocking in at $3$ recruits for every $10,000$ such individuals. The Egypt-specific propensity of recruitment is dramatically lower than that of Tunisia, again highlighting the role of contextual effects. In Tunisia, the most recruitable profile  (loosely characterised as a young, high-status, Sunni male who has university education, is unmarried and is not currently studying) has a probability of recruitment equivalent to $0.04$ on the log-odds scale. This profile is over $335$ times as likely as the average Tunisian Sunni male to be recruited, highlighting that though recruitment is still relatively rare in the population, is significantly concentrated in the top recruitment profiles. This is also supported by Figure \ref{figure::predicted_probabilities_tunisia_prob}, which shows only a handful of profiles have predicted probabilities above $\frac{1}{100}$.

\begin{spacing}{1}
\begin{table}[H]
\centering
\scalebox{0.62}{
\begin{tabular}{c|ccccc|cccc}
  \hline
  \hline
 Rank & Married & Student & University Edu & Low-Status & Age & $\hat{\mbox{P}}(r = 1 \mid X = x)$ & Predicted Rate & $\frac{\hat{\mbox{P}}(r = 1 \mid X = x)}{\hat{\mbox{P}}(r = 1 \mid X = \bar{X})}$ & $\mbox{logit}\left(\hat{\mbox{P}}(r = 1 \mid X = x)\right)$ \\ 
  \hline
1 &       0 &       0 &       1 &       0 &      18 & 0.043680 & 437/10000 & 335.663 & -3.086 \\ 
  2 &       1 &       0 &       1 &       0 &      18 & 0.031493 & 315/10000 & 242.502 & -3.426 \\ 
  3 &       0 &       0 &       1 &       0 &      26 & 0.028925 & 289/10000 & 222.013 & -3.514 \\ 
  4 &       1 &       0 &       1 &       0 &      26 & 0.021177 & 212/10000 & 159.752 & -3.833 \\ 
  5 &       0 &       0 &       0 &       0 &      18 & 0.017397 & 174/10000 & 137.069 & -4.034 \\ 
  6 &       1 &       0 &       0 &       0 &      18 & 0.012615 & 126/10000 & 98.539 & -4.360 \\ 
  7 &       0 &       0 &       0 &       0 &      26 & 0.011647 & 116/10000 & 88.370 & -4.441 \\ 
  8 &       0 &       0 &       1 &       0 &      33 & 0.009503 & 95/10000 & 73.103 & -4.647 \\ 
  9 &       0 &       1 &       1 &       0 &      18 & 0.008662 & 87/10000 & 68.471 & -4.740 \\ 
  10 &       1 &       0 &       0 &       0 &      26 & 0.008327 & 83/10000 & 64.331 & -4.780 \\ 
   \hline
   \hline
\end{tabular}
}
\caption{Top $10$ recruitable theoretical profiles according to the Tunisia `Worm's Eye' model. Profiles are ordered by predicted probability of recruitment net of sampling protocol. $10$ ages are evaluated, starting at $18$ (to avoid non-existent profiles) and ending at the largest observed age ($86$). The last four columns represent respectively: i. the predicted probability of recruitment; ii. the predicted rate of recruitment per $10,000$ people; iii. the predicted odds of recruitment, relative to the `average' profile; iv. the log-odds of recruitment}
\label{table::predicted_probabilities_tunisia_1pct}
\end{table}
\end{spacing}{}

\begin{spacing}{1}
\begin{table}[H]
\centering
\scalebox{0.62}{
\begin{tabular}{c|ccccc|cccc}
  \hline
  \hline
 Rank & Married & Student & University Edu & Low-Status & Age & $\hat{\mbox{P}}(r = 1 \mid X = x)$ & Predicted Rate & $\frac{\hat{\mbox{P}}(r = 1 \mid X = x)}{\hat{\mbox{P}}(r = 1 \mid X = \bar{X})}$ & $\mbox{logit}\left(\hat{\mbox{P}}(r = 1 \mid X = x)\right)$ \\ 
  \hline
1 &       1 &       1 &       1 &       1 &      18 & 0.000287 & 3/10000 & 156.721 & -8.155 \\ 
  2 &       0 &       1 &       1 &       1 &      18 & 0.000258 & 3/10000 & 141.021 & -8.264 \\ 
  3 &       1 &       1 &       1 &       1 &      25 & 0.000173 & 2/10000 & 97.486 & -8.660 \\ 
  4 &       0 &       1 &       1 &       1 &      25 & 0.000155 & 2/10000 & 87.536 & -8.772 \\ 
  5 &       1 &       0 &       1 &       1 &      18 & 0.000113 & 1/10000 & 62.422 & -9.091 \\ 
  6 &       1 &       1 &       1 &       0 &      18 & 0.000112 & 1/10000 & 62.529 & -9.096 \\ 
  7 &       1 &       1 &       1 &       1 &      31 & 0.000108 & 1/10000 & 59.223 & -9.130 \\ 
  8 &       0 &       0 &       1 &       1 &      18 & 0.000101 & 1/10000 & 56.978 & -9.199 \\ 
  9 &       0 &       1 &       1 &       0 &      18 & 0.000098 & 1/10000 & 53.336 & -9.229 \\ 
  10 &       0 &       1 &       1 &       1 &      31 & 0.000097 & 1/10000 & 52.977 & -9.244 \\ 
   \hline
   \hline
\end{tabular}
}
\caption{Top 10 recruitable theoretical profiles according to the Egypt `Worm's Eye' model. Profiles are ordered by predicted probability of recruitment net of sampling protocol. $10$ ages are evaluated, starting at $18$ (to avoid non-existent profiles) and ending at the largest observed age ($78$). The last four columns represent respectively: i. the predicted probability of recruitment; ii. the predicted rate of recruitment per $10,000$ people; iii. the predicted odds of recruitment, relative to the `average' profile; iv. the log-odds of recruitment}
\label{table::predicted_probabilities_egypt_1pct}
\end{table}
\end{spacing}{}

\section{Conclusion}
Extreme forms of political behaviour are rarely ever committed by more than a tiny subsection of any given national population. Despite their small size, these groups often have an outsized influence on state and international politics. \textit{Because} of their small size, extremists are particularly hard to study using conventional statistical methods and research designs.

To address this, we propose that extremism researchers adopt a new variant of the case-control design taken from epistemology. This approach allows us to combine survey techniques with ecological forms of analysis, allowing for meaningful comparisons with the underlying populations from which recruits are drawn. To implement this, we solve a number of statistical problems when explaining rare and extreme forms of political behaviour. In particular, we build on existing contaminated case-control designs to demonstrate: 1) how best to incorporate area-level random effects when the number of recruits for a given unit is small; 2) how to account for spatial autocorrelation in this setup; 3) how to regularize coefficients to guard against separation; and 4) provide the \texttt{extremeR} software package so that extremism researchers working on a range of different cases can easily apply our models. Following this approach, we are able to robustly recover individual fixed effects for relevant demographic characteristics net of context, as well as the proportion of unexplained variance attributable to unobserved spatial effects. Finally, we provide simulations to help benchmark our approach and inform the model selection of future researchers.

We demonstrate the potential of our technique with data on ISIS recruits across nine countries in the MENA. Here, the paper makes an important empirical contribution to the literature on Islamist extremism. We find confirmation that recruits tend to have a high social status and are more likely to be highly educated -- and we have mixed support for the relative deprivation hypothesis. Despite the considerable advances our approach represents, we are still constrained by the sparsity of information available for certain countries in our sample. Future work might extend our analysis by introducing contextual information across all countries for which have a sufficient number of records for recruits.

Introducing finer-grain contextual data is especially appropriate given that we find a large proportion of variance in the incidence of recruitment to Islamist extremism is attributable to unexplained ecological and spatial confounding. A broader implication here is that future work should focus on obtaining precise estimates of potential local-level confounders. Often, we have only limited data on the characteristics of individual cases, and the characteristics of controls are naturally constrained by the richness of the information we have for our cases. In this regard, introducing richer ecological measures, which is often more easily accessible, could yield important insights, by illuminating the role of contextual factors net of individual-level drivers. Here, promising avenues include the measurement of sub- and cross-national recruitment networks (e.g., \citealt{rosenblatt_localism_2020}), as well as the use of nontraditional data sources to derive granular, time-varying ecological data (e.g., \citealt{chi_micro-estimates_2021}). 

\pagebreak

\bibliography{isis.bib}
\pagebreak

\renewcommand \thepart{}
\renewcommand \partname{}
\appendix
\pagenumbering{arabic}
\setcounter{page}{1}
\renewcommand*{\thepage}{A\arabic{page}}
\addcontentsline{toc}{section}{Appendix} 
\part{Appendix} 
\parttoc 
\thispagestyle{empty}

\clearpage
\pagenumbering{arabic}
\counterwithin{table}{section}
\counterwithin{figure}{section}
\begin{spacing}{1}

\newpage	
	



\section{Independent variable details}\label{appendix:independent_variables}

\begin{table}[htbp]
\label{betab}
\footnotesize
\caption{Individual-level variable codings across border documents and survey data}
\begin{tabularx}{\linewidth}{>{\hsize=0.4\hsize}X >{\hsize=0.4\hsize}X >{\hsize=0.6\hsize}X}   
Variable & Border Documents & ABIII\\
\midrule
coledu & 1 if Education level mentions "university" & 1 if q1003 $>$5 (or $>$4 for Tunisia; $>$6 for Yemen)\\
age & Date of entry - Date of birth & q1001\\
married & 1 if Marital status is "married" & q1010\\
student & 1 if Occupation prior to arrival is "student" & q1004 = 3 (Student)\\
lowstat & 1 if Occupation prior to arrival is agricultural or manual/unemployed & q1004 = 5 (Unemployed) or q1010 = 4/5 (Agricultural or manual worker)\\
\bottomrule
\end{tabularx}
\end{table}

\begin{table}[htbp]
\caption{Egypt district-level covariates}
\label{egytab}
\footnotesize
\begin{tabularx}{\linewidth}{>{\hsize=0.4\hsize}X >{\hsize=0.4\hsize}X >{\hsize=0.6\hsize}X}   
Variable & Details & Source\\
\midrule
Population density & number of individuals in district/district area in km\textasciicircum{}2 & 2006 Census \\
Population & number of individuals in district aged 10 or over & 2006 Census \\
\% Christian & percentage of individuals in district recorded as Christian & 2006 Census \\
\% University & percentage of individuals in district who are university educated & 2006 Census \\
\% Agriculture & percentage individuals employed in agriculture denominated by total active population & 2006 Census\\
\addlinespace
\% Mursi & percent of total votes in district for Muhammad Mursi in the first round of the 2012 presidential election & El-Masry and Ketchley (2021) \\
Unemployment rate & number individuals aged without employment denominated by total active population & 2006 Census\\
Killed at Rabaa & number of deaths of individuals from district at the 2013 Rabaa Massacre (square-rooted) & Ketchley and Biggs (2017)\\
Post-revolutionary protest & number of protests recorded in district in 12 months after Jan 25 Revolution (square-rooted) & Barrie and Ketchley (2019)\\
\bottomrule
\end{tabularx}
\end{table}

\begin{table}[htbp]
\small
\caption{Tunisia district-level covariates}
\label{tuntab}
\footnotesize
\begin{tabularx}{\linewidth}{>{\hsize=0.4\hsize}X >{\hsize=0.4\hsize}X >{\hsize=0.6\hsize}X}   
Variable & Details & Source\\
\midrule
Population & number of individuals in district aged 10 or over & 2014 Census\\
Population density & number of individuals in district aged 10 or over/district area in km\textasciicircum{}2 & 2014 Census\\
\% University & percentage population with higher education certificate denominated by total population & 2014 Census\\
\% Agriculture & percentage individuals employed in agriculture denominated by total active population aged 15 or over & 2014 Census\\
Unemployment rate & number individuals aged 18-59 without employment denominated by total active population aged 18-59 & 2014 Census\\
\addlinespace
Graduate unemployment rate & number individuals with higher education certificate without employment denominated by total active population aged 18-59 & 2014 Census\\
\% Ennahda 2011 & percentage of total votes in district for Ennahdha in 2011 election & INS Tunisia\\
\% Ennahdha 2014 & percentage of total votes in district for Ennahdha in 2014 election & INS Tunisia\\
Post-revolutionary protests & number of protests recorded in district in 12 months after Jan 14 Revolution (square-rooted) & Barrie and Ketchley (2019)\\
Distance to Libya & distance to Libyan border from centroid of target district (square-rooted) & NA\\
\bottomrule
\end{tabularx}
\end{table}

\newpage
\section{\texttt{Stan} listings}

\begin{lstlisting}[caption={\texttt{Stan} Data Declaration Block. },captionpos=t,label={lst:stan_dist}]
data{

     int<lower = 1> n;         // total number of observations
     int<lower = 1> p;         // number of covariates in design matrix
     int<lower = 0> y[n];         // vector of labels
     matrix[n, p] X;         // design matrix
         
     int<lower = 1> small_area_id[n];         // small-area id
     int<lower = 1> N_small_area;         // number of small areas
         
     int<lower = 1> N_small_area_edges;         // number of edges in the spatial process
     int<lower=1, upper=N_small_area> node1_small_area[N_small_area_edges]; // node1[i] adjacent to node2[i]
     int<lower=1, upper=N_small_area> node2_small_area[N_small_area_edges]; // node1[i] adjacent to node2[i]
         
     real scaling_factor;         // scaling factor derived from the adjacency matrix
         
     int<lower = 1> large_area_id[n];         // large-area ids
     int<lower = 1> N_large_area;         // number of large-areas
         
     vector[N_large_area] log_offset;         // log-scale offset
         
     matrix[2,N_large_area] theta;         // Pr(Y = 1 | r = 1, s = 1)
         
     }
\end{lstlisting}

\begin{lstlisting}[caption={\texttt{Stan} Parameters Declaration Block.},captionpos=t,label={lst:stan_dist1}]
parameters{

           // cauchy prior for individual-level coefficients expressed as scale mixture of gaussian density functions
           vector[p] aux_a;         // central component
           vector<lower = 0>[p] aux_b;         // scale component


           vector[N_small_area] phi;         // small-area unstructured effects
           vector[N_small_area] psi;         // small-area spatial effect

           real<lower = 0,upper = 1> lambda;         // mixing prior on spatial component

           real<lower = 0> sigma_gamma;         // small-area effect scale

           vector[N_large_area] eta;         // large-area unstructured effect

           real<lower = 0> sigma_eta;         // large-area effect scale
         
           }
\end{lstlisting}

\pagebreak

\begin{lstlisting}[caption={\texttt{Stan} Transformed Parameters Block.},captionpos=t,label={lst:stan_dist2}]
transformed parameters{

                       vector[p] beta = aux_a ./ sqrt(aux_b);         // individual-effect prior

                       vector[n] mu;         // expected propensity of recruitment 
         
                       vector[N_small_area] gamma = (sqrt(1-lambda) * phi + sqrt(lambda / scaling_factor) * psi)*sigma_gamma;         
                       // convolved small -area effect

                       mu = log_offset[large_area_id] + eta[large_area_id]*sigma_eta + gamma[small_area_id] + X * beta;         
                       // linear function of the logit -scale propensity to be a recruit
         
                       }
\end{lstlisting}

\begin{lstlisting}[caption={\texttt{Stan} Model Declaration Block. },captionpos=t,label={lst:stan_dist3}]
model{

      aux_a ~ normal(0,1);            // prior on the centrality of the cauchy prior
      aux_b[1] ~ gamma(0.5,100*0.5);            // prior on intercept-scale
      aux_b ~ gamma(0.5,0.5);            // prior on individual covariate scales
        
      target += -0.5 * dot_self(psi[node1_small_area] - psi[node2_small_area]);            
      // ICAR prior
        
      phi ~ normal(0,1);            // unstructured random effect on small -area 
      sum(psi) ~ normal(0, 0.01 * N_small_area);            
      // soft sum -to-zero , equivalent to mean(psi) ~ normal (0 ,0.01)
        
      lambda ~ beta(0.5,0.5);            // mixing weight prior
        
      sigma_gamma ~ normal(0,1);            // prior small-area scale
        
      eta ~ normal(0,1);            // prior large-area effect
        
      sigma_eta ~ normal(0,1);            // prior large-area scale
        
      // likelihood
      for (i in 1:n) { 
            target += log_mix(1-inv_logit(mu[i]),
                              bernoulli_lpmf(y[i] | theta[1,large_area_id[i]]),
                              bernoulli_lpmf(y[i] | theta[2,large_area_id[i]]));          
                              // labels distributed as mixture of bernoulli distributions
                     }
      }
\end{lstlisting}

\pagebreak

\section{Simulation study}\label{appendix:simulation_study}
In what follows we present a more detailed view of the setup and results of the simulation study. First, we create a data-generating function to draw sample-datasets generated according to the mechanism implied by either the rare-events or contaminated case-control model. We reduce the data generating process to its essence for simplicity: a single continuous covariate $x_i$ is considered, and large-area effects are dropped. Small-area effects are simulated according to a random intrinsic conditionally auto-regressive process from one of three widely-used maps.\footnote{$\mathcal{M} = \{ \mbox{\texttt{scotland\_lipcancer}},\mbox{\texttt{newyork\_lukemia}},\mbox{\texttt{pennsylvania\_lungcancer}} \}$}, available from the \texttt{R} package \texttt{SpatialEpi} \citep{kim2010r}. This enables the random sampling of ICAR effects whilst preserving a plausible geography (i.e. neighbourhood structure and distance between areal units). Pseudo-algorithm \ref{alg:dgf} describes the steps taken to generate the simulated data. 

\begin{algorithm}
\caption{A pseudo algorithm displaying the steps taken by the data generating function to generate a random sample of data.}\label{alg:dgf}
\begin{algorithmic}
\Require \State sample size: \hspace{95pt} $n \in [100,2000]$  
\State population prevalence: \hspace{40pt} $\pi \in \left[\frac{1}{1000000},\frac{1}{2}\right]$
\State expected sample prevalence:  \hspace{12.5pt} $\hat{\pi} \in [0.01,0.99]$  
\State global auto-correlation: \hspace{37.5pt} $I \in (0,1)$ 
\State map: \hspace{124pt} $\mathcal{M} \in \{ \mbox{\texttt{scotland}},\mbox{\texttt{newyork}},\mbox{\texttt{pennsylvania}} \}$\\
\State (0.) derive key quantities directly from inputs:\\

\hspace{30pt}i. expected number of case-labelled records:\\ \hspace{60pt}$ n_1 \leftarrow \mbox{ } n \times \hat{\pi}$\\
\hspace{30pt}ii. expected number of unlabelled records:\\\hspace{60pt} $ n_u  \leftarrow \mbox{ }  n-n_1$\\
\hspace{30pt}iii. relative prob. of sampling a case v. control:\\ \hspace{60pt}$ \frac{P_1}{P_0}  \leftarrow \mbox{ }  \frac{(n_1+\pi \times n_u)/\pi}{n_u}$\\
\hspace{30pt}iv. prob. of sampling a case-labelled record conditional on being a true control:\\ \hspace{60pt}$ \theta_0  \leftarrow \mbox{ }  0$\\
\hspace{30pt}v. prob. of sampling a case-labelled record conditional on being a true case:\\\hspace{60pt}$ \theta_1  \leftarrow \mbox{ }  \frac{n_1}{n_1+\pi \times n_u}$\\

\State (1.) sample area effects on selected map: \hspace{17.5pt} $\bm{\gamma} \sim \mbox{ICAR}(\mathcal{M})$ 
\State (2.) sample initial value for intercept: \hspace{30pt} $\beta_1 \sim N(0,1)$
\State (3.) sample covariate value: \hspace{83.5pt}  $x_i \sim N(0,1)$
\State (4.) sample covariate effect: \hspace{81.5pt}  $\beta_2 \sim N(0,1)$
\State (5.) optimise intercept to meet specified sample prevalence: \\
 \hspace{232.5pt} $\beta^{\star}_1 \leftarrow  \operatorname*{arg\,max}_{\beta_1} f(\hat{\pi};\beta_1)$
\State (6.) calculate latent recruitment propensity: \hspace{2pt} $\bm{\mu} \leftarrow \mbox{log}\left(\frac{P1}{P0}\right) + \beta_1 + \bm{x}\beta_2 + \bm{\gamma}$
\State (7.) calculate recruitment propensity: \hspace{36.5pt} $\bm{\rho}  \leftarrow \mbox{\texttt{inv\_logit}}(\bm{\mu})$
\State (8.) sample recruitment status: \hspace{70pt} $\bm{r}  \sim \mbox{Bernoulli}(\bm{\rho})$
\State (9.) sample labels: \hspace{135pt} $\bm{y} \sim \mbox{Bernoulli}(\theta_{\bm{r}}) $

\end{algorithmic}
\end{algorithm}
We simulate \texttt{n.sims} = $200$ datasets\footnote{In practice we simulate datasets in two stages: first we examine the model performance by sampling $100$ draws from a `rare event' process ($\pi \in [1/1000000,1/10]$); then, in a second stage, we sample another $100$ draws from a process with less extreme prevalence ($\pi \in [1/10,1/2]$). This is done to evaluate performance in two different scenarios -- extreme (rare-event) v. non-extreme -- and ensure a large-enough sample size to capture salient dynamics in both.} using the data-generating function. The inputs to the function (highlighted under the `Required' header in the pseudo-code) are sampled at random from uniform distributions conforming to the specified range for each input - in the case of the maps, a map is chosen at random amongst the three candidates.

\begin{figure}[ht]
    \centering
    \includegraphics[scale = 0.4]{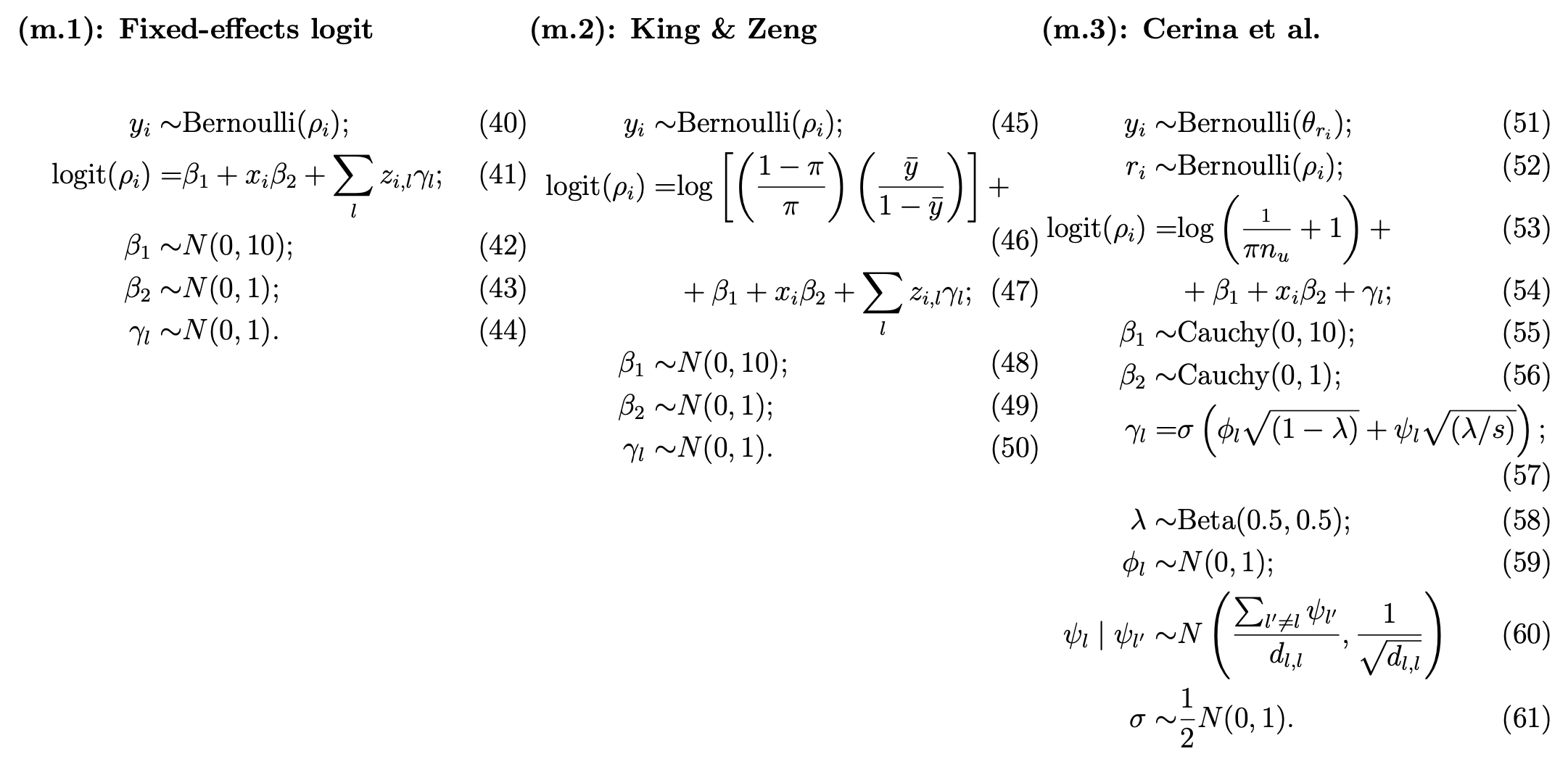}
    \caption{Hierarchical formulation of the three competing models considered in the simulation study.}
    \label{eq:sims.model}
\end{figure}

A second step is to define the models analysed in this simulation study. The candidate models are: \textbf{m.1} -- a simple fixed-effects logistic regression, where the area-effects are also estimated via fixed-effects; \textbf{m.2} -- similar to \textbf{m.1}, but importantly augmented with the use on an offset (prior-correction) a-la \citep{king2001logistic}; \textbf{m.3} -- an essential version of our rare-events, Bayesian contaminated-controls model with a BYM2 area-effects prior. The models are detailed in Figure \ref{eq:sims.model}.

In order to fit the $600$ models necessary for this simulation study, we have to `live dangerously'\footnote{As others have done before when frequently fitting complex \texttt{Stan} models on large datasets -- see \citep{lauderdale2020model} for an example running multiple short chains.}, and lower our expectations over the stringent convergence properties of any given model. What we are interested in is the stability of the simulation results, and this is an aggregate set of quantities which is relatively robust to the semi-convergence of any given model. We therefore run each model in \texttt{Stan}, with the following settings: $\mbox{\texttt{n.cores}}=4$; $\mbox{\texttt{n.chains}}=4$; $\mbox{\texttt{n.thin}}=4$; $\mbox{\texttt{n.iter}}=4000$; $\mbox{\texttt{n.warmup}}=\frac{2}{3} \mbox{\texttt{n.iter}}$; all other settings are set to the \texttt{Stan} default. This gives us posterior samples which have relatively small effective sample-sizes, but are nevertheless able to give us reliable central-estimates for the parameters of interest -- proof of this is that the results from the simulation study are replicable over multiple samples. Note that very rarely the chains will diverge for m.3 under high-levels of contamination. When this happens, we drop these simulations from the analysis and re-run the model.\\

Finally, we define the parameters of the comparison. The simulations are intended to investigate the ability of competing models to estimate the following quantities of interest:  $\bm{\mu}^\star = \bm{\mu} - \mbox{log}\left(\frac{P_1}{P_0}\right)$, the latent propensity to be a recruit; $\beta_1$, the baseline propensity to be a recruit; $\beta_2$, the effect of simulated covariate $\bm{x}$; $\bm{\gamma}$, the set of area-level effects which contribute to the latent propensity.  $\bm{\mu}^\star$ is a good summary metric of performance on all of these dimensions, so our primary inference refers to this quantity, though see the Appendix for a more detailed analysis. The models are scored only on their point-estimates, as an evaluation of uncertainty is computationally unfeasible due to the large number of MCMC iterations necessary to obtain convergent estimates of the second-moment for all these parameters.

\begin{figure}[ht]
    \centering
    \includegraphics[scale = 0.5]{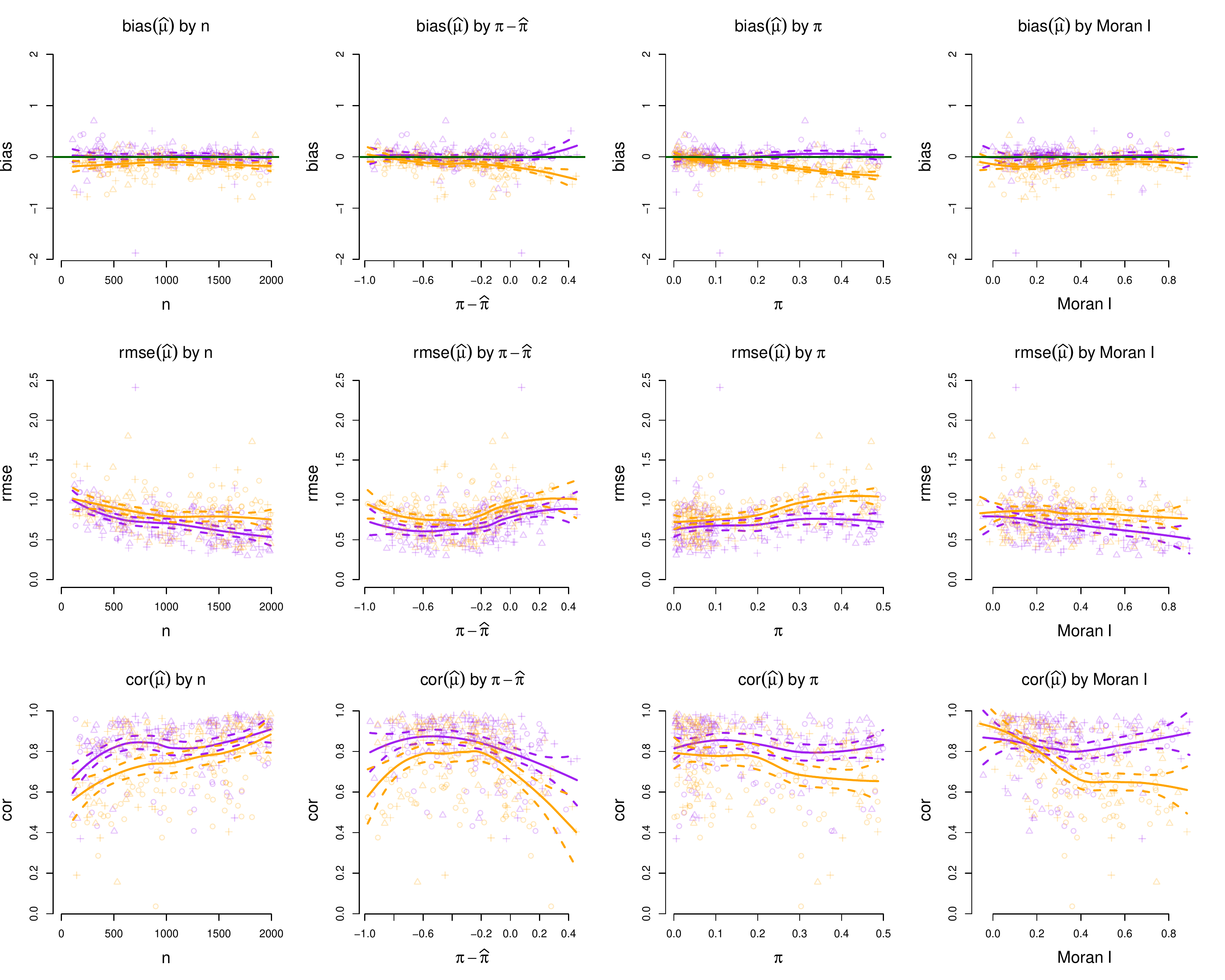}
    \caption{Results of the simulation study, comparing the performance of our model (m.3, in purple) and the more traditional rare-events logistic regression with prior-correction for the intercept \citep{king2001logistic} (m.2, in orange) in estmating the true latent propensity $\mu^\star = \mu - \mbox{log}(\frac{P_1}{P_0})$.}
    \label{fig:simulations_KingZeng}
\end{figure}

The models are generating parameter estimates $\hat{f}$ to approximate the true simulated parameters $f$; they are scored on three dimensions: i. $\mbox{bias}(\hat{f}) = \frac{1}{n}\sum_i \hat{f}_i - f_i$ ; ii. Root-mean-square-error $\mbox{RMSE}(\hat{f}) = \frac{1}{n}\sum_i (\hat{f}_i - f_i)^2$; iii. Pearson correlation coefficient $r(\hat{f}) = \frac{\sum_i (\hat{f}_i - \bar{\hat{f}}_i) (f_i - \bar{f}_i)}{\sqrt{\sum_i (\hat{f}_i - \bar{\hat{f}}_i)^2 \sum_i (f_i - \bar{f}_i)^2}}$. The bias tells us the average direction of the estimation error; the RMSE tells us about the average magnitude of the error, penalising large deviations more heavily than smaller-ones; the Pearson correlation tells us about the ability of the model to correctly order (rank) the parameters. Figure \ref{fig:simulations_KingZeng} presents a comparison of m.2 and m.3 in their ability to estimate latent propensity $\bm{\mu}^\star$, for each scoring function (on the y-axis) across key characteristics of the data (on the x-axis). A comparison including m.1 is initially omitted here as the scale of the errors in m.1 is so large that it makes it visually impossible to distinguish between the (otherwise substantial) differences in m.2 and m.3 performance. A complete plot including m.1 is available below.\\

\begin{figure}
    \centering
    \includegraphics[scale = 0.5]{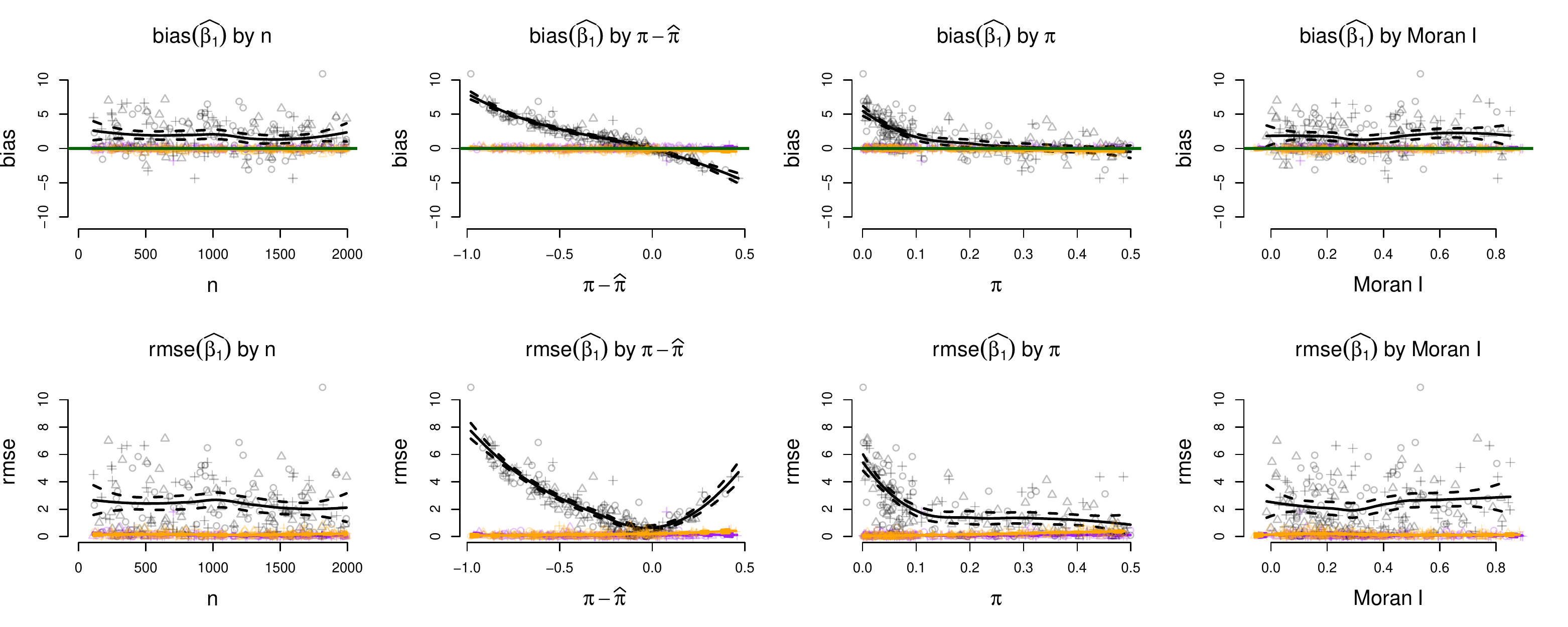}
    \caption{Results from the simulation study, capturing the ability of the simple fixed effects model (m.1, in black), the King \& Zeng model (m.2, in orange) and our proposed approach (m.3, in purple) to estimate the true intercept $\beta_1$.}
    \label{fig:simulations_intercept_full}
    \includegraphics[scale = 0.5]{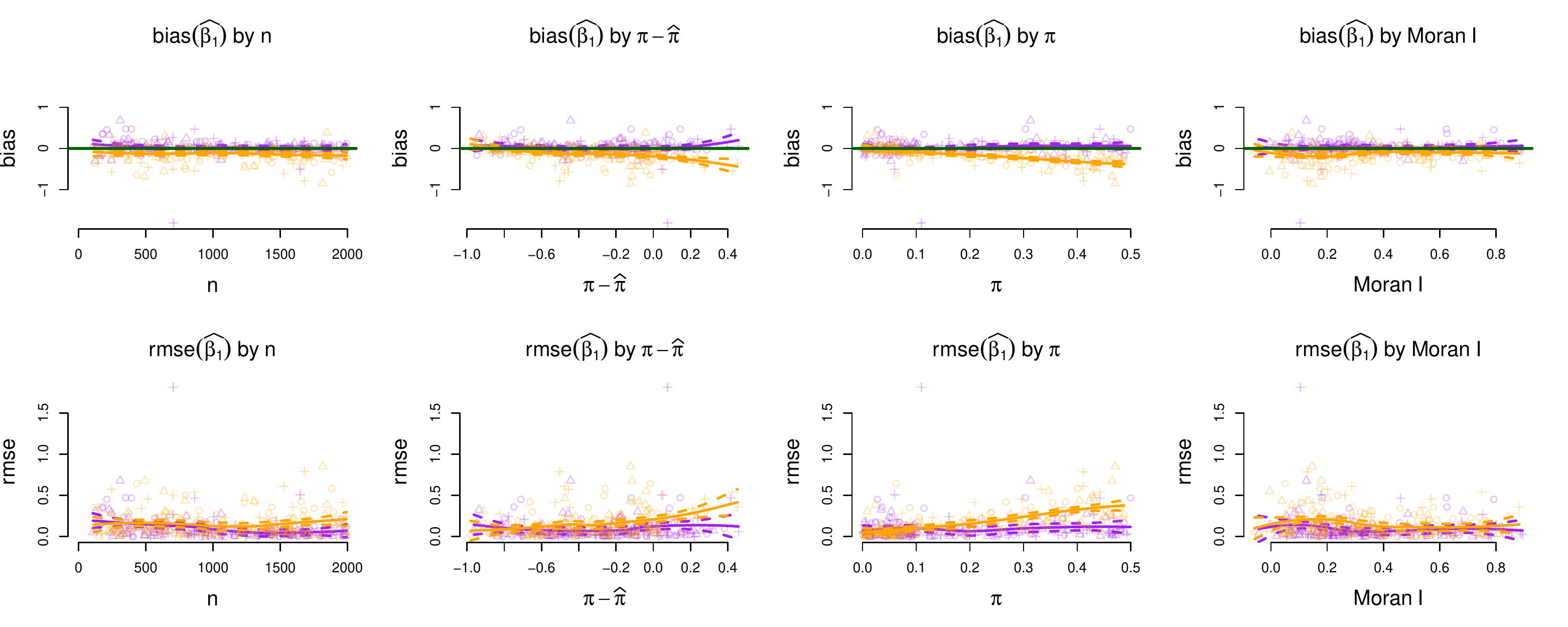}
    \caption{Results from the simulation study, capturing the ability of  the King \& Zeng model (m.2, in orange) and our proposed approach (m.3, in purple) to estimate the true intercept $\beta_1$.}
    \label{fig:simulations_intercept}
\end{figure}

\begin{figure}
    \centering    
    \includegraphics[scale = 0.5]{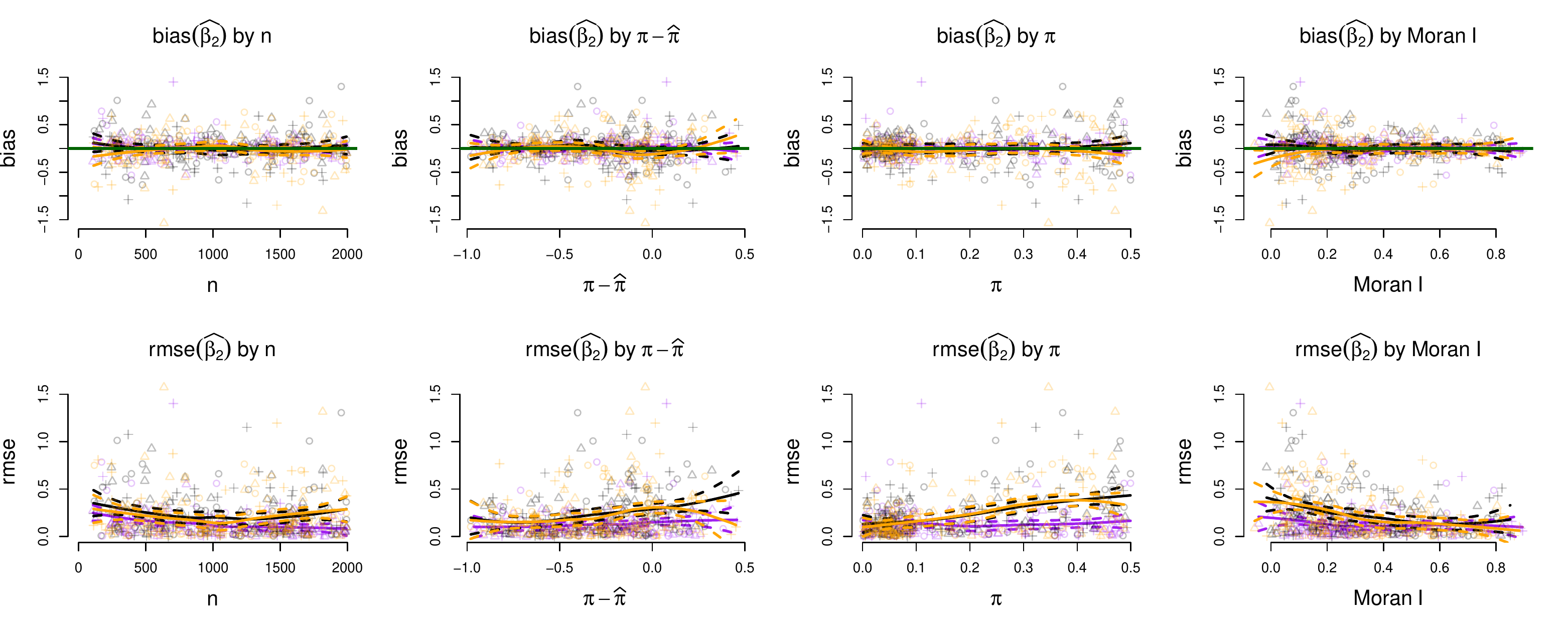}
    \caption{Results from the simulation study, capturing the ability of the simple fixed effects model (m.1, in black), the King \& Zeng model (m.2, in orange) and our proposed approach (m.3, in purple) to estimate the true covariate effect $\beta_2$.}
    \label{fig:simulations_covariate_full}
    \includegraphics[scale = 0.5]{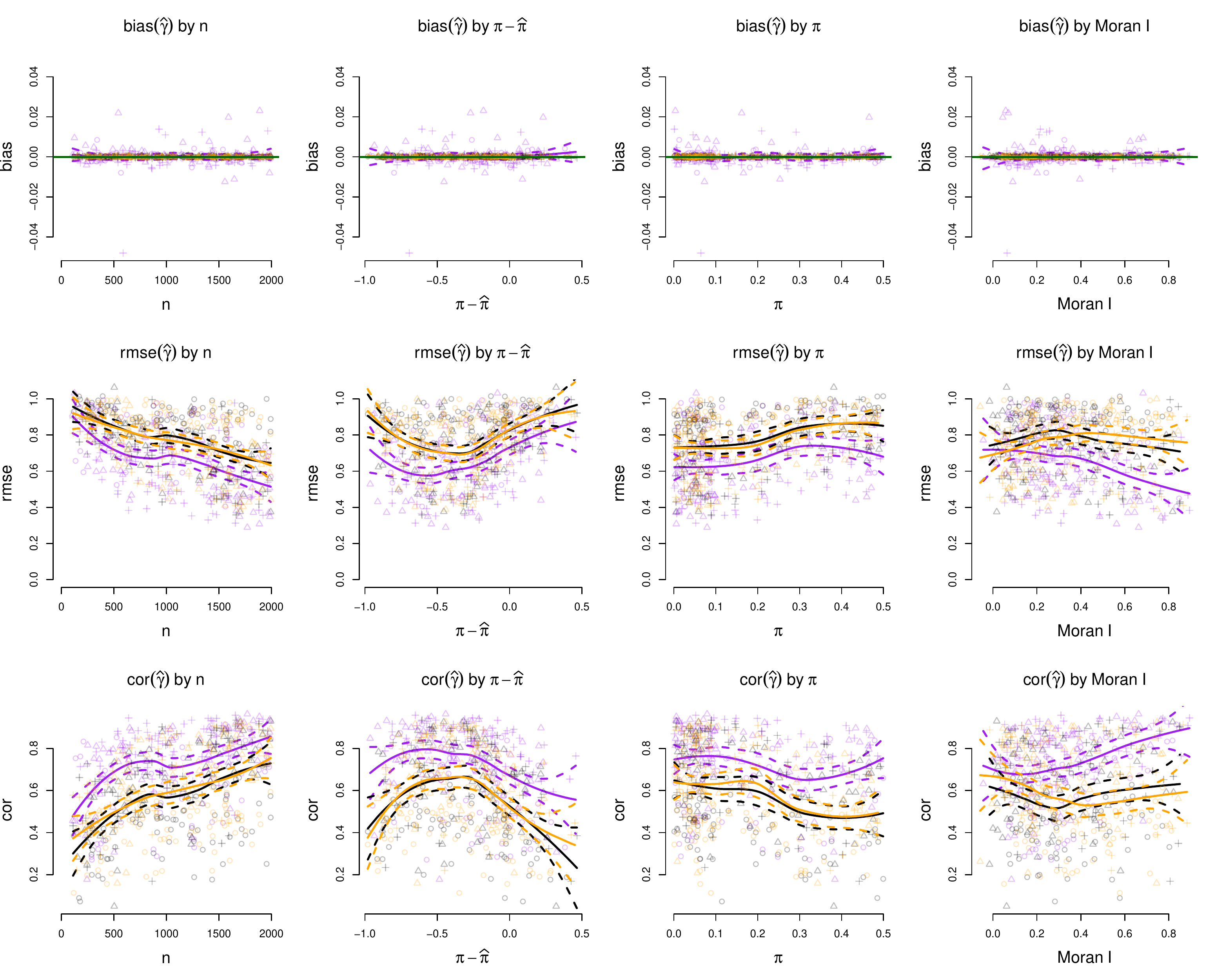}
    \caption{Results from the simulation study, capturing the ability of the simple fixed effects model (m.1, in black), the King \& Zeng model (m.2, in orange) and our proposed approach (m.3, in purple) to estimate the true area-level effects $\bm{\gamma}$.}
    \label{fig:simulations_gamma_full}
\end{figure}

A visual analysis of Figure \ref{fig:simulations_KingZeng} presents two clear dimensions in which our model advances the literature: i) m.3 is superior at moderate levels of prevalence ($\pi > 0.1$), a feat obtained thanks to the contamination layer of the model; ii) m.3 is superior under moderate-to-high levels of spatial auto-correlation ($I>0.2$), due to the BYM2 spatial component. Related to the first advantage, we note that as the discrepancy between population and sample prevalence becomes positive ($\pi - \hat{\pi} > 0$, the upward bias which m.3 suffers from as a result of contamination is significantly more contained than the downward bias which characterises m.2 as a result of a non-contaminated offset, again highlighting another robustness advantage, pertaining to the relationship between sample and population prevalence. Moreover, Figures \ref{fig:simulations_intercept} and \ref{fig:simulations_covariate_full}, which present the ability of m.2 and m.3 to estimate respectively the correct intercept parameter $\beta_1$ and the covariate effect $\beta_2$, also paint a favourable picture. The ability of our model to perform under high levels of prevalence affords significant reductions in bias an RMSE, in both $\beta_1$ and $\beta_2$, already at moderate levels of contamination. Figure \ref{fig:simulations_gamma_full} compares models in their ability to estimate the correct area-level effect. Though all three models are, unsurprisingly, unbiased, m.3 is clearly more precise (lower RMSE) and and better at ordering areas according to their propensity (higher Pearson correlation), in the presence of spatial auto-correlation. 

\begin{figure}[t!]
    \centering
    \includegraphics[scale = 0.4]{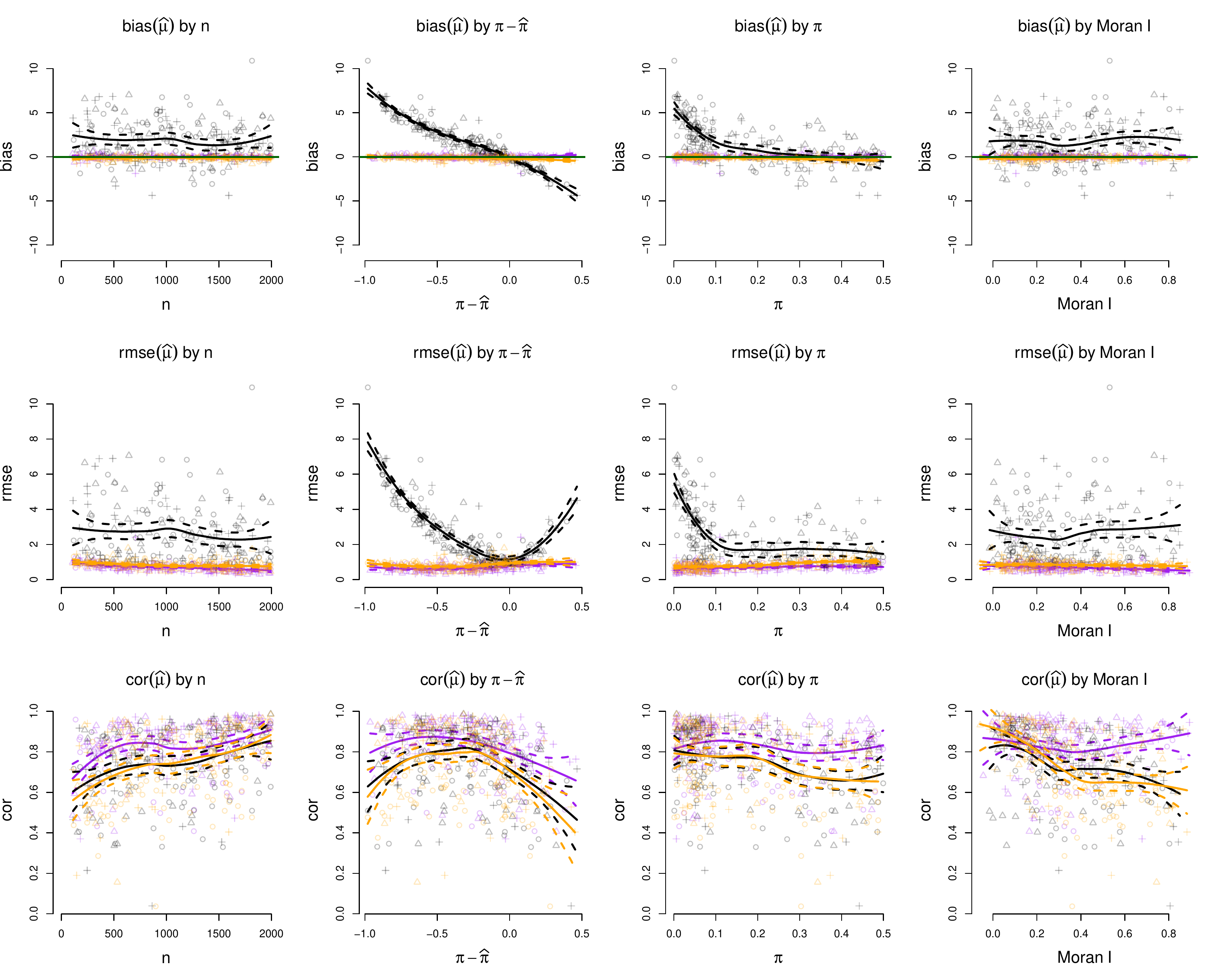}
    \caption{Results from the simulation study, capturing the ability of the simple fixed effects model (m.1, in black), the King \& Zeng model (m.2, in orange) and our proposed approach (m.3, in purple) to estimate the latent propensity of recruitment for each record in our sample $\bm{\mu}^\star$.}
    \label{fig:simulations_mu_full}
\end{figure}

Figure \ref{fig:simulations_mu_full} presents the scoring of models in their ability to predict latent propensity $\bm{\mu}$; Figure \ref{fig:simulations_intercept_full} displays the models' performance in estimating the baseline propensity $\beta_1$, with Figure \ref{fig:simulations_intercept} zooming-in to a comparison between out proposed model and the rare-events logit by King \& Zeng; Figure \ref{fig:simulations_intercept_full} shows model performance in estimating covariate effect $\beta_2$; Figure \ref{fig:simulations_gamma_full} presents a comparison with respect to the estimation of area-level effects $\bm{\gamma}$.

\pagebreak

\section{Convergence diagnostics}

To ensure absolute convergence of all model parameters we run our model with extremely conservative settings: 
\texttt{n.iter} $> 10,000$, \texttt{n.warmup} $>9,000$,\footnote{For the `Bird's Eye model, we set \texttt{n.iter} $= 10,000$ and \texttt{n.warmup} $= 9,000$, and ran the model over $8$ chains spread over $8$ cores, thnning by a factor of $8$ -- whilst for the Worm's Eye models we can afford a larger number of iterations -- \texttt{n.iter}$= 25,000$ and \texttt{n.warmup} $= 22,500$, runnng $4$ chains spread over $4$ cores, and thinning by a factor of $4$.} \texttt{n.chains} $> 4$; \texttt{n.thin} = \texttt{n.cores} = \texttt{n.chains}; \texttt{max\_treedepth} $= 25$, \texttt{adapt\_delta} = $0.99$. Note that the Worm's Eye models take around $12$ hours to run for Egypt, $24$ hours for Tunisia, whilst the Bird's Eye model takes $48$ hours. As a final note, it's worth highlighting that convergence of point estimates for the individual-level covariates happens under far more laxed estimates, and exploratory versions of this model can be fit under $1$ hour in all cases.

\begin{figure}[H]
    \centering
    \includegraphics[scale = 0.5]{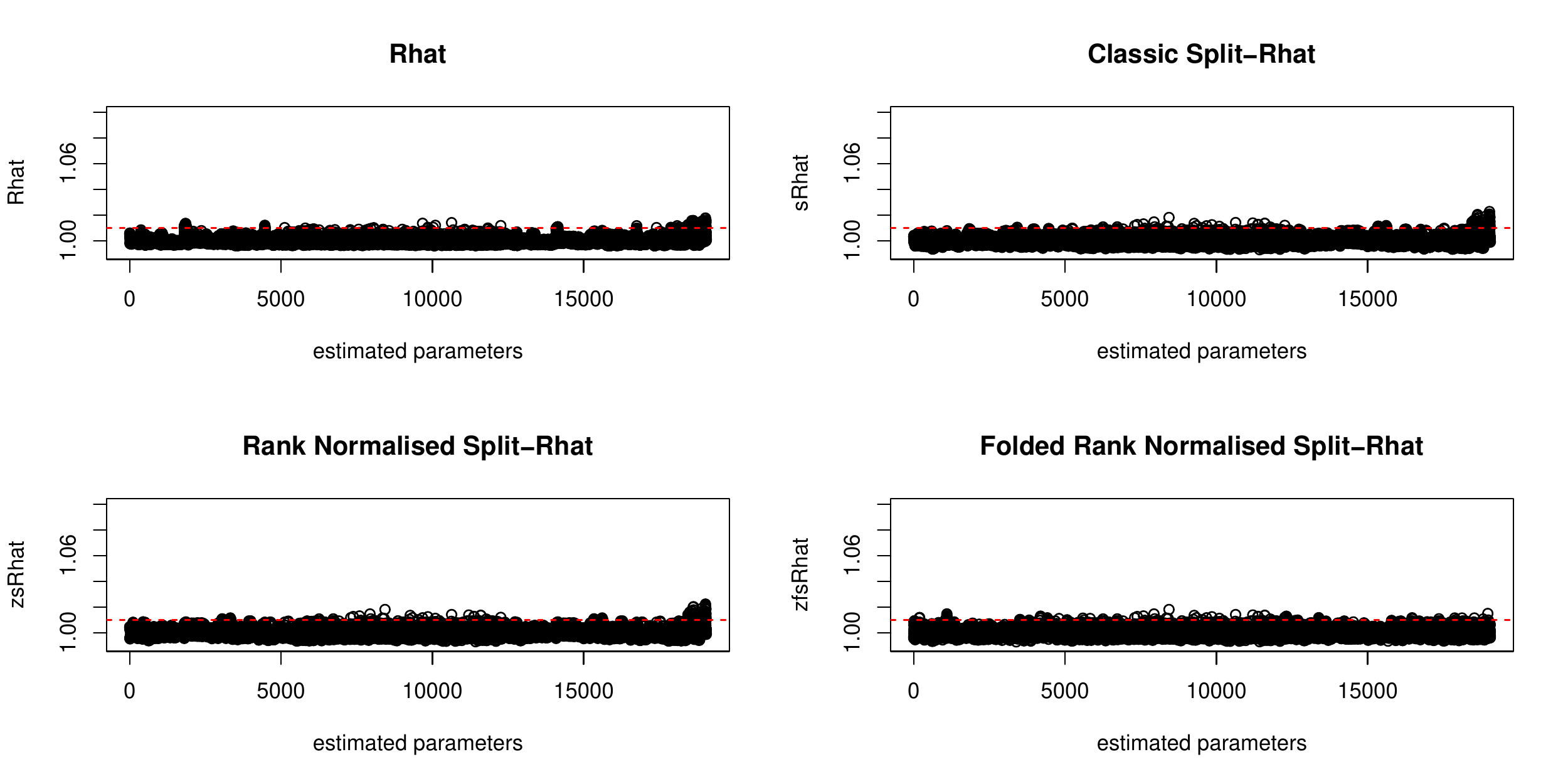}
    \caption{Gelman-Rubin Statistics for the \emph{Bird's Eye} model.} 
    \label{fig:bird_rhat_all}
\end{figure}

\begin{figure}[H]
    \centering
    \includegraphics[scale = 0.5]{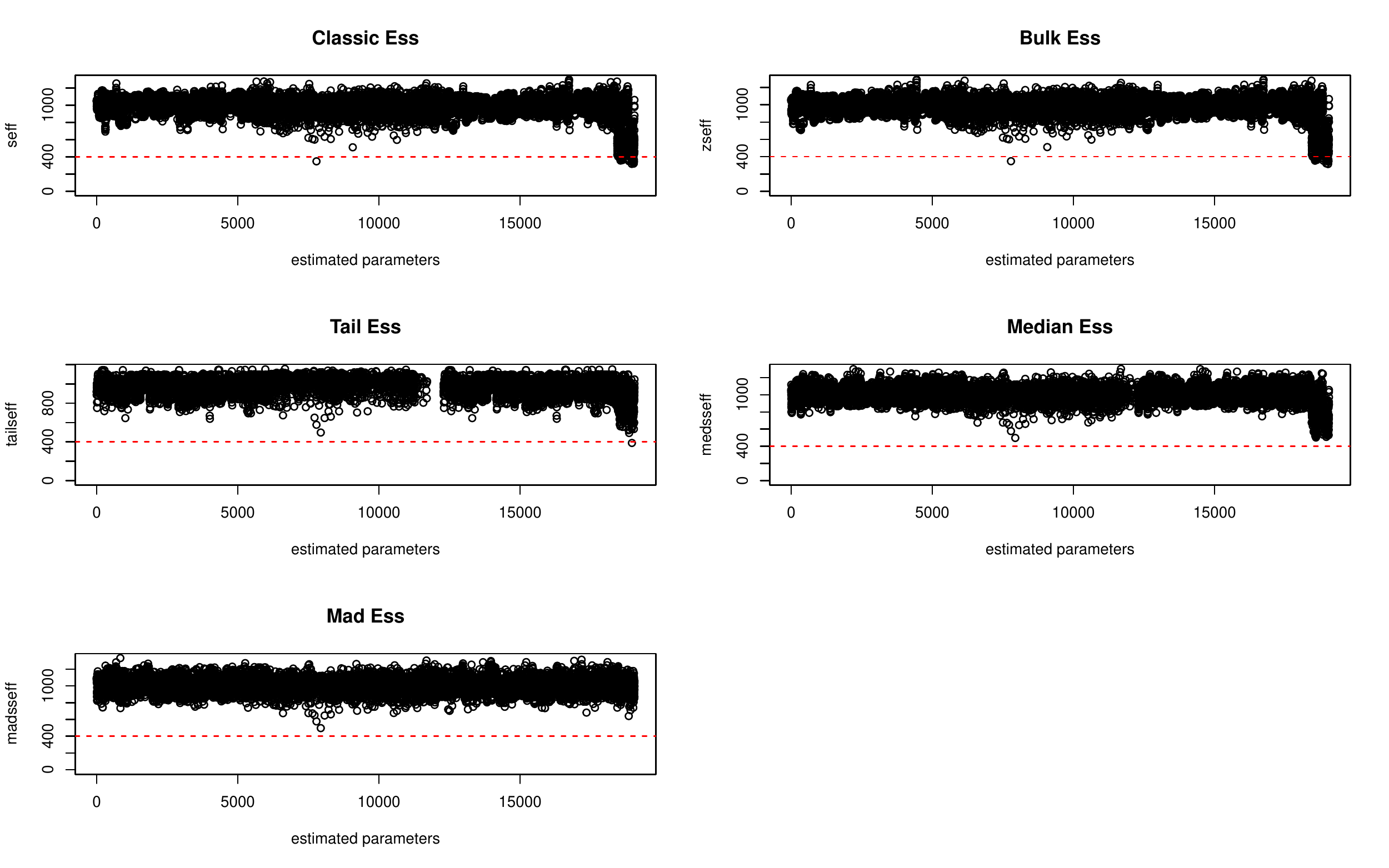}
    \caption{Effective sample-size (ESS) for the parameters of the \emph{Bird's Eye} model.} 
    \label{fig:bird_ESS_all}
\end{figure}

\begin{figure}[H]
    \centering
    \includegraphics[scale = 0.65]{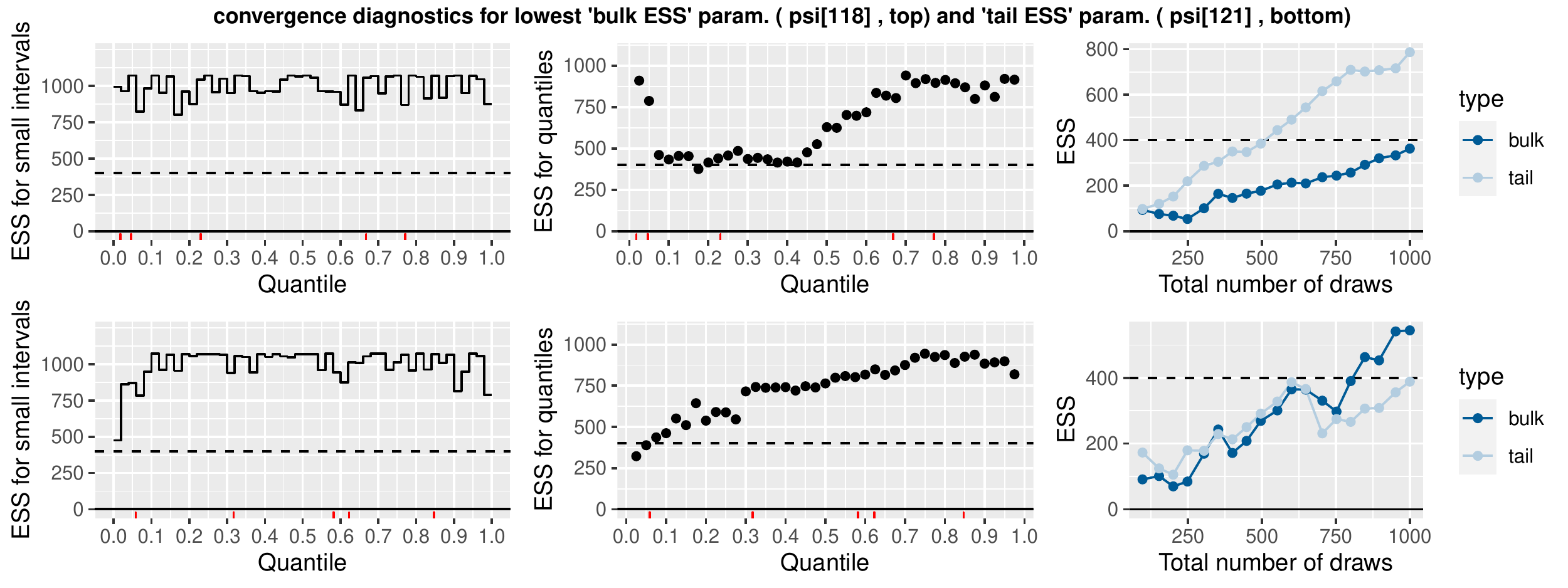}
    \caption{Convergence dynamics for the parameters with the lowest bulk (top) and tail (bottom) ESS, for the \emph{Bird's Eye} model. The quantile plots show satisfactory ESS for every section of the posterior distribution, whilst the positive and close-to-linear gradient in the `total number of draws' plot suggests ESS would improve further by drawing more samples -- a sign that the posterior is well-explored.} 
    \label{fig:bird_min.ESS_all}
\end{figure}

\begin{landscape}
\begin{figure}[H]
    \centering
    \includegraphics[scale = 0.5]{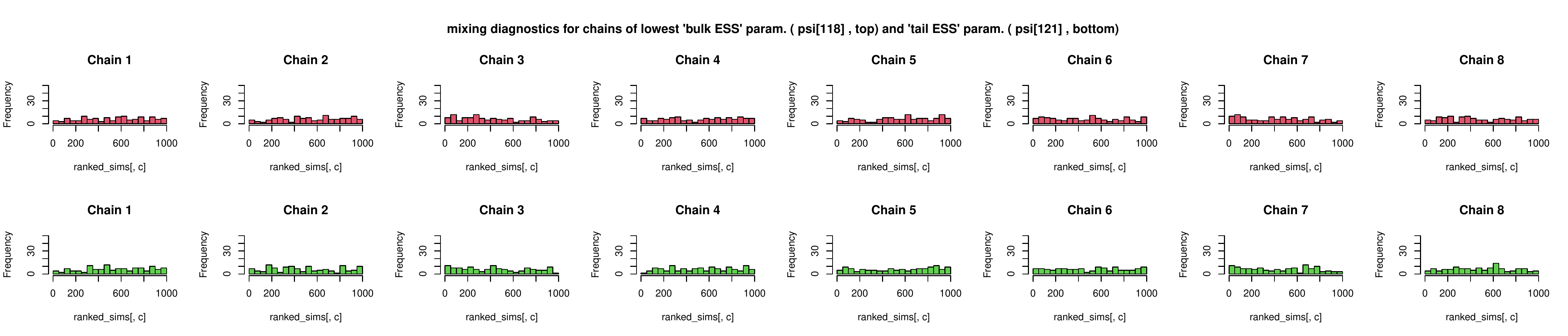}
    \caption{Histogram of the ranked posterior draws for the parameters with the lowest bulk (top) and tail (bottom) ESS, for the  \emph{Bird's Eye} model. This plot is evidence of reasonably good mixing also for our `least convergent' parameters, as the ranked draws from each chain could be reasonably drawn from a uniform distributions.} 
    \label{fig:bird_mix.ESS_all}
\end{figure}
\end{landscape}

\begin{figure}[H]
    \centering
    \includegraphics[scale = 0.5]{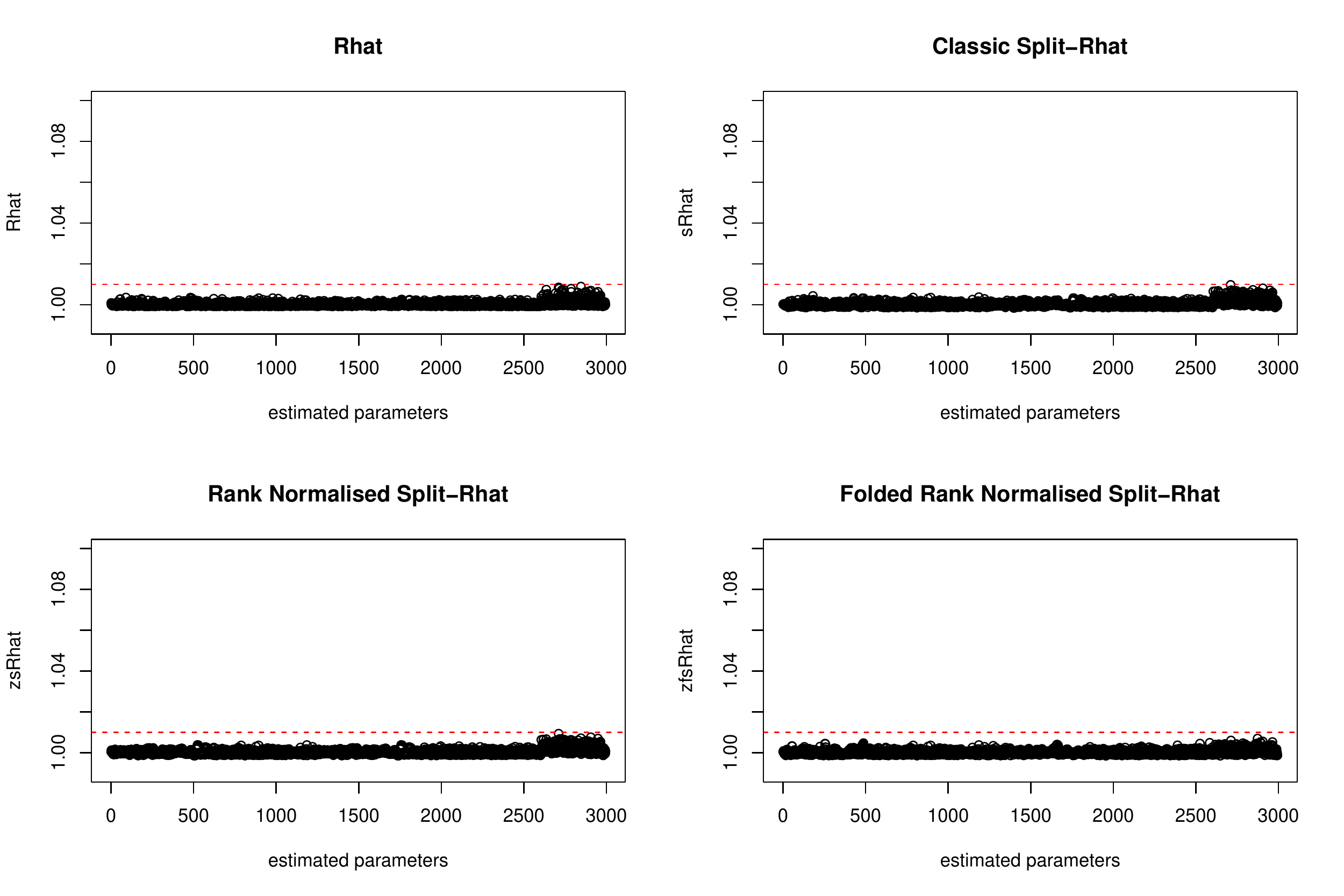}
    \caption{Gelman-Rubin Statistics for the Egypt \emph{Worm's Eye} model.} 
    \label{fig:egypt_rhat_all}
    \includegraphics[scale = 0.5]{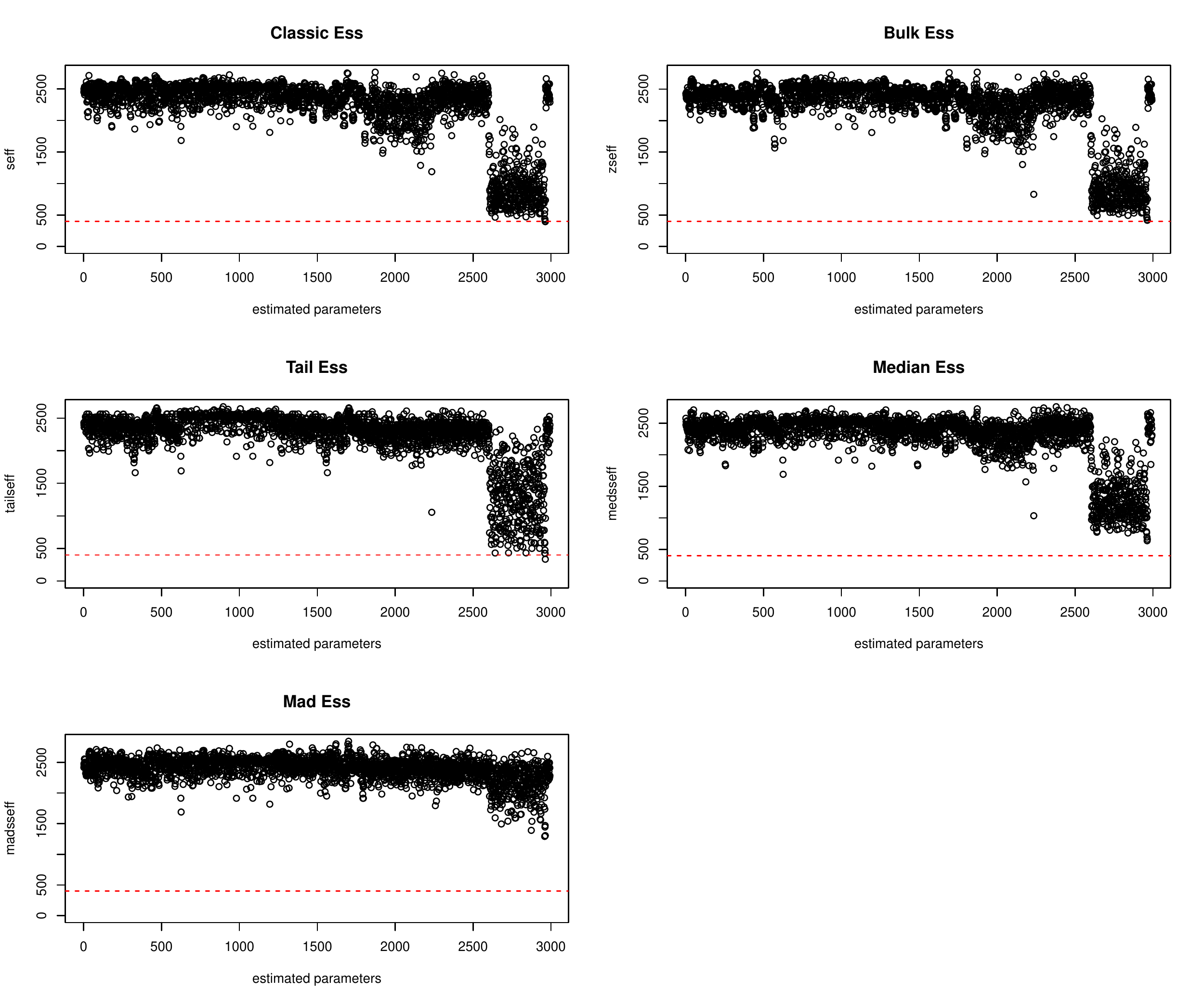}
    \caption{Effective sample-size (ESS) for the parameters of the Egypt \emph{Worm's Eye} model.} 
    \label{fig:egypt_ESS_all}
\end{figure}

\begin{figure}[H]
    \centering
    \includegraphics[scale = 0.5]{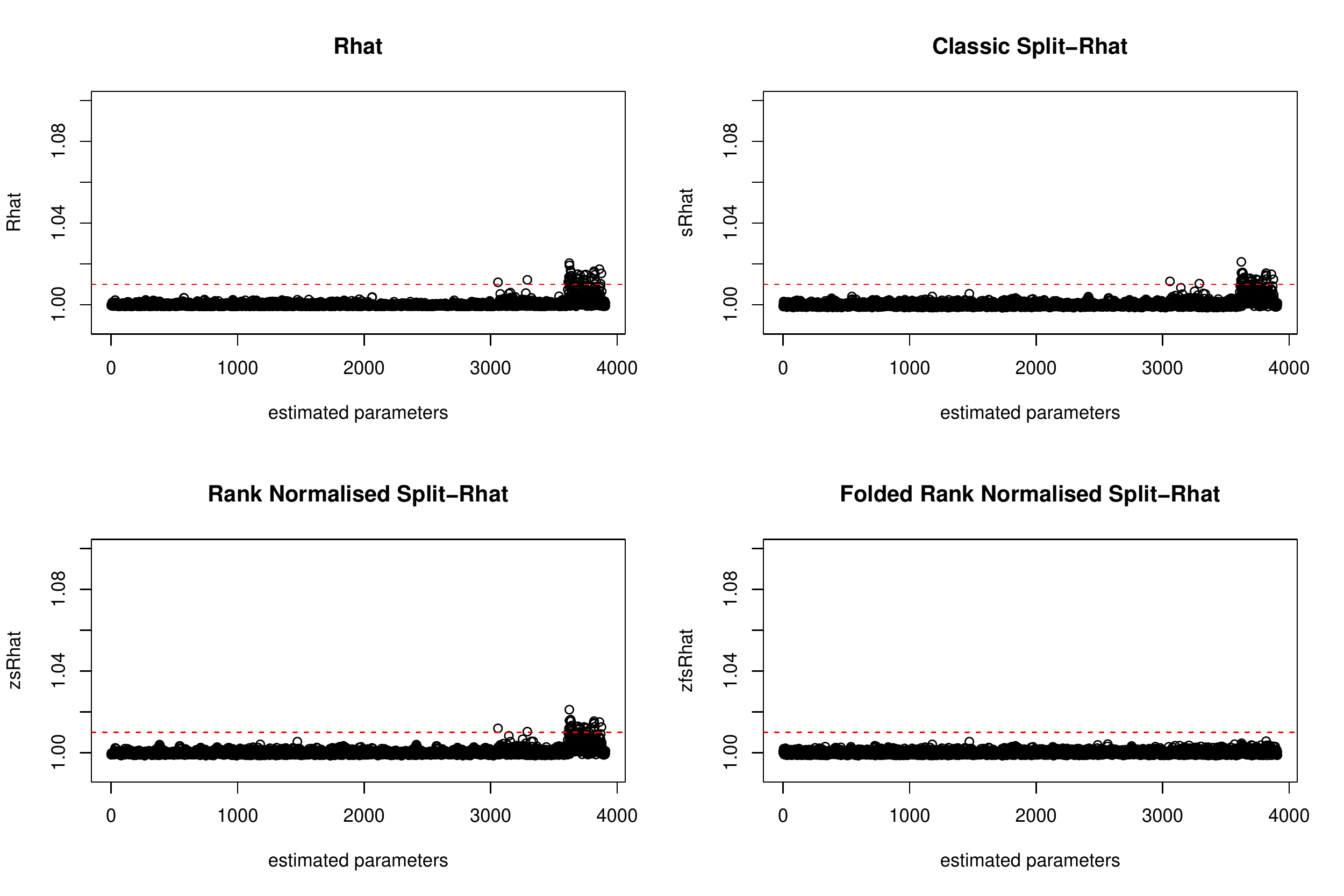}
    \caption{Gelman-Rubin Statistics for the Tunisia \emph{Worm's Eye} model.} 
    \label{fig:tunisia_rhat_all}
    \includegraphics[scale = 0.5]{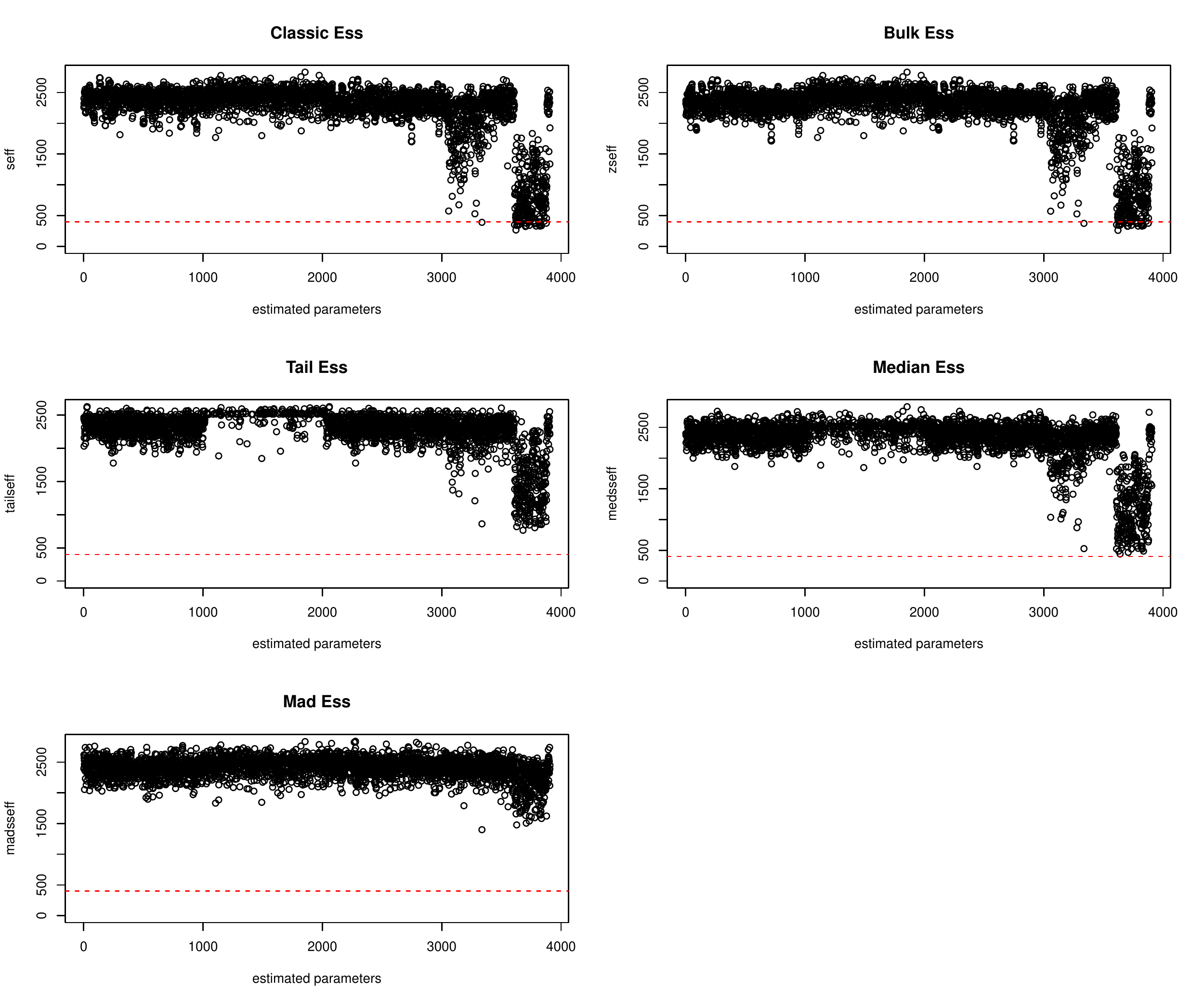}
    \caption{Effective sample-size (ESS) for the parameters of the Tunisia \emph{Worm's Eye} model.} 
    \label{fig:tunisia_ESS_all}
\end{figure}

\newpage

\section{Posterior densities of regression coefficients}
\begin{figure}[H]
    \centering
    \includegraphics[scale = 0.85]{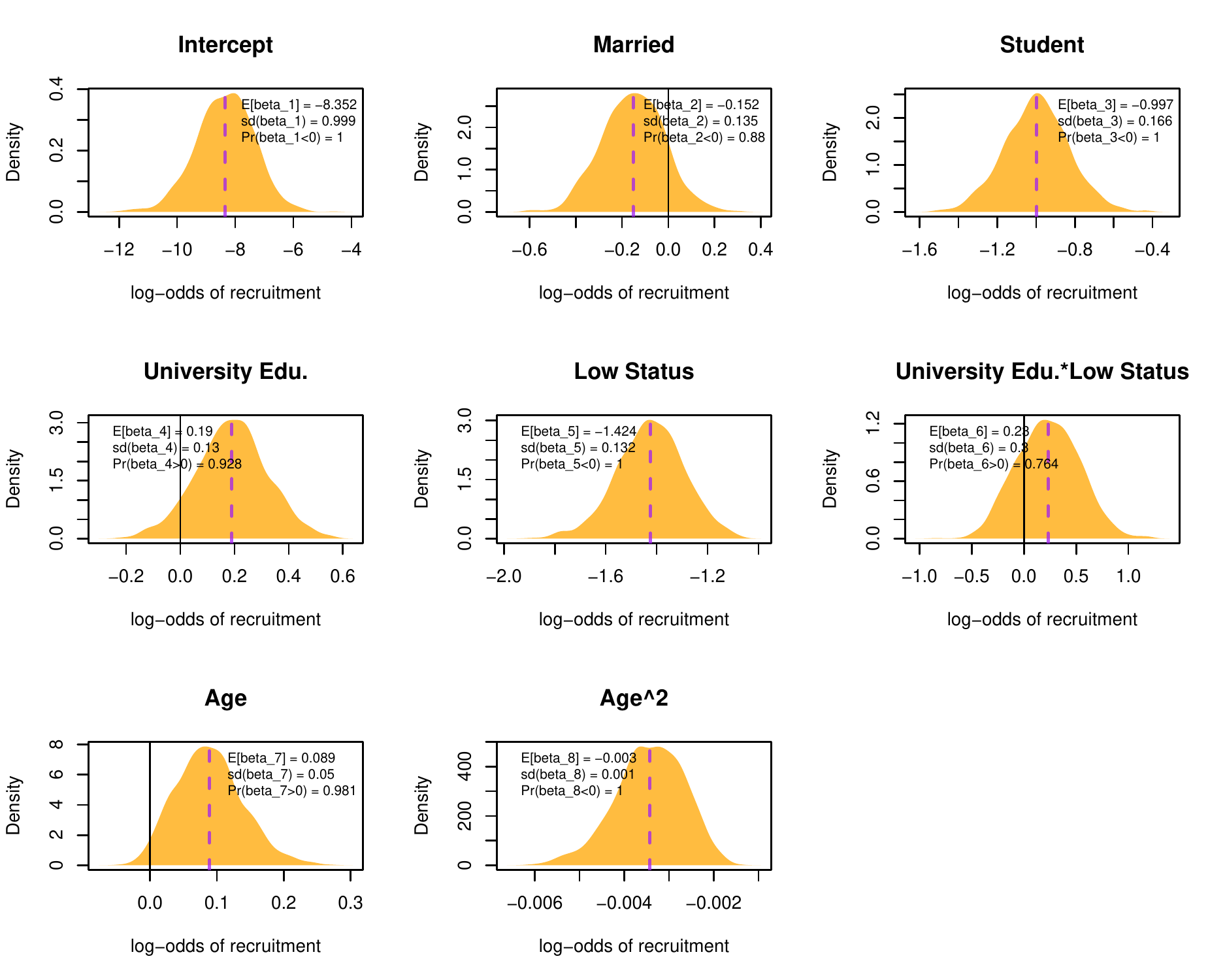}
    \caption{Posterior density of individual-level fixed-effect coefficients for the `Bird's Eye' model. These effects are presented on the original, non-standardized scale.}
    \label{fig:bird_fixedeff_original}
\end{figure}

\begin{figure}[H]
    \centering
    \includegraphics[scale = 0.85]{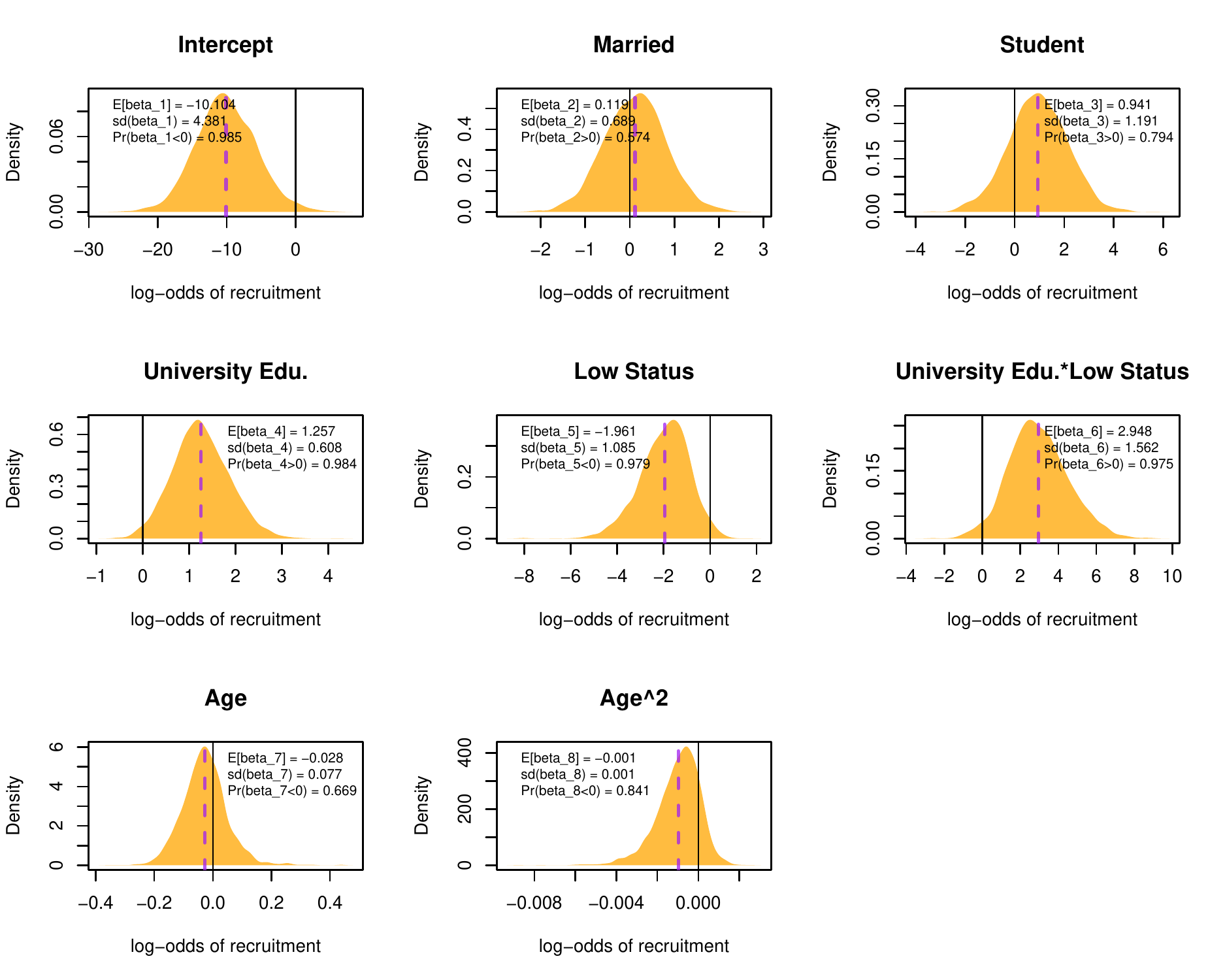}
    \caption{Posterior density of individual-level fixed-effect coefficients for the Egypt `Worm's Eye' model. These effects are presented on the original, non-standardized scale.}
    \label{fig:worm_fixedeff_egypt_ind_original}
\end{figure}

\begin{figure}[H]
    \centering
    \includegraphics[scale = 0.85]{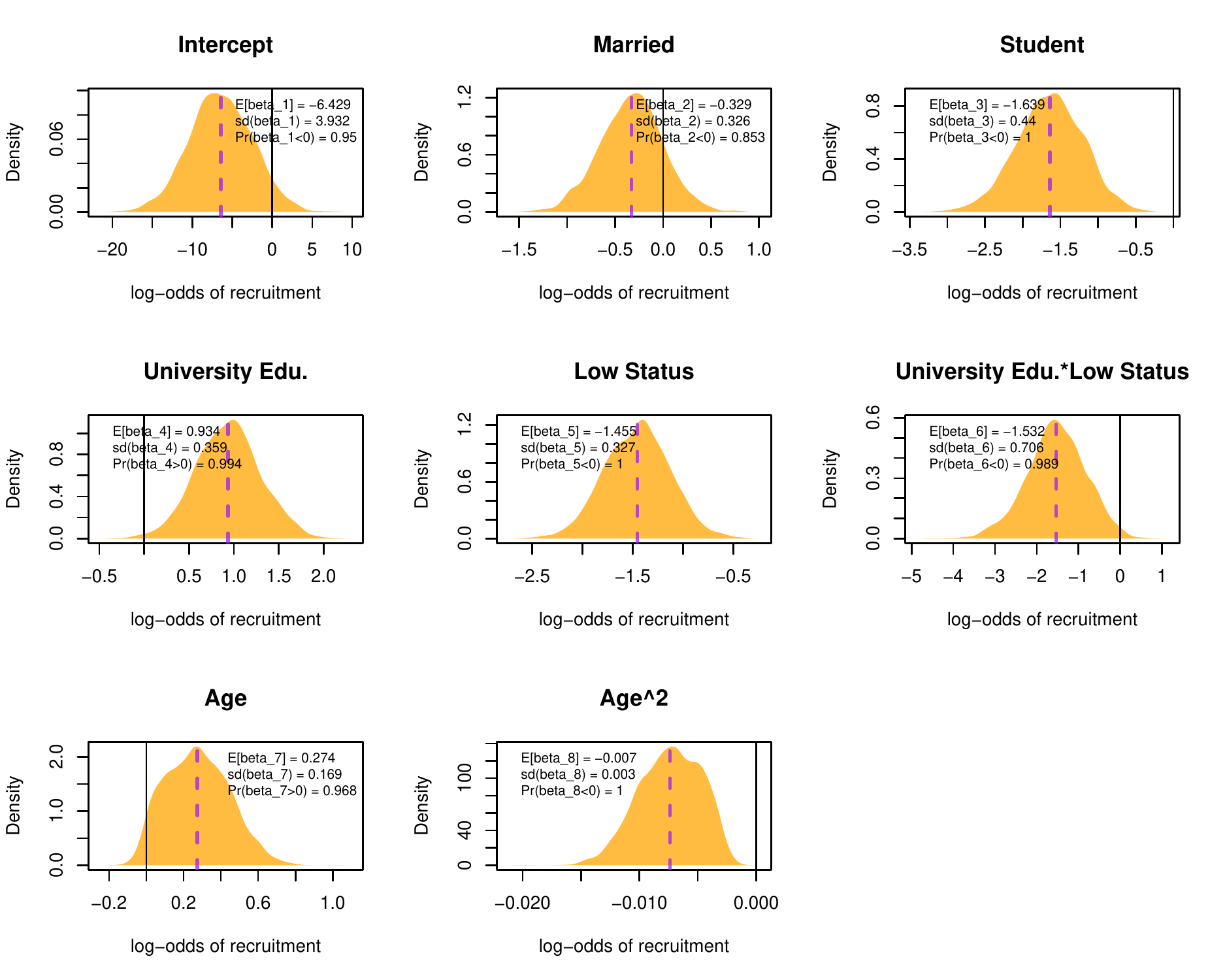}
    \caption{Posterior density of individual-level fixed-effect coefficients for the Tunisia `Worm's Eye' model. These effects are presented on the original, non-standardized scale.}
    \label{fig:worm_fixedeff_tunisia_ind_original}
\end{figure}

\begin{figure}[H]
\centering
\begin{subfigure}{0.8\textwidth}
    \includegraphics[width=1\linewidth]{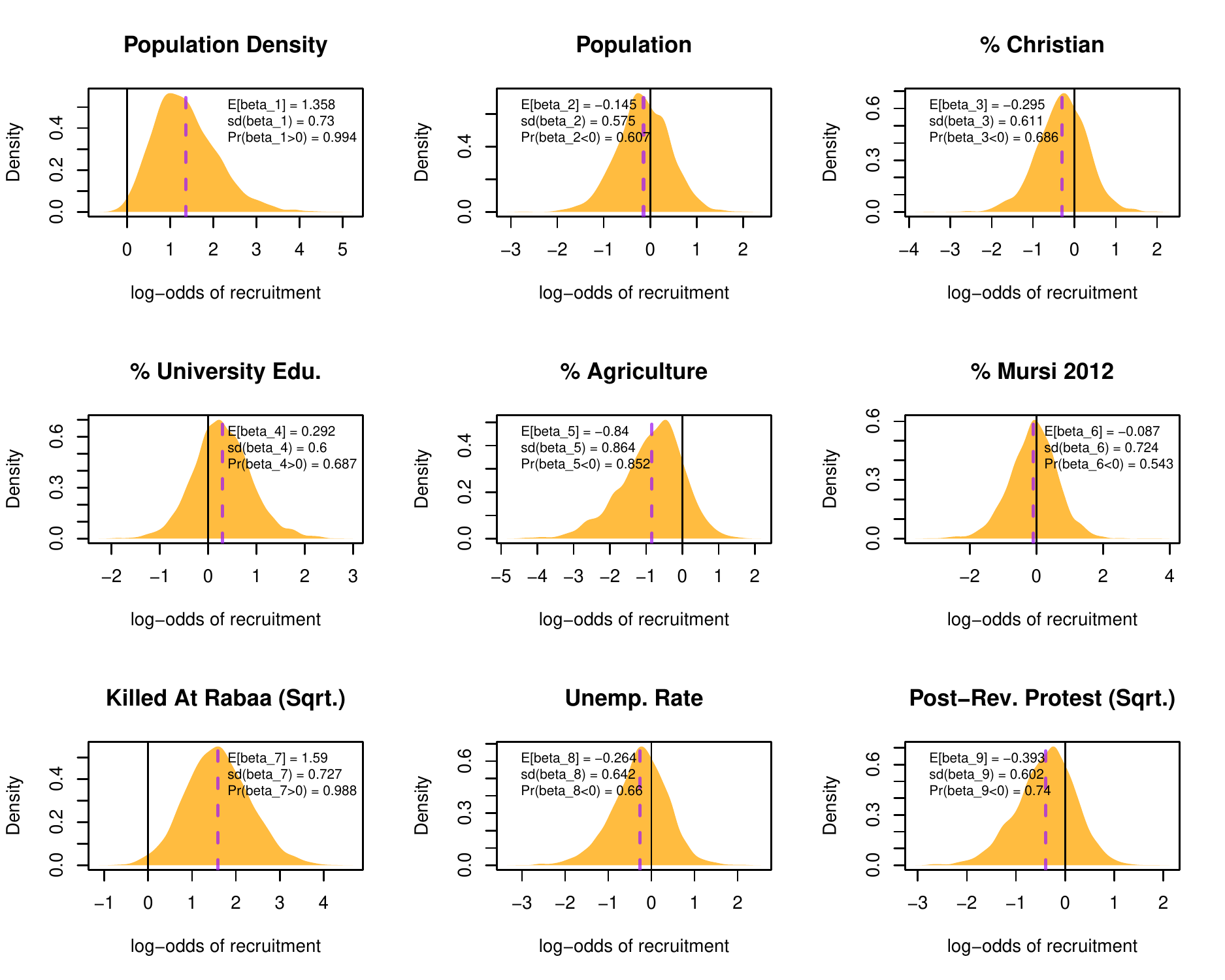}
    \caption{Egypt}
    \label{fig:worm_fixedeff_egypt}
    \end{subfigure}

\begin{subfigure}{1\textwidth}
    \includegraphics[width=1\linewidth]{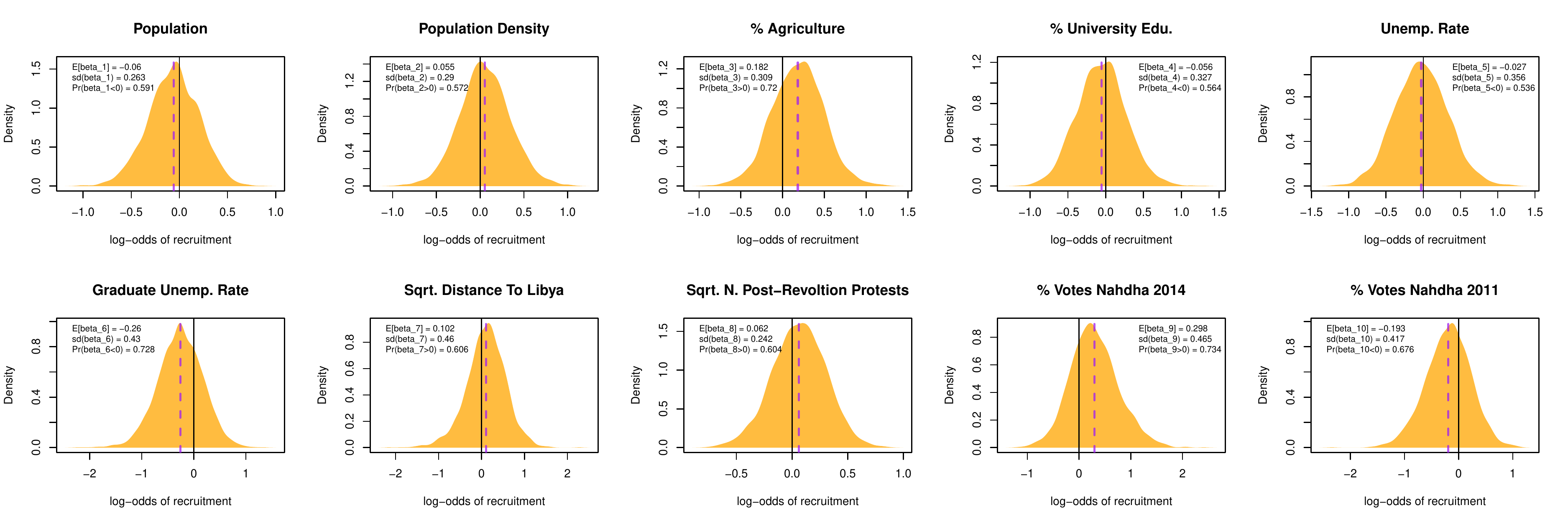}
    \caption{Tunisia}
    \label{fig:worm_fixedeff_tunisia}
    \end{subfigure}
    \caption{Posterior density of district-level fixed-effect coefficients for the \emph{Worm's Eye} models.}
\end{figure}

\begin{figure}[H]
\centering
\begin{subfigure}{0.8\textwidth}
    \includegraphics[width=1\linewidth]{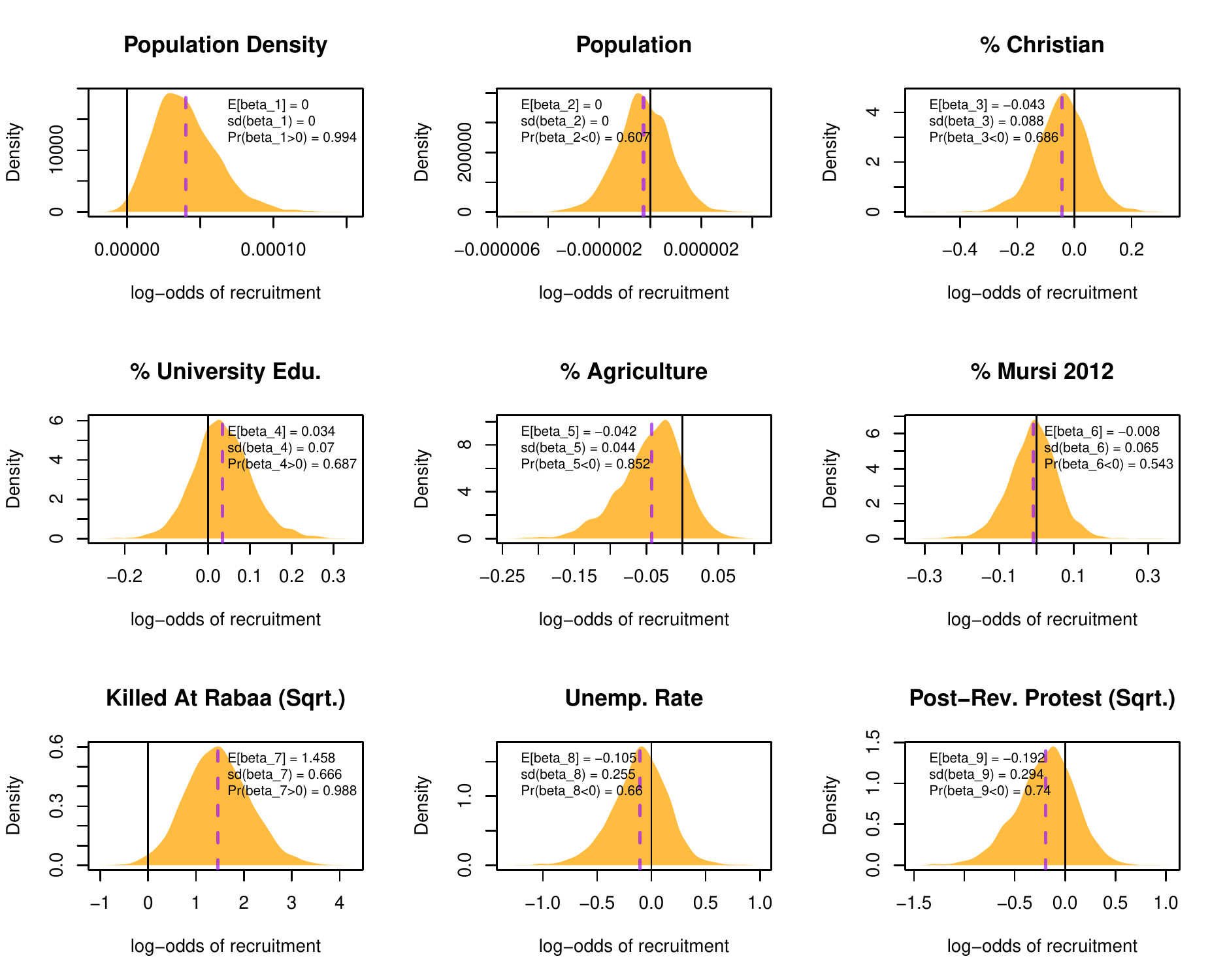}
    \caption{Egypt}
    \label{fig:worm_fixedeff_egyptl_original}
    \end{subfigure}

\begin{subfigure}{1\textwidth}
    \includegraphics[width=1\linewidth]{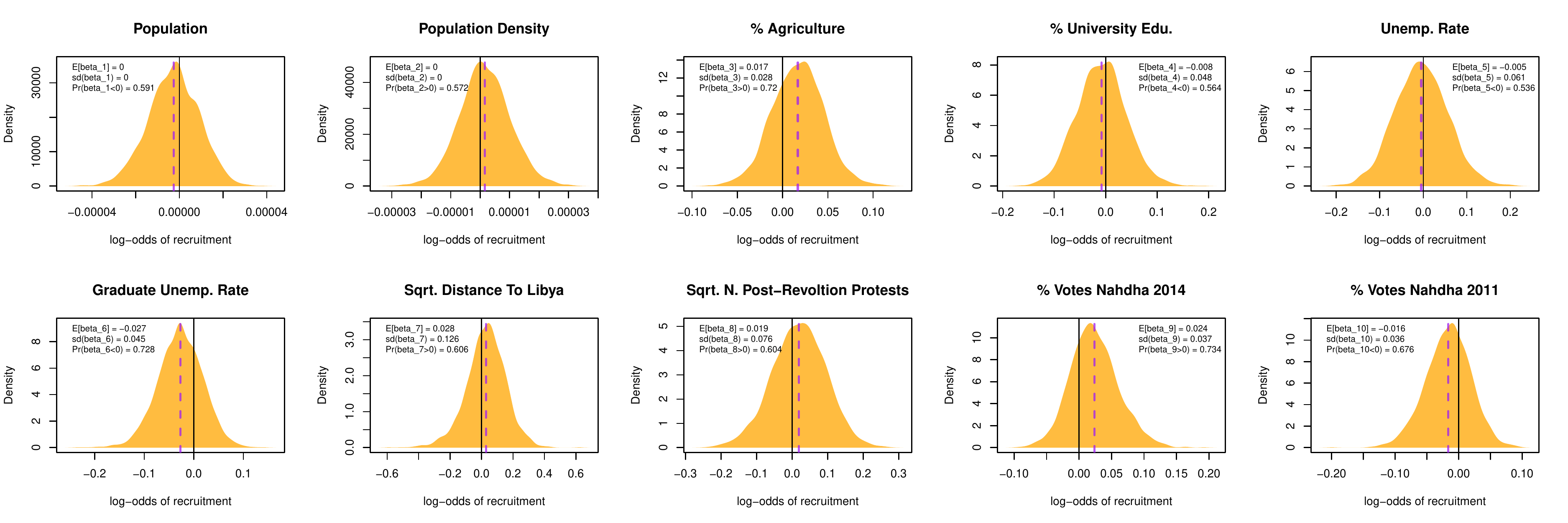}
    \caption{Tunisia}
    \label{fig:worm_fixedeff_tunisia_original}
    \end{subfigure}
    \caption{Posterior density of district-level fixed-effect coefficients for the \emph{Worm's Eye} models. These effects are presented on the original, non-standardized scale.}
\end{figure}

\pagebreak
\section{Relative Deprivation Effects}
\begin{figure}[H]
    \centering
    \includegraphics[scale = 0.45]{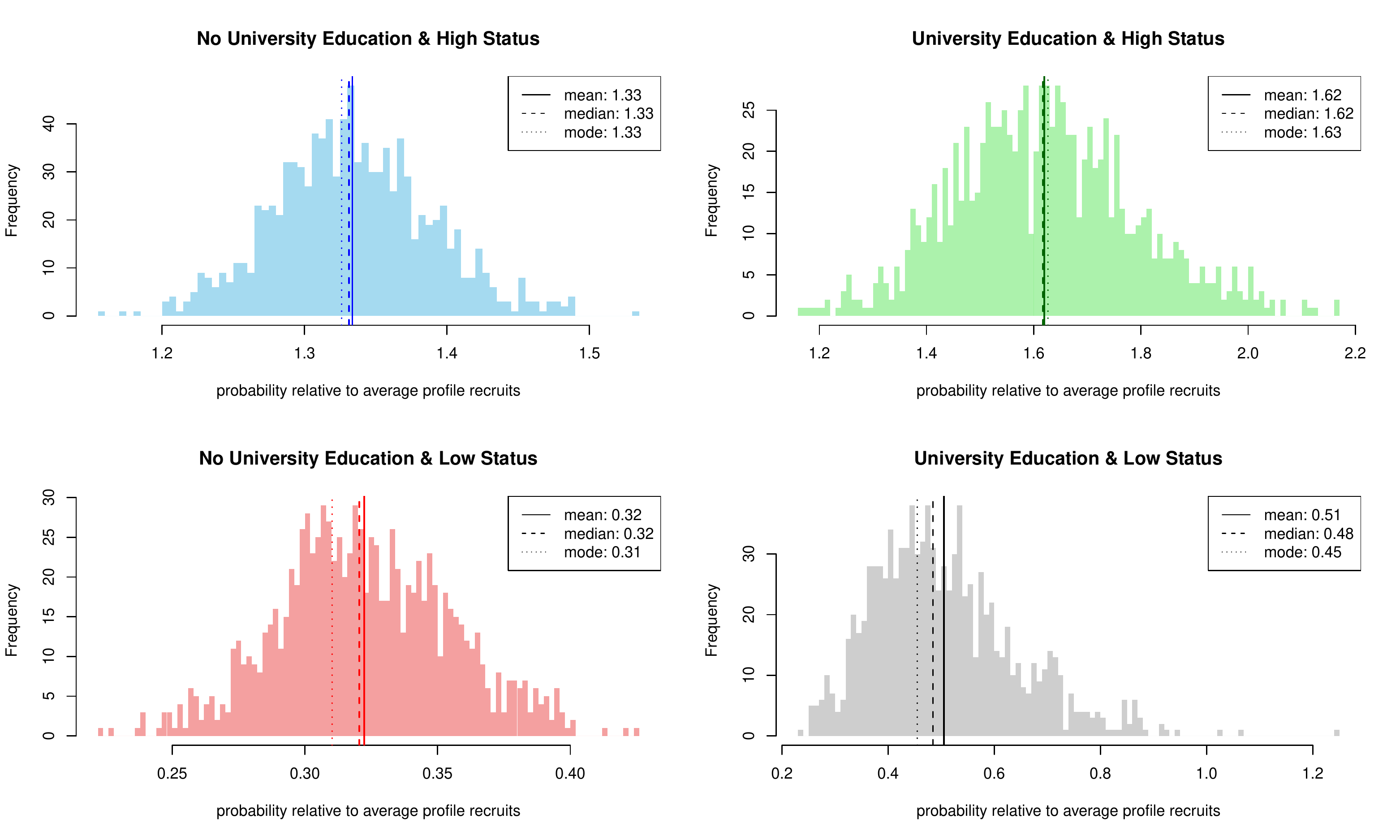}
    \caption{\footnotesize{Predicted propensity of recruitment for relative-deprivation profiles according to the `Bird’s Eye' model. The effects are presented as odds relative to the `average' recruitment profile.}}
    \label{figure::bird_relative_deprivation_relprob}
    \centering
    \includegraphics[scale = 0.45]{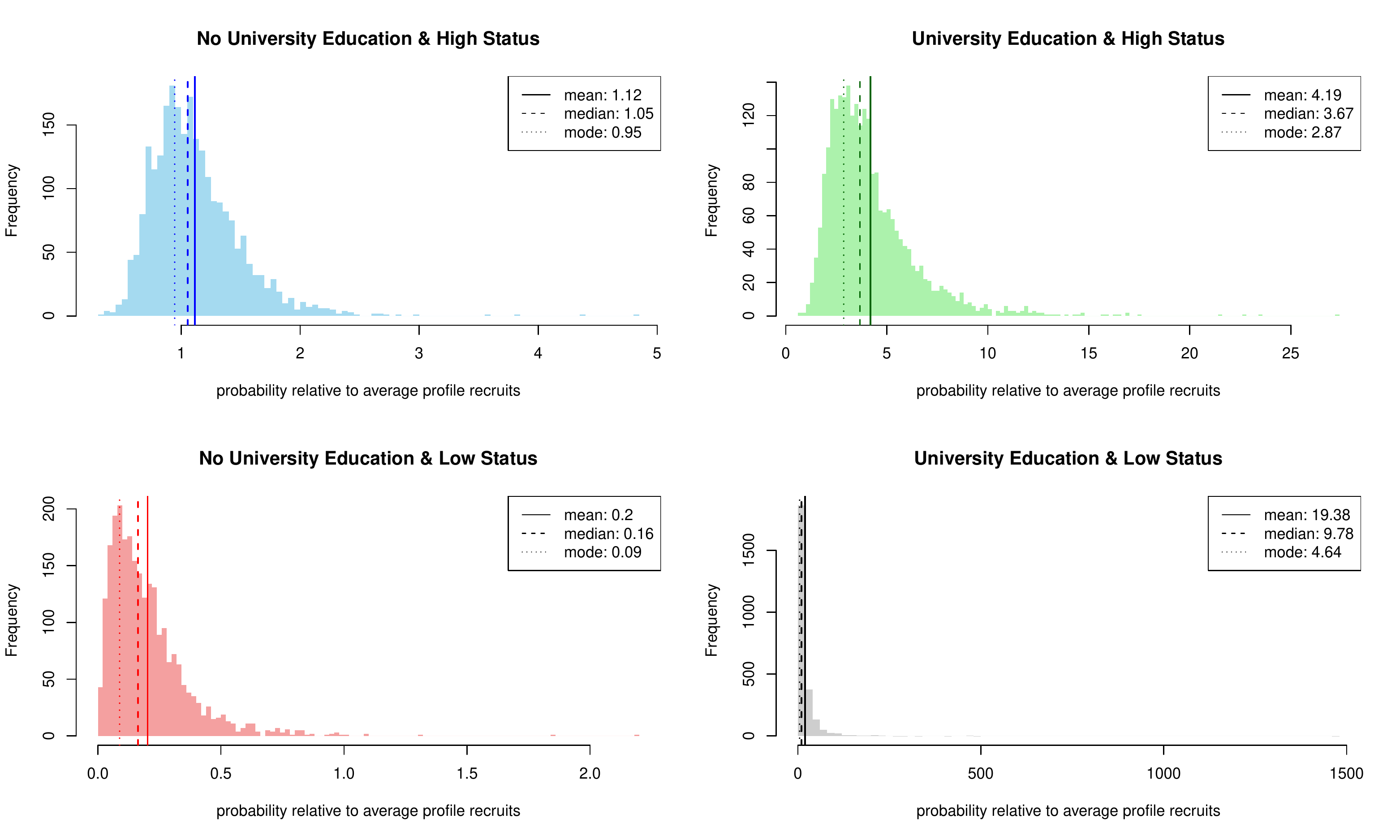}
    \caption{\footnotesize{Predicted propensity of recruitment for relative-deprivation profiles according to the Egypt `Worm’s Eye' model. The effects are presented as odds relative to the `average' recruitment profile.}}
    \label{figure::egypt_relative_deprivation_relprob}
\end{figure}

\begin{figure}[htp]
    \centering
    \includegraphics[scale = 0.45]{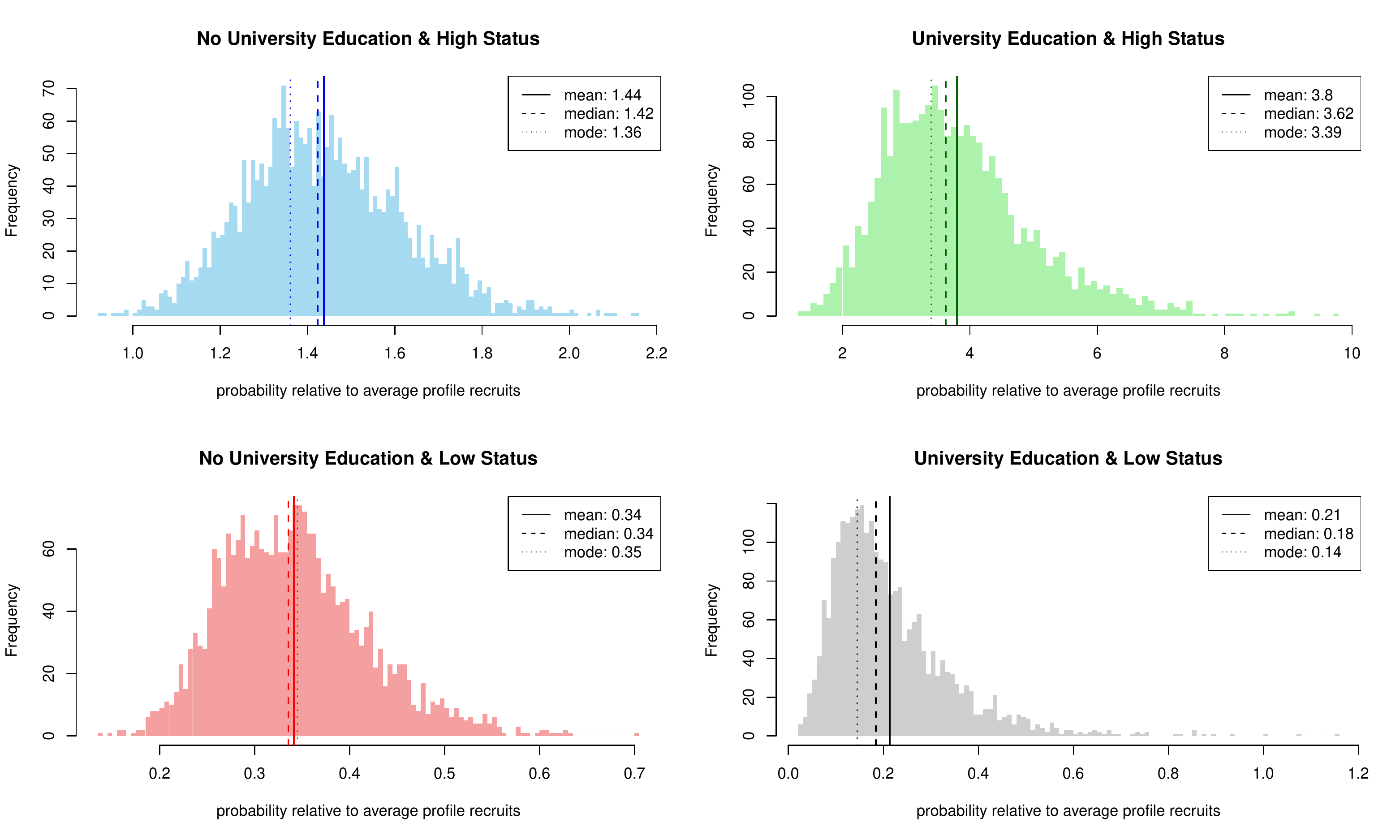}
    \caption{\footnotesize{Predicted propensity of recruitment for relative-deprivation profiles according to the Tunisia `Worm’s Eye' model. The effects are presented as odds relative to the `average' recruitment profile.}}
    \label{figure::tunisia_relative_deprivation_relprob}
    \centering
    \includegraphics[scale = 0.45]{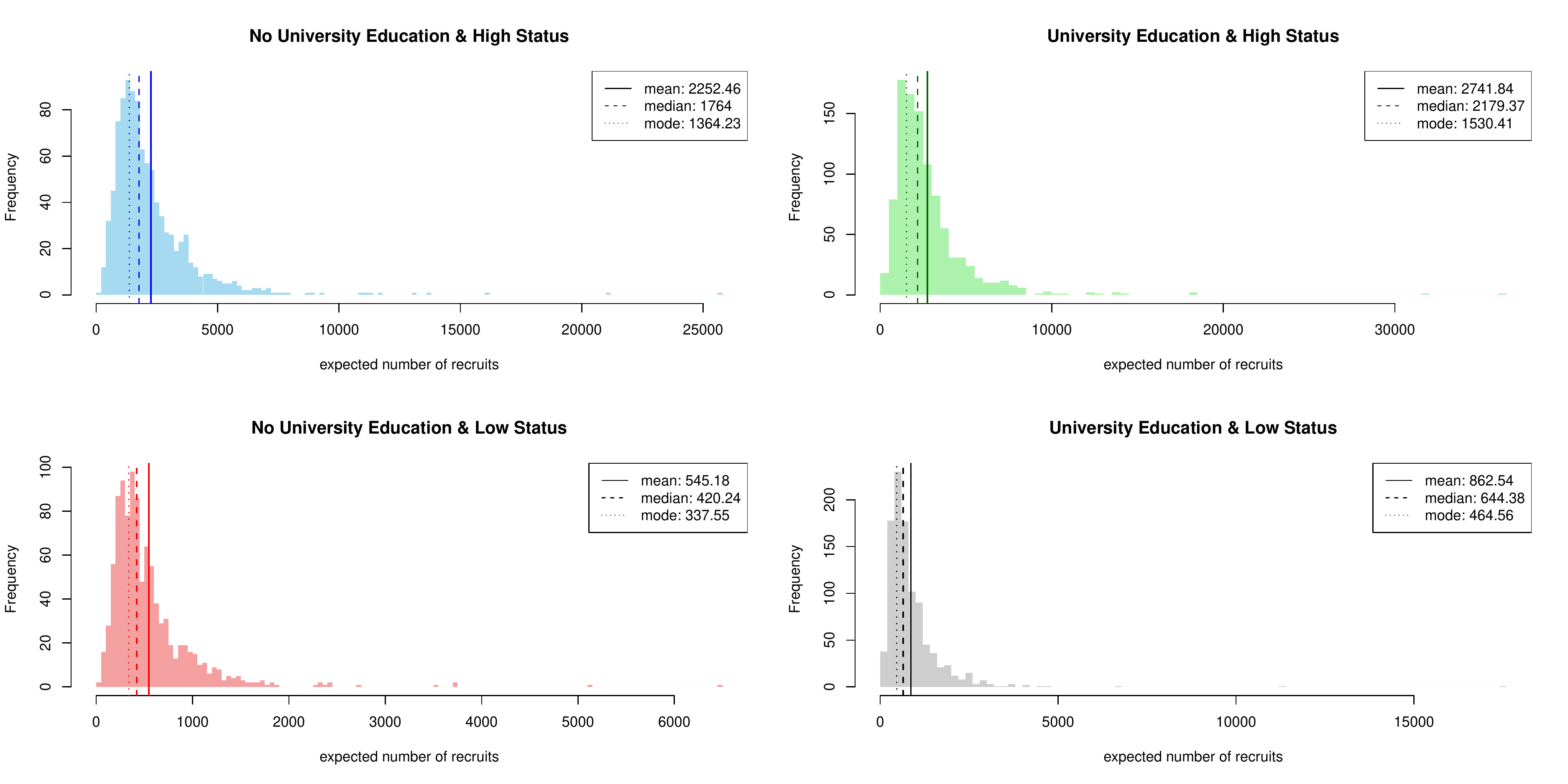}
    \caption{\footnotesize{Predicted propensity of recruitment for relative-deprivation profiles according to the `Bird’s Eye' model. The effects are presented as predicted counts under the assumption that everyone in the population is an `average profile', and only changing the profile's relative deprivation status.}}
    \label{figure::bird_relative_deprivation_n}
\end{figure}

\begin{figure}[htp]
    \centering
    \includegraphics[scale = 0.45]{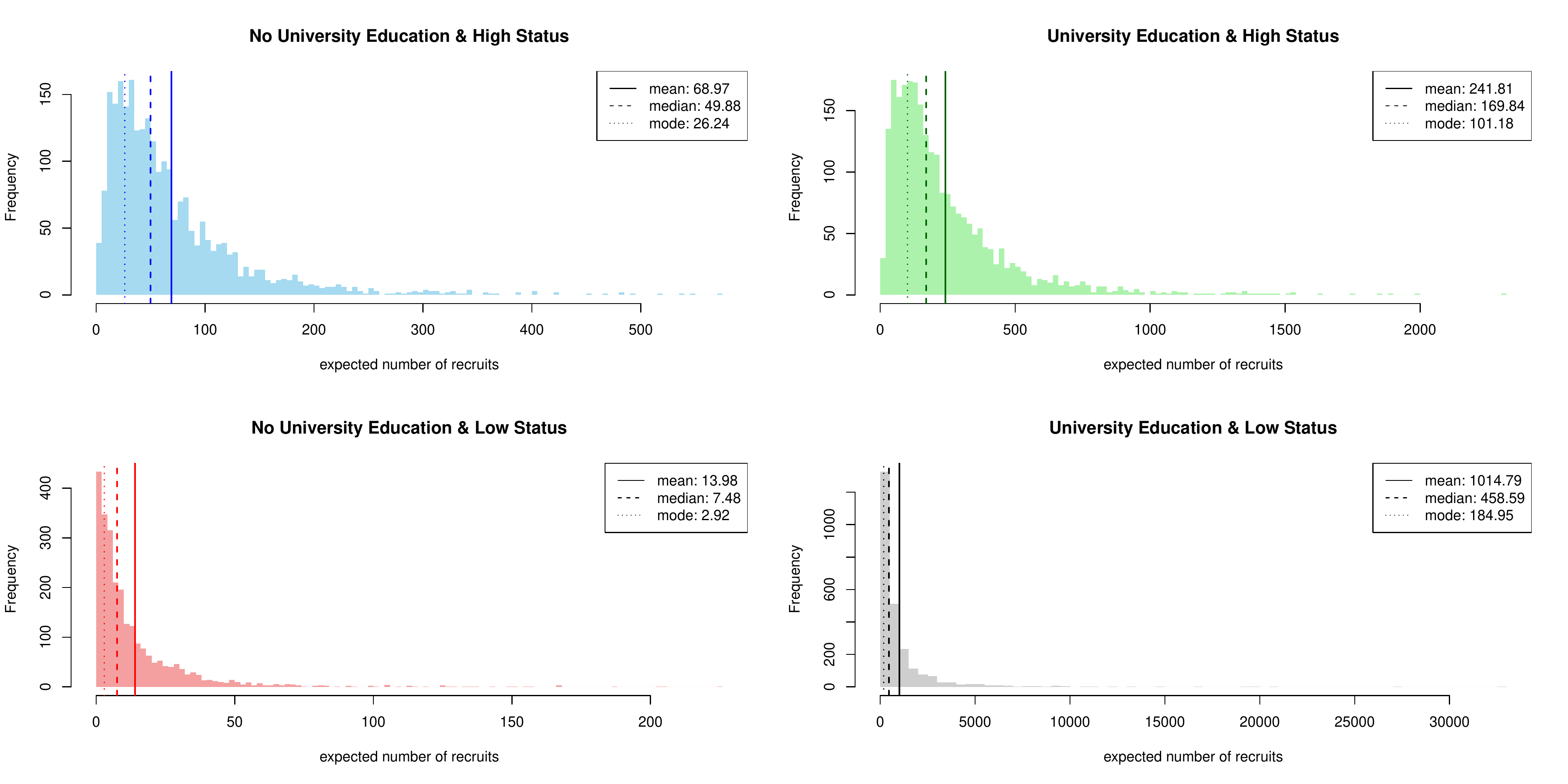}
    \caption{\footnotesize{Predicted propensity of recruitment for relative-deprivation profiles according to the Egypt `Worm’s Eye' model. The effects are presented as predicted counts under the assumption that everyone in the population is an `average profile', and only changing the profile's relative deprivation status.}}
    \label{figure::egypt_relative_deprivation_n}
    \centering
    \includegraphics[scale = 0.45]{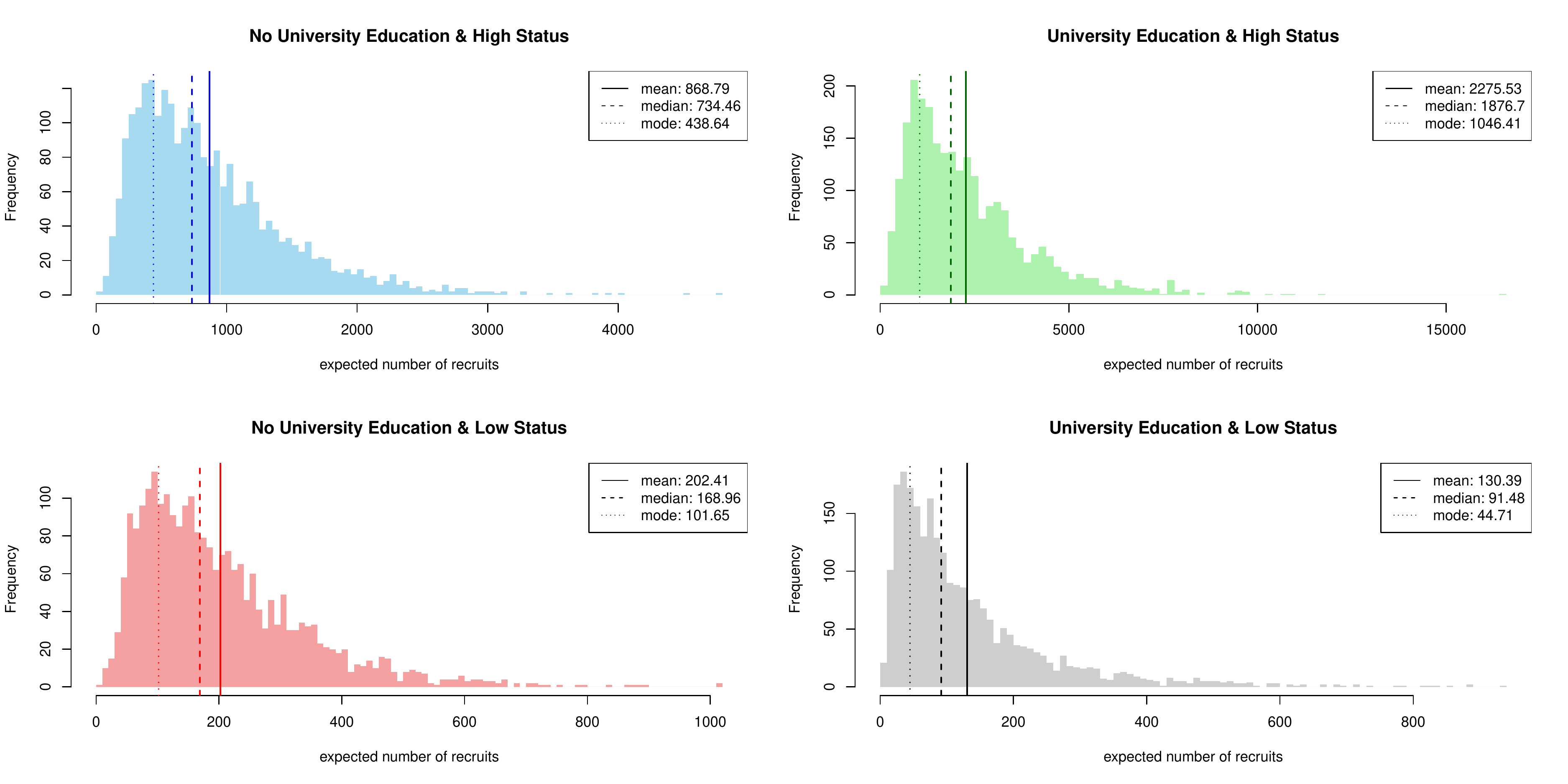}
    \caption{\footnotesize{Predicted propensity of recruitment for relative-deprivation profiles according to the Tunisia `Worm’s Eye' model. The effects are presented as predicted counts under the assumption that everyone in the population is an `average profile', and only changing the profile's relative deprivation status.}}
    \label{figure::tunisia_relative_deprivation_n}
\end{figure}

\pagebreak
\section{Residual Area-Level Analysis}
\begin{figure}[htp]
\centering
\includegraphics[scale = 0.5]{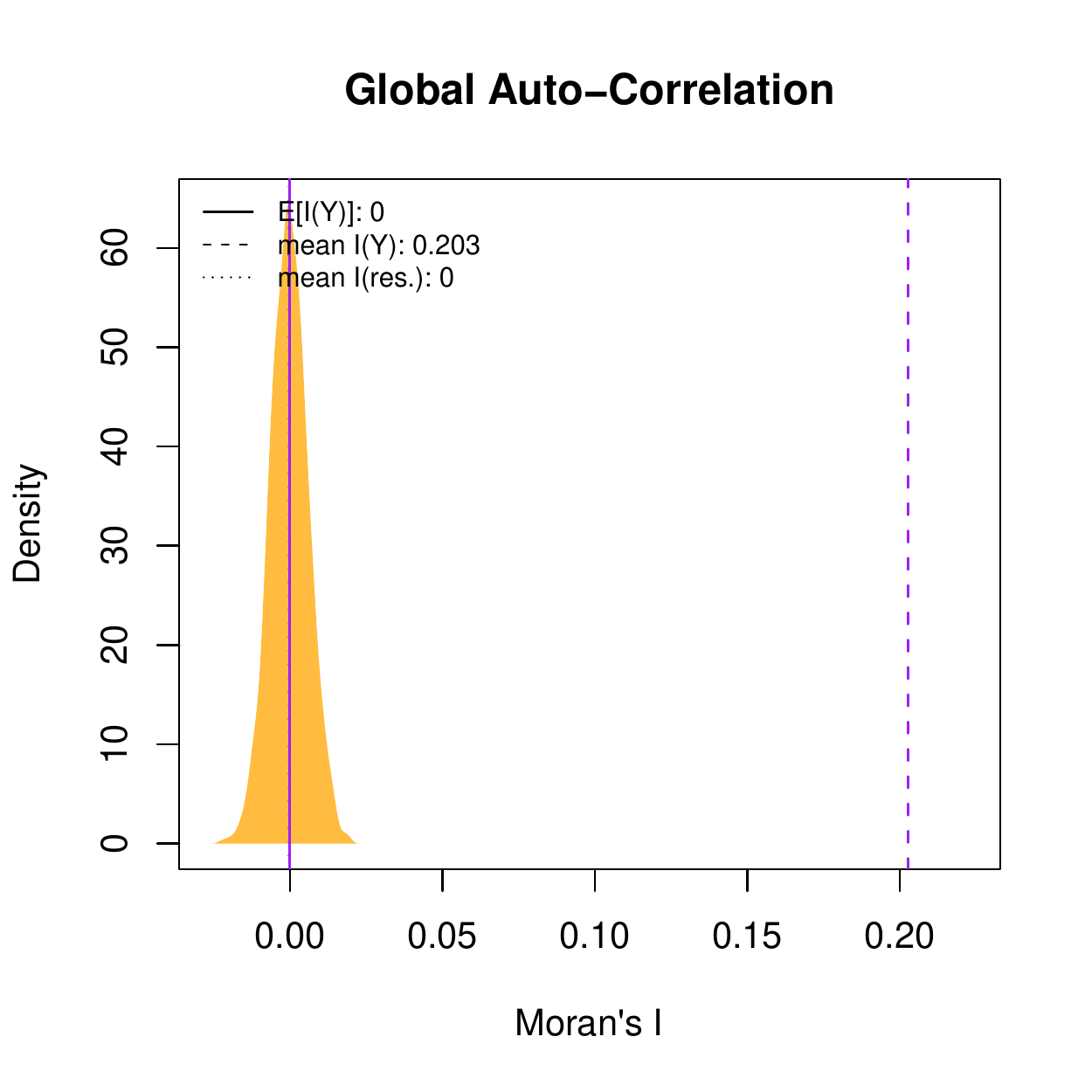}
\caption{\footnotesize{Posterior distribution of Moran's I, a coefficient of global spatial auto-correlation. The adjacency matrix implied by Figure \ref{fig:bird_fullyconnected_all} is used as the weight matrix. $I(Y)$ indicates the coefficient value prior to spatial modeling; $I(res)$ shows the complete nullification of auto-correlation as a result of the ICAR prior. The expected value under the null distribution, $E[I(Y_{null})]$, is calculated as $\frac{-1}{(n_1 + n_u) - 1}$.}} \label{fig::bird_moranI}
\end{figure}

\begin{figure}[t!]
\begin{subfigure}{0.5\textwidth}
\includegraphics[width=\linewidth]{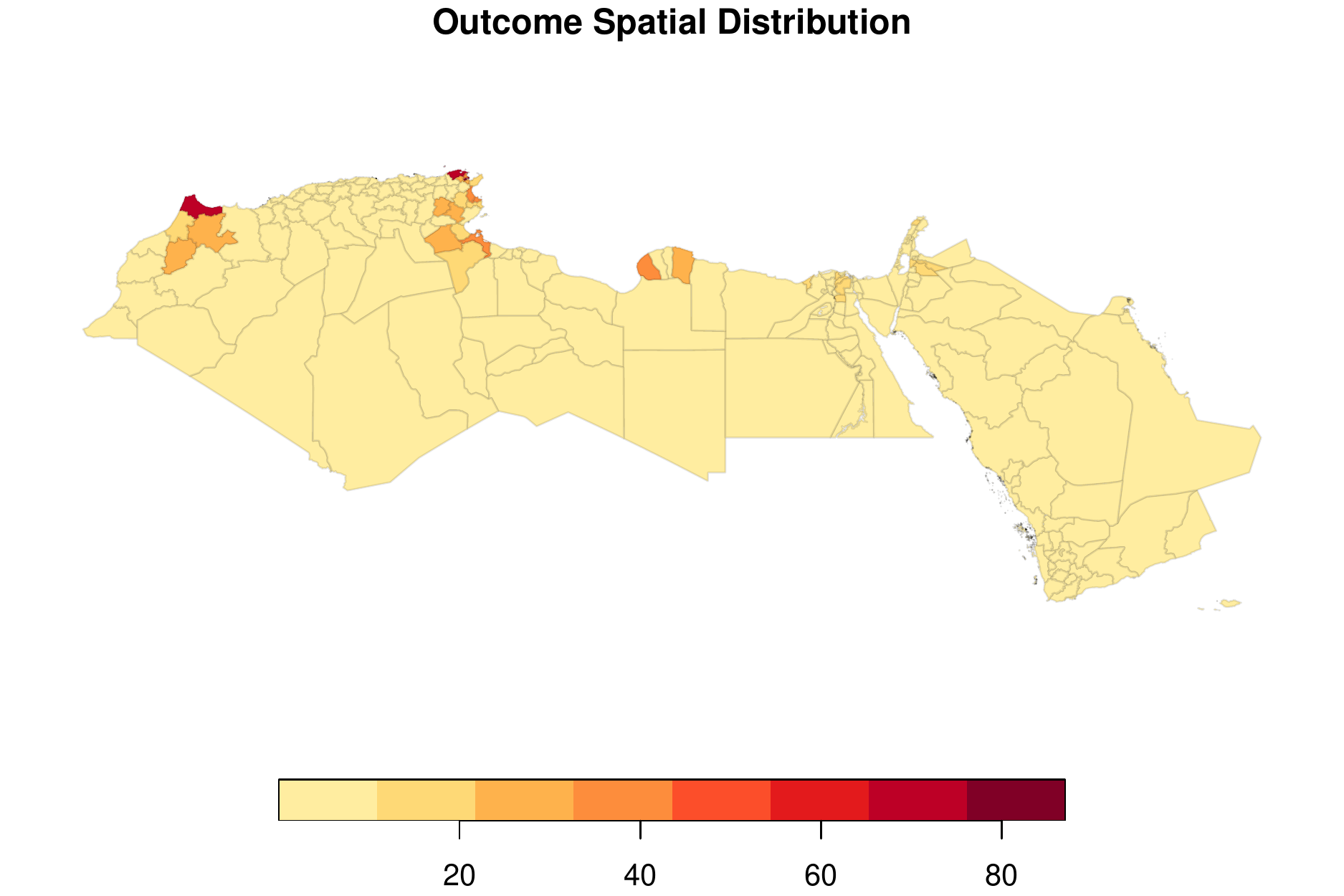}
\caption{} 
\label{fig::bird_spatial_outcome}
\end{subfigure}\hspace*{\fill}
\begin{subfigure}{0.5\textwidth}
\includegraphics[width=\linewidth]{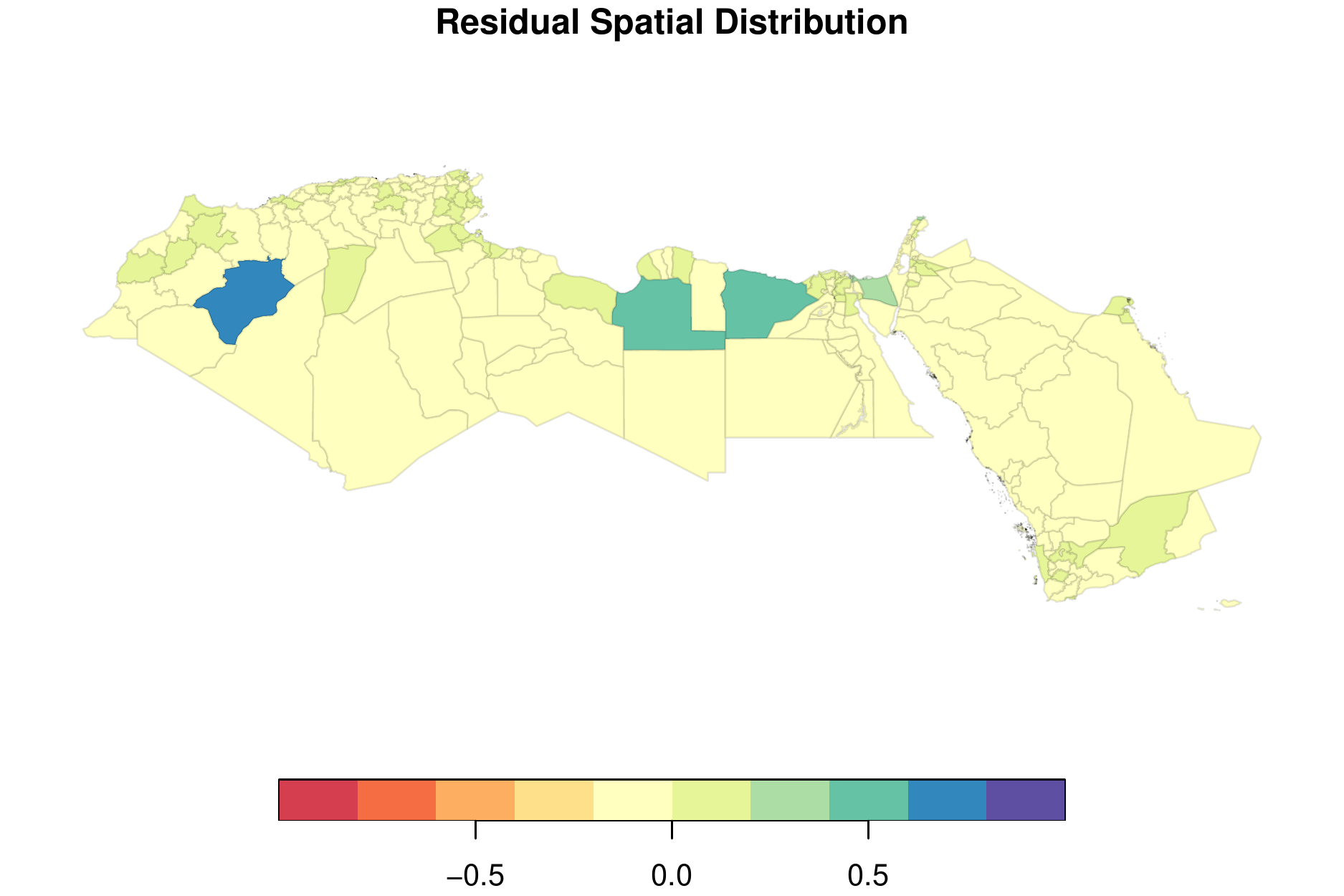}
\caption{} 
\label{fig::bird_spatial_residual}
\end{subfigure}
\caption{\footnotesize{Spatial distribution of observations (a) and residuals (b) at the Governorate level.} 
} \label{fig::bird_spatial_outcome_residual}
\end{figure}

We are satisfied that the spatial pattern implied by the adjacency matrix derived from the fully connected graph is completely extracted from the residuals, as shown by the relatively uniform color pallet of the rightmost map in Figure \ref{fig::bird_spatial_outcome_residual}, and most importantly the posterior distribution of the residuals' Moran's I in Figure \ref{fig::bird_moranI}, which is normally distributed around the expected null value. 

Figure \ref{fig::bird_spatial_outcome_residual} displays the observed number of recruits per area, along with the residual for each governorate.

In \ref{fig::bird_spatial_residual} the residual is calculated as follows: take $z = {z_1,...,z_L}$ to be the subset of individuals $i \in z_l$, who belong to small-area $l$; take $s = {1,...,S}$ to be the index of posterior sample draws; then $\mbox{res}_l = \frac{1}{S} \sum_s \left[ \frac{1}{\sum_i \mathds{1}(i \in z_l)}  \sum_{i \in z_l} \left( y_{i} - \hat{y}_{i,s} \right) \right]$. A first concern is the presence of spatial autocorrelation in the recruitment data, which could bias individual-level coefficients. The spatial distribution in Figure \ref{fig::bird_spatial_outcome} seem to suggest the possibility of spillover effects around high-density coastal areas. This is confirmed by the Moran's I ($I(Y)$), which shows statistically significant spatial auto-correlation.\footnote{As I(Y) is an observed, not modeled, quantity, it carries no uncertainty around it; it is reasonable to assume that the distribution of the $I(Y)$ would be the same as that of the $I(res)$ in terms of its shape and variance, and only differ as a result of the mean paramater. This is what is commonly assumed under standard hypothesis testing. Hence, it is easy to see that by applying the extremely narrow simulated variance around the $I(Y)$ dotted line, there would be a $0$ probability of that distribution crossing the $E[I(Y)]$ line, and hence we can say the $I(Y)$ is  highly significant. Calculating the significance of  $I(Y)$ in frequentist terms, using the \texttt{ape} package, reveals a p-value of $0$. We plot and describe our calculation for Moran's I in Appendix Figure \ref{fig::bird_moranI}}

\begin{figure}[htp]
    \centering
    \includegraphics[scale = 0.5]{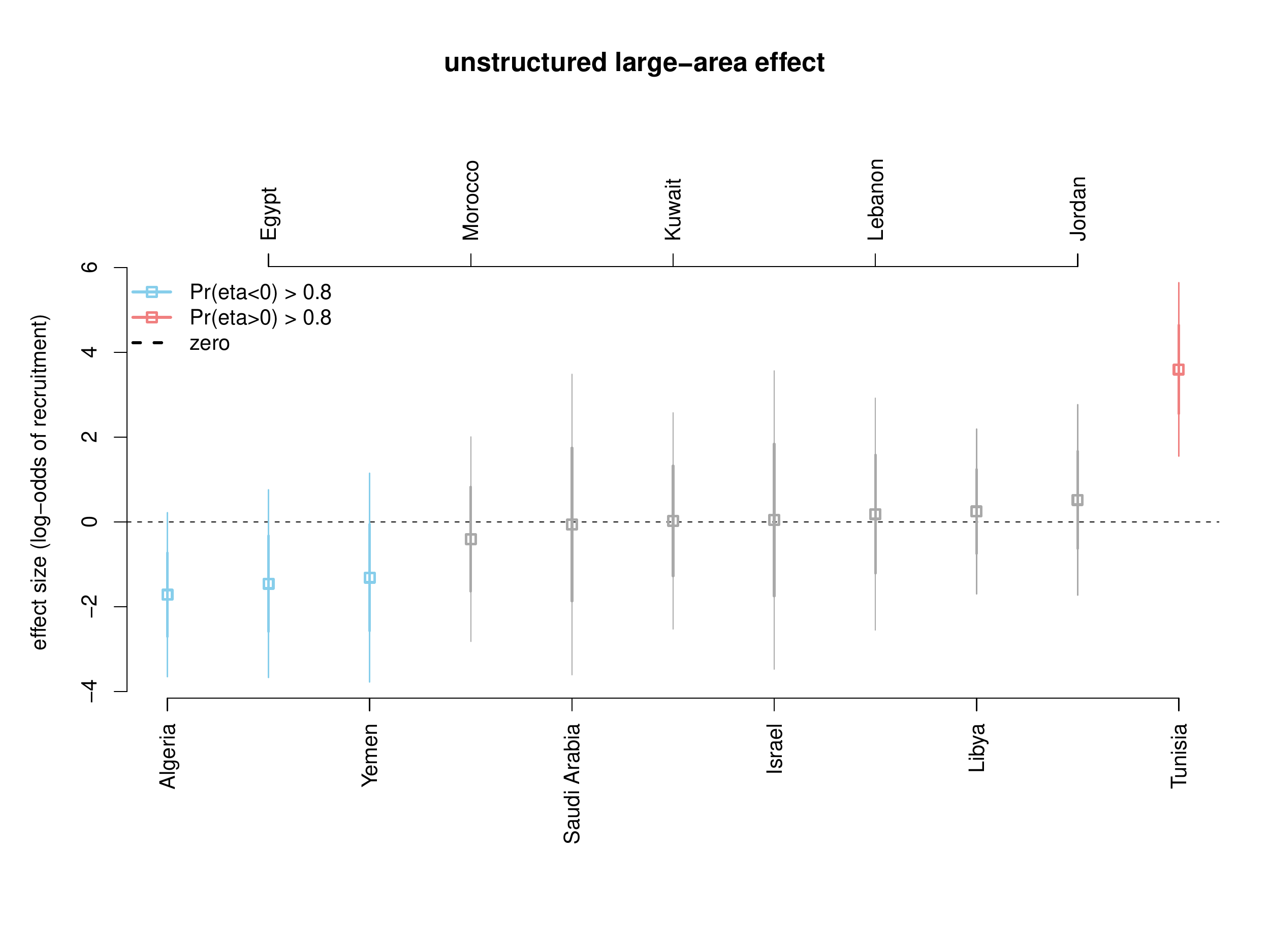}
    \caption{\footnotesize{Unstructured large-area effect $\eta$ for the `Bird's Eye' model.}}
    \label{fig:bird_gov}
\end{figure}

\begin{landscape}

\begin{figure}[htp]
    \flushleft
    \includegraphics[scale = 0.475]{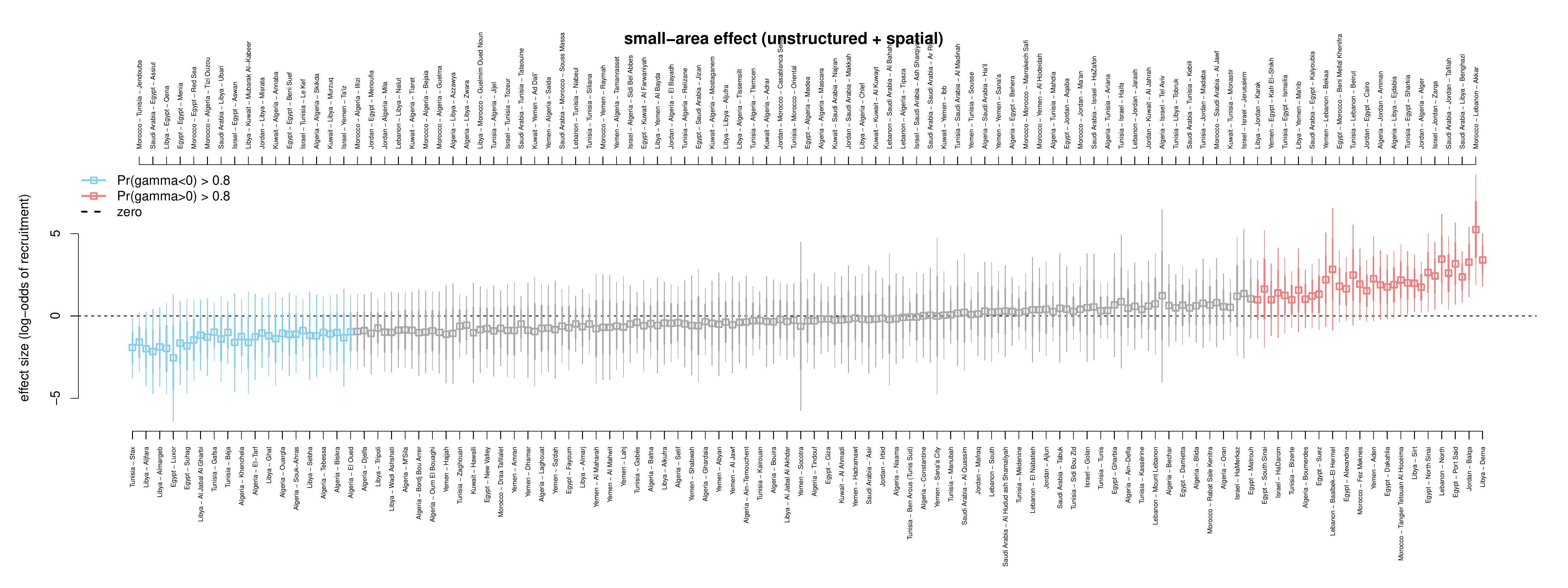}
   \caption{\footnotesize{BYM2 small-area effect $\gamma$ for the `Bird's Eye' model.}}
    \label{fig:bird_dist}
\end{figure}
\end{landscape}

We display below the spatial distribution of the point estimates for Governorate and Country-level random effects in Figure \ref{fig:bird_spatial_effects}. The corresponding prediction intervals for country and governorate effects are 
shown in Figure \ref{fig:bird_gov} for country-effects and Figure \ref{fig:bird_dist} for Governorate effects. It is worth noting that part of the reason for heightened recruitment propensity around the eastern Governorates could be the increasing proximity to Syria and the ISIS caliphate itself, as well as higher proportions of refugees from destabilized regions of Syria, and in general more potential for pro-ISIS unobservable network-dynamics. We see a strong unexplained effect in Tunisia, highlighting unobserved but systematic variance in favour of recruitment 
, while Algeria, Egypt and Yemen show significant unexplained negative effects on recruitment over and above their spatial and unstructured Governorate-level variance.

\begin{figure}[!b] 

\begin{subfigure}{0.5\textwidth}
\includegraphics[width=\linewidth]{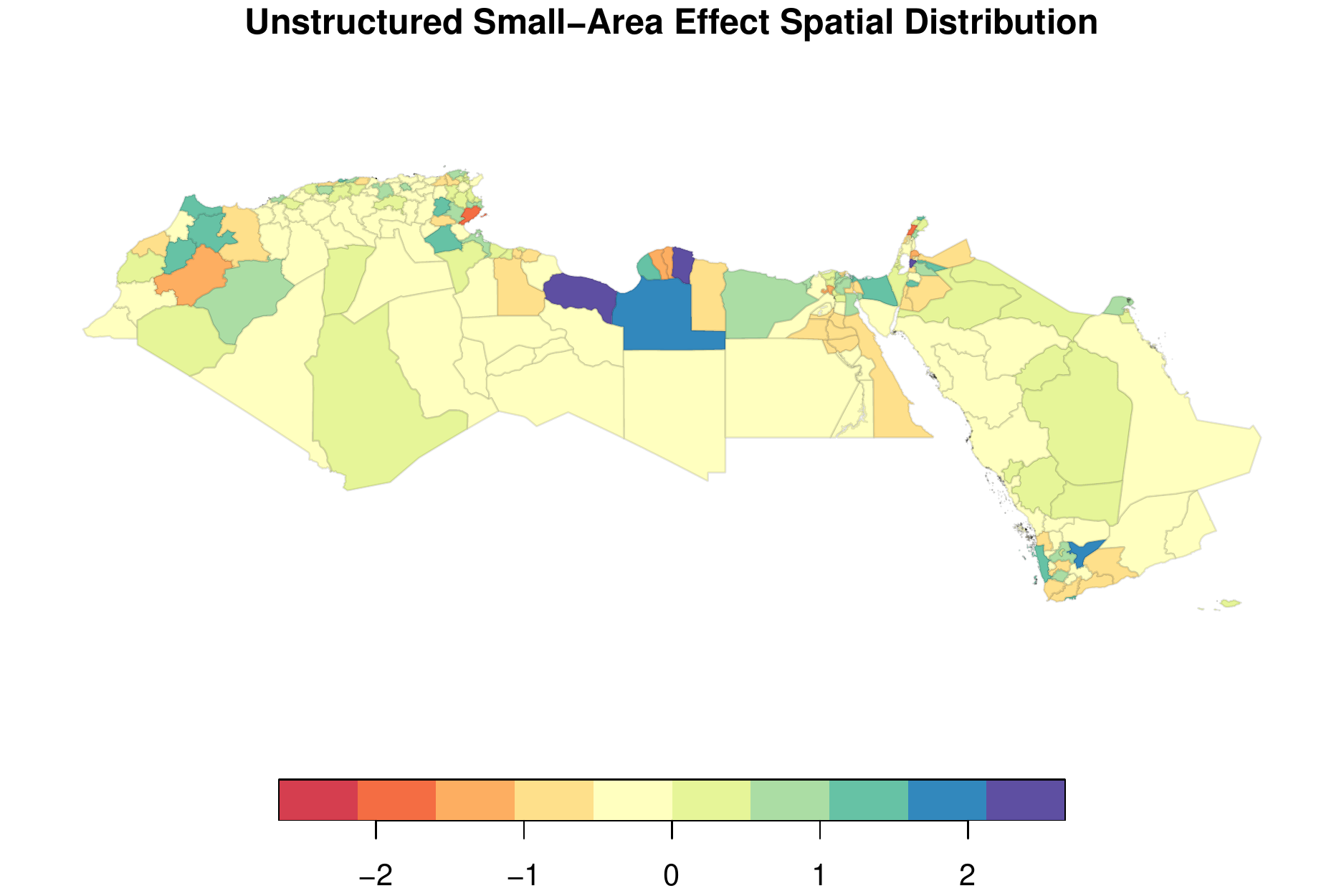}
\caption{} 
\end{subfigure}\hspace*{\fill}
\begin{subfigure}{0.5\textwidth}
\includegraphics[width=\linewidth]{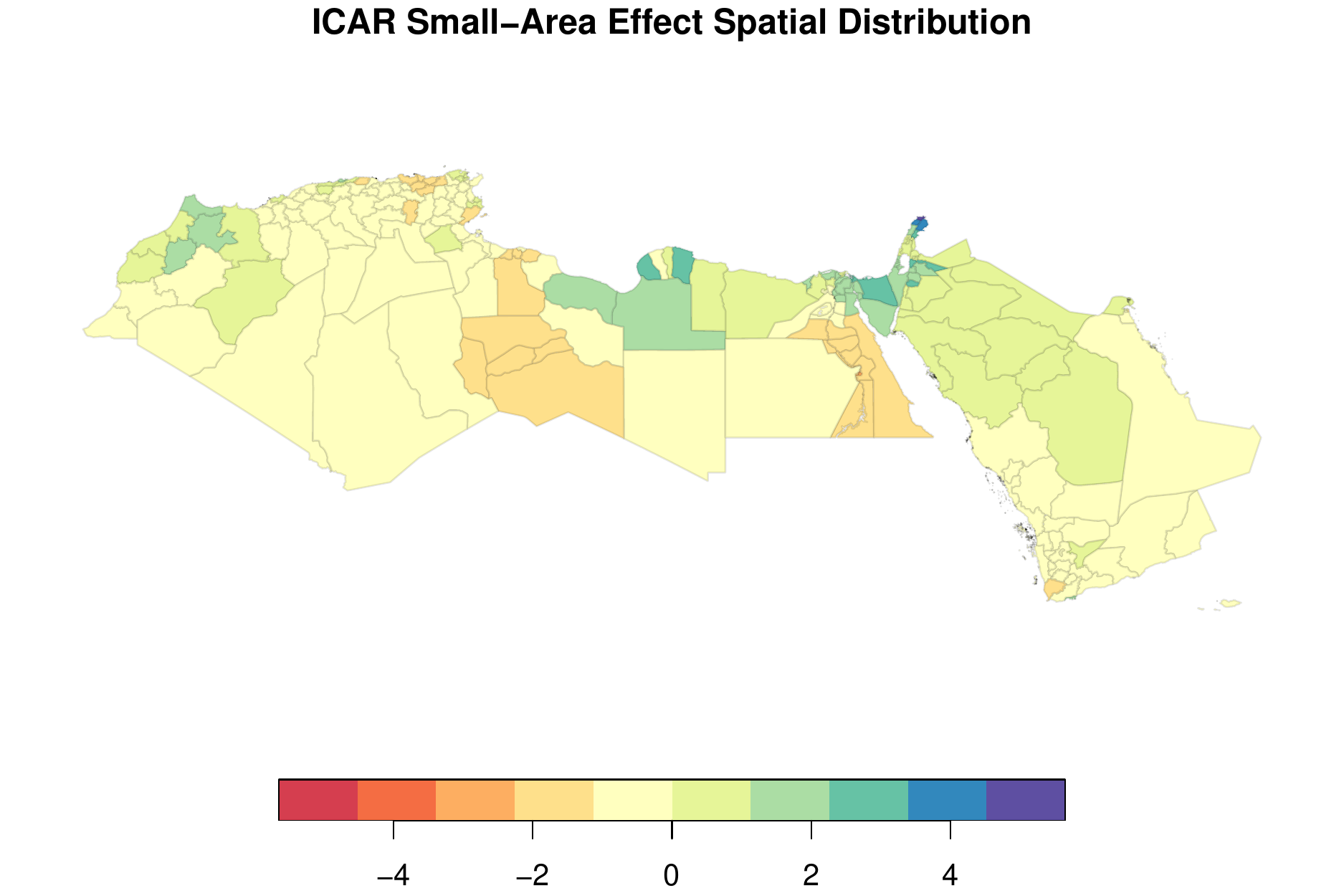}
\caption{} 
\end{subfigure}

\end{figure}%
\begin{figure}[ht]\ContinuedFloat
\begin{subfigure}{0.5\textwidth}
\includegraphics[width=\linewidth]{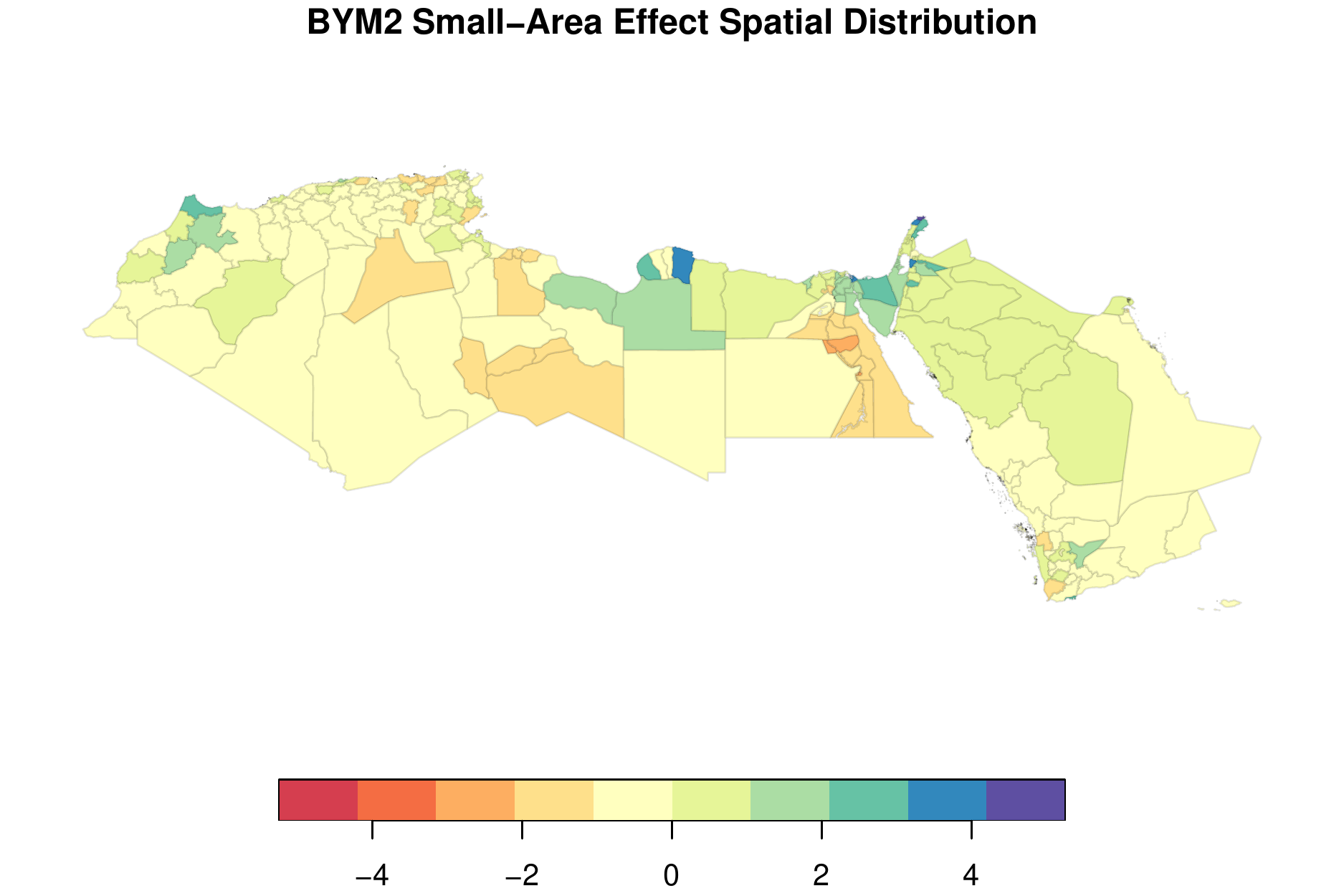}
\caption{} 
\end{subfigure}\hspace*{\fill}
\begin{subfigure}{0.5\textwidth}
\includegraphics[width=\linewidth]{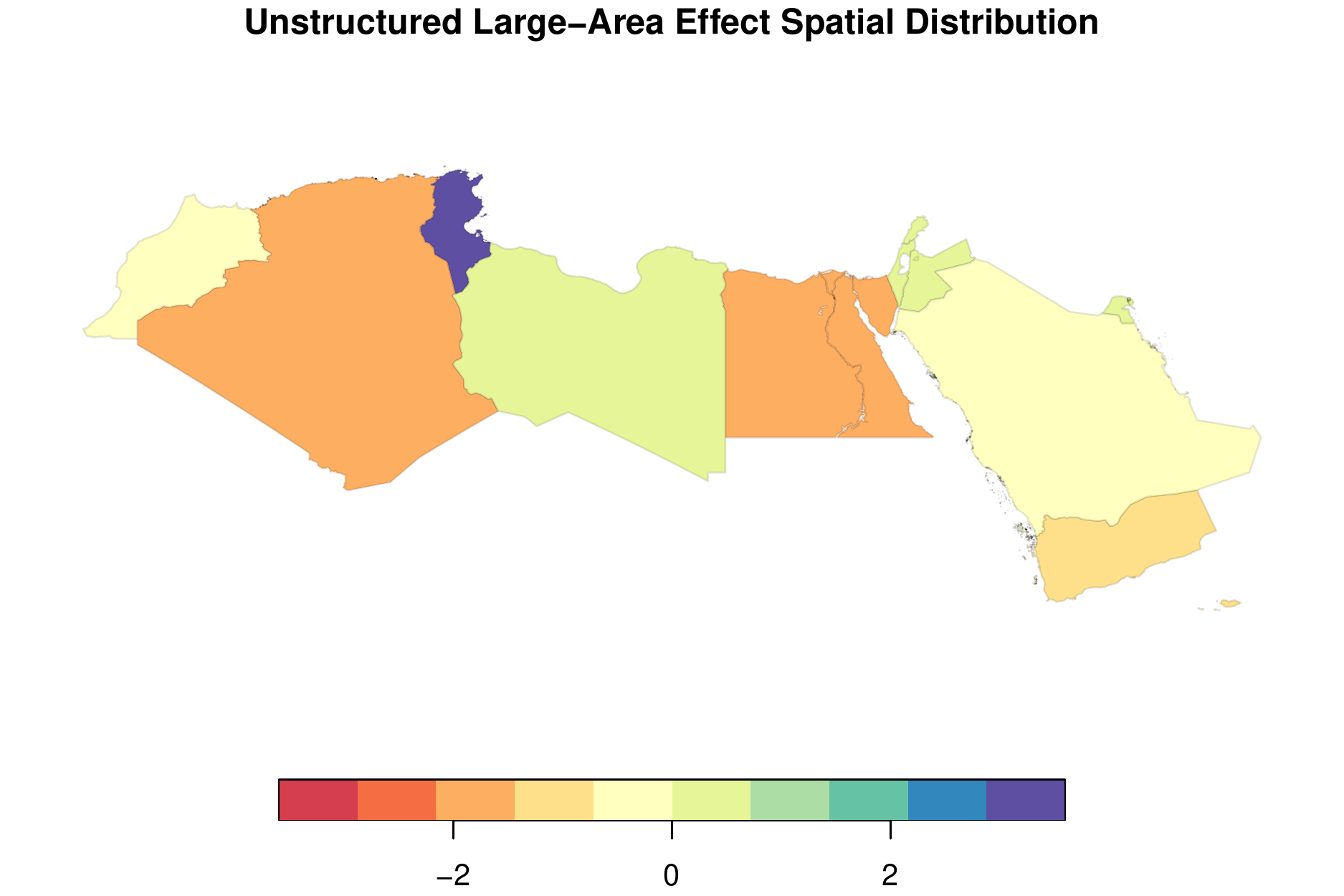}
\caption{} 
\end{subfigure}
\caption{\footnotesize{Spatial distribution of: (a) the unstructured Governorate-level effect - $\phi$; (b) the spatial Governorate level effect - $\psi$; (c) the total Governorate effect - $\gamma = \sigma(\phi\sqrt{1-\lambda)} + \psi\sqrt{\lambda/s})$; (d) the unstructured Country effect - $\eta$.}} \label{fig:bird_spatial_effects}
\end{figure}

\pagebreak

\begin{figure}[htp]
\begin{subfigure}{0.5\textwidth}
\includegraphics[width=\linewidth]{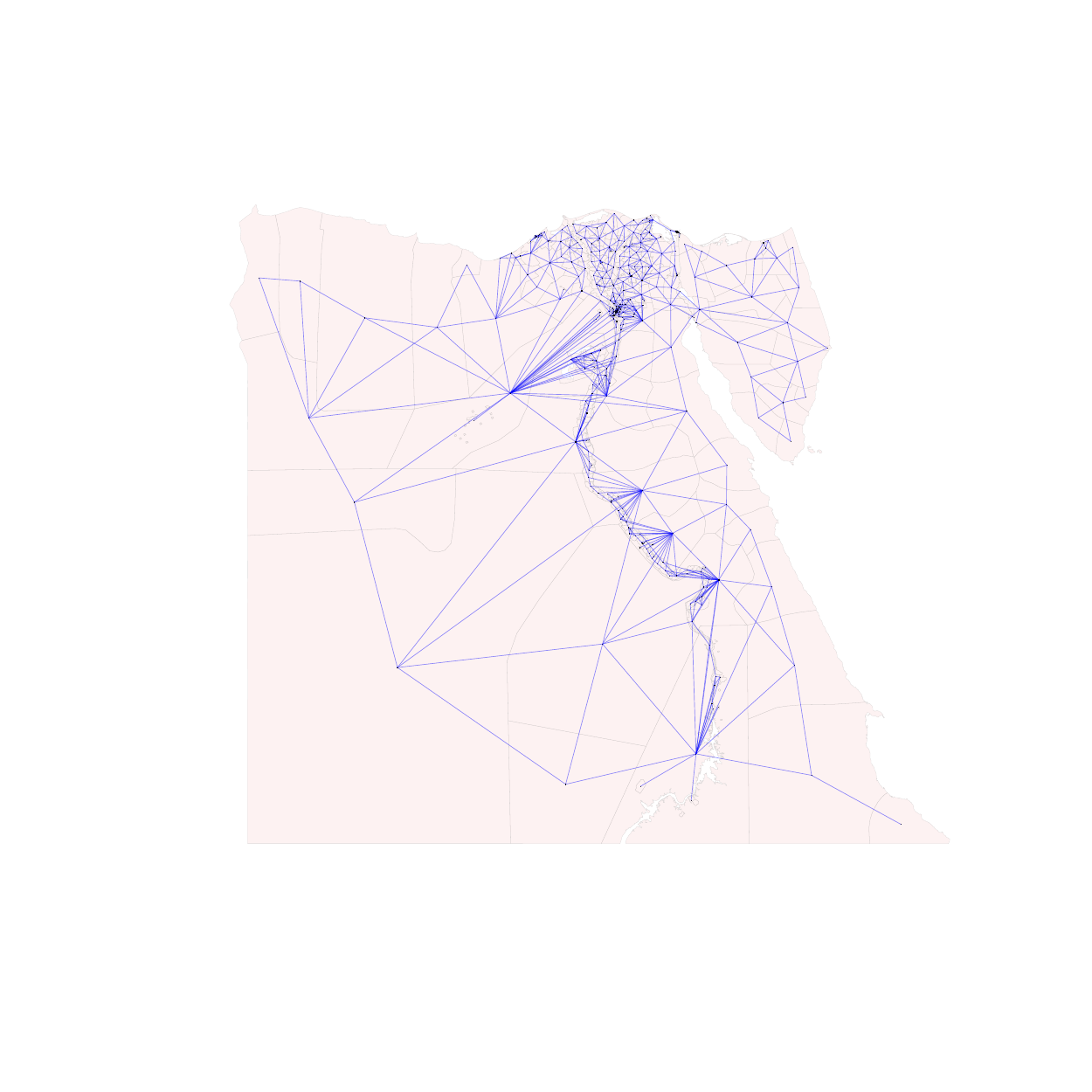}
\caption{} \label{fig::worm_fullyconnected_egypt}
\end{subfigure}\hspace*{\fill}
\begin{subfigure}{0.5\textwidth}
\includegraphics[width=\linewidth]{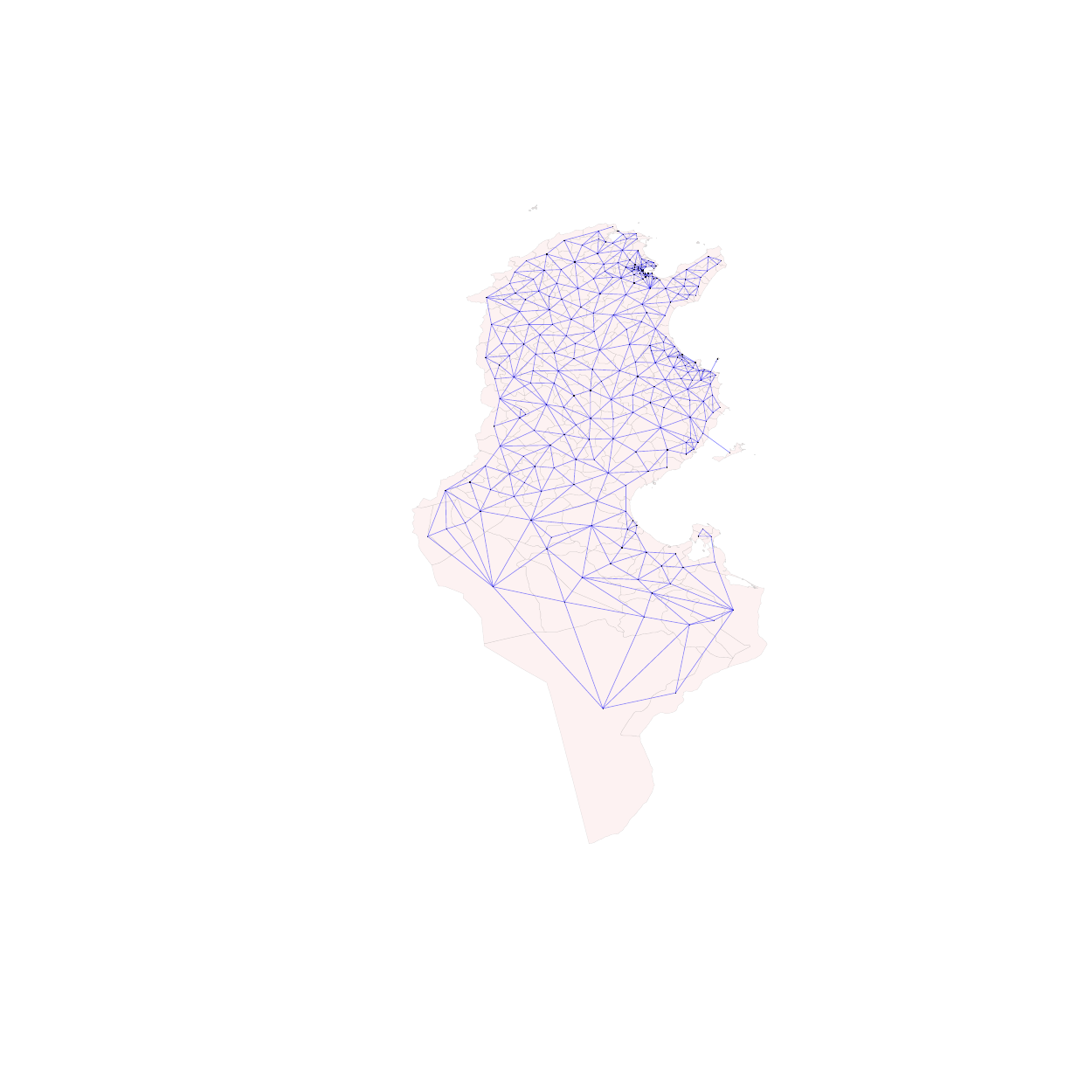}
\caption{} \label{fig::bird_fullyconnected_tunisia}
\end{subfigure}
\caption{\footnotesize{Fully-connected graphs of (a) Egypt and (b) Tunisia at the District level.}} \label{fig::worm_fullyconnected}
\end{figure}

Figure \ref{fig::worm_fullyconnected} presents the Egypt and Tunisia fully-connected graphs used to derive the district-level adjacency matrices fed to the ICAR model. Again, a small number of adjustments were made to connect islands and ensure full-connectivity.

\begin{figure}[t!]
\begin{subfigure}{0.5\textwidth}
\includegraphics[width=\linewidth]{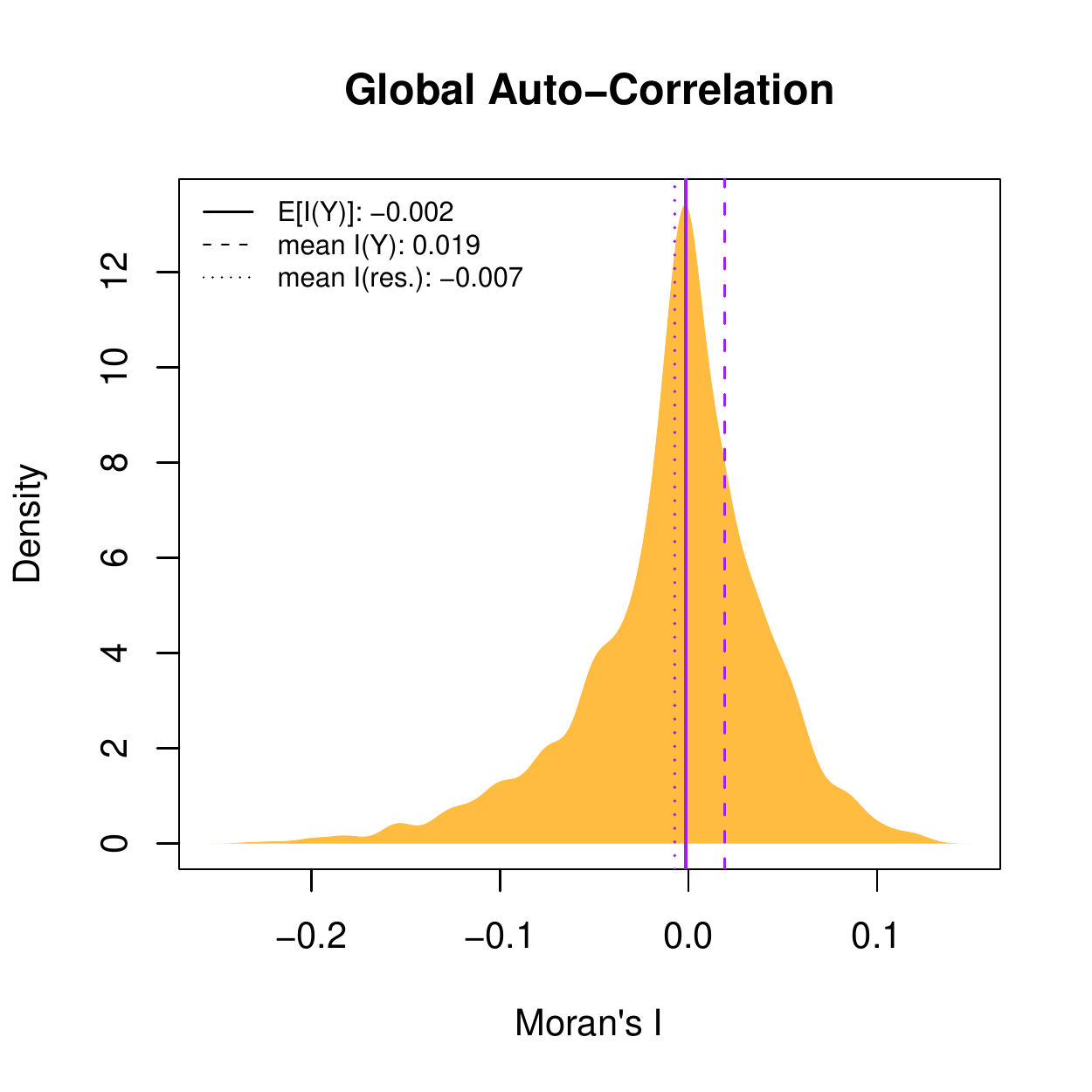}
\caption{} 
\end{subfigure}\hspace*{\fill}
\begin{subfigure}{0.5\textwidth}
\includegraphics[width=\linewidth]{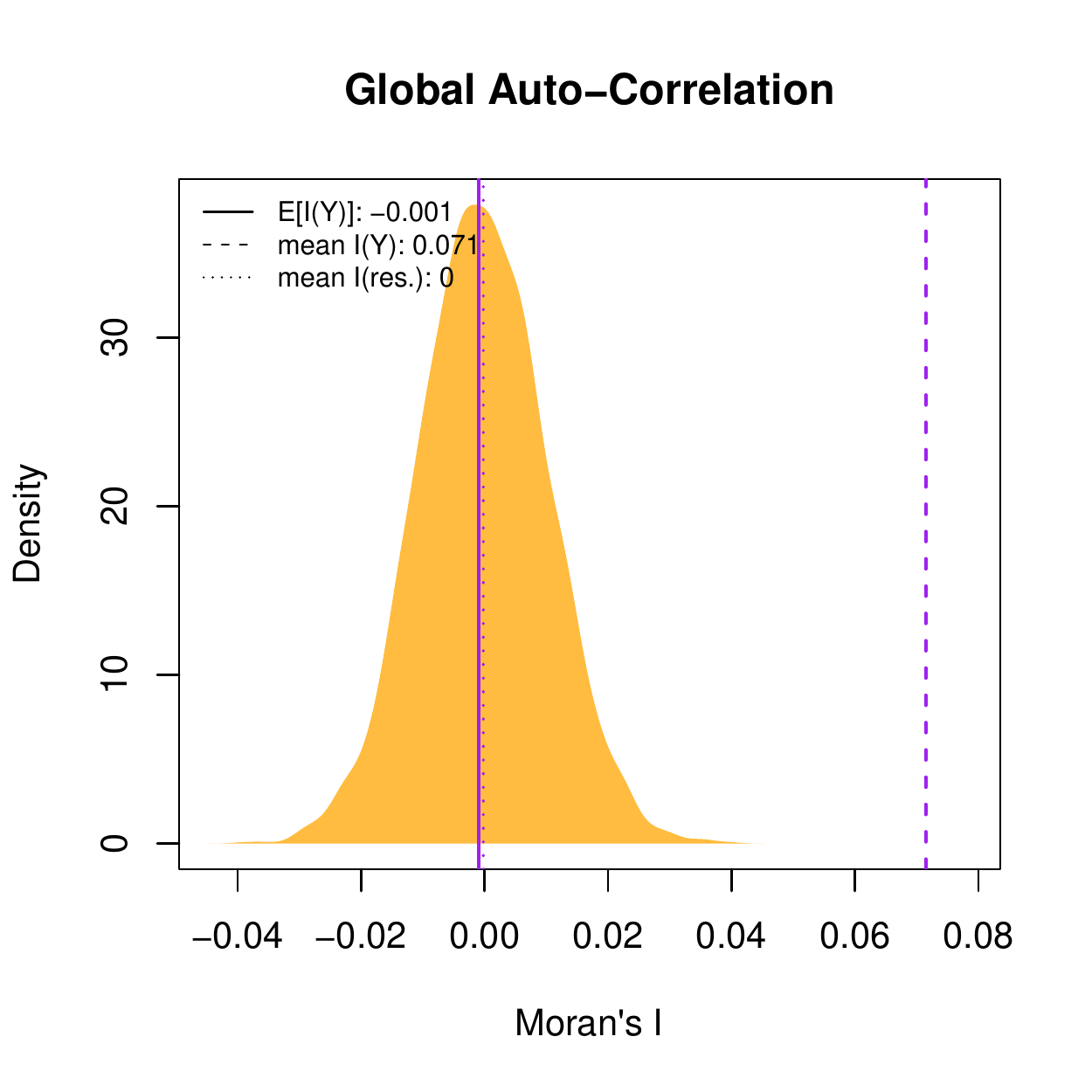}
\caption{} 
\end{subfigure}
\caption{\footnotesize{Posterior distribution of Moran's I for Egypt (a) and Tunisia (b). The adjacency matrices implied by Figure \ref{fig::worm_fullyconnected} are used as the weight matrices.}} \label{fig::worm_moranI}
\end{figure}

\begin{figure}
\begin{subfigure}{0.525\textwidth}
\includegraphics[width=\linewidth]{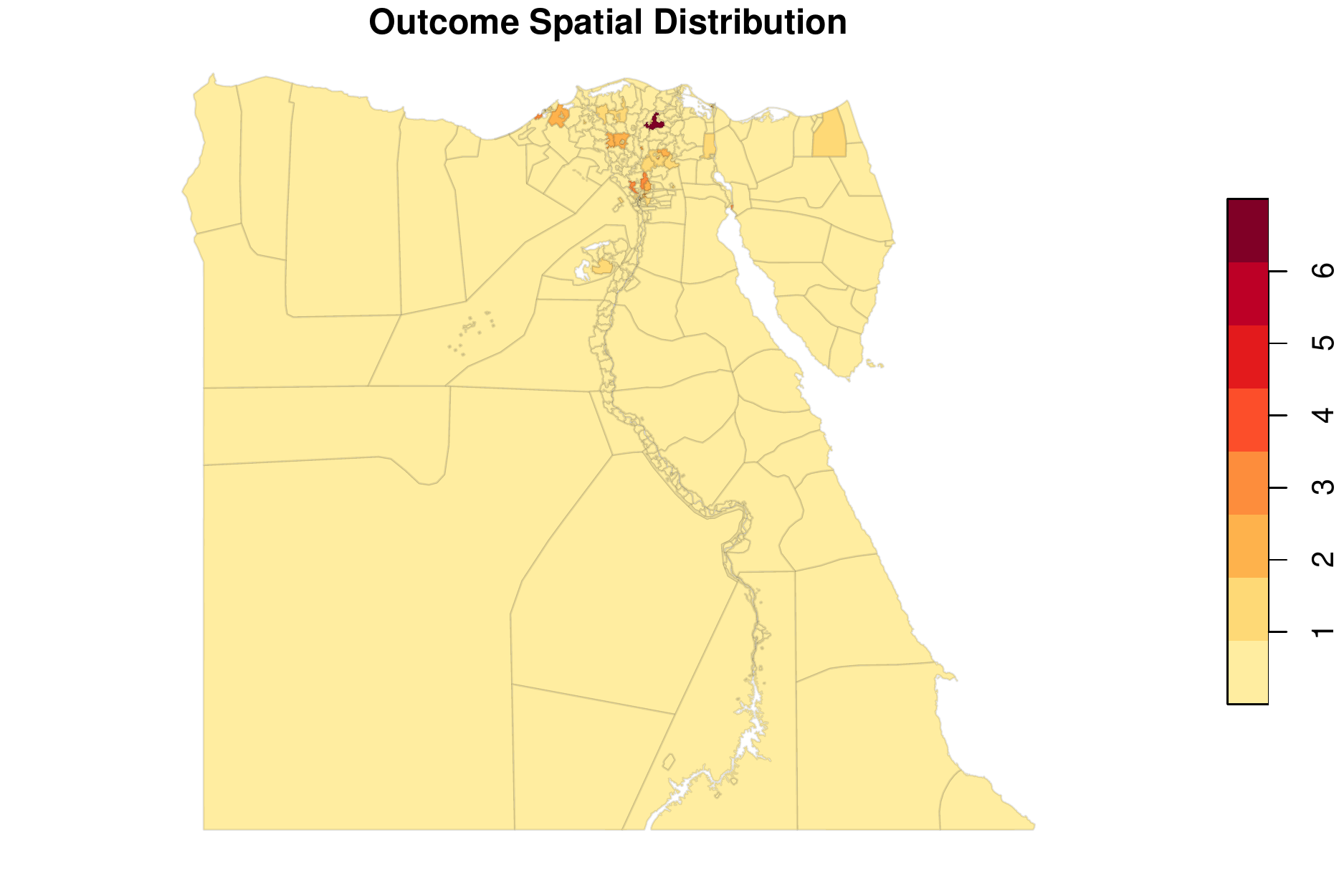}
\caption{} 
\end{subfigure}\hspace*{\fill}
\begin{subfigure}{0.525\textwidth}
\includegraphics[width=\linewidth]{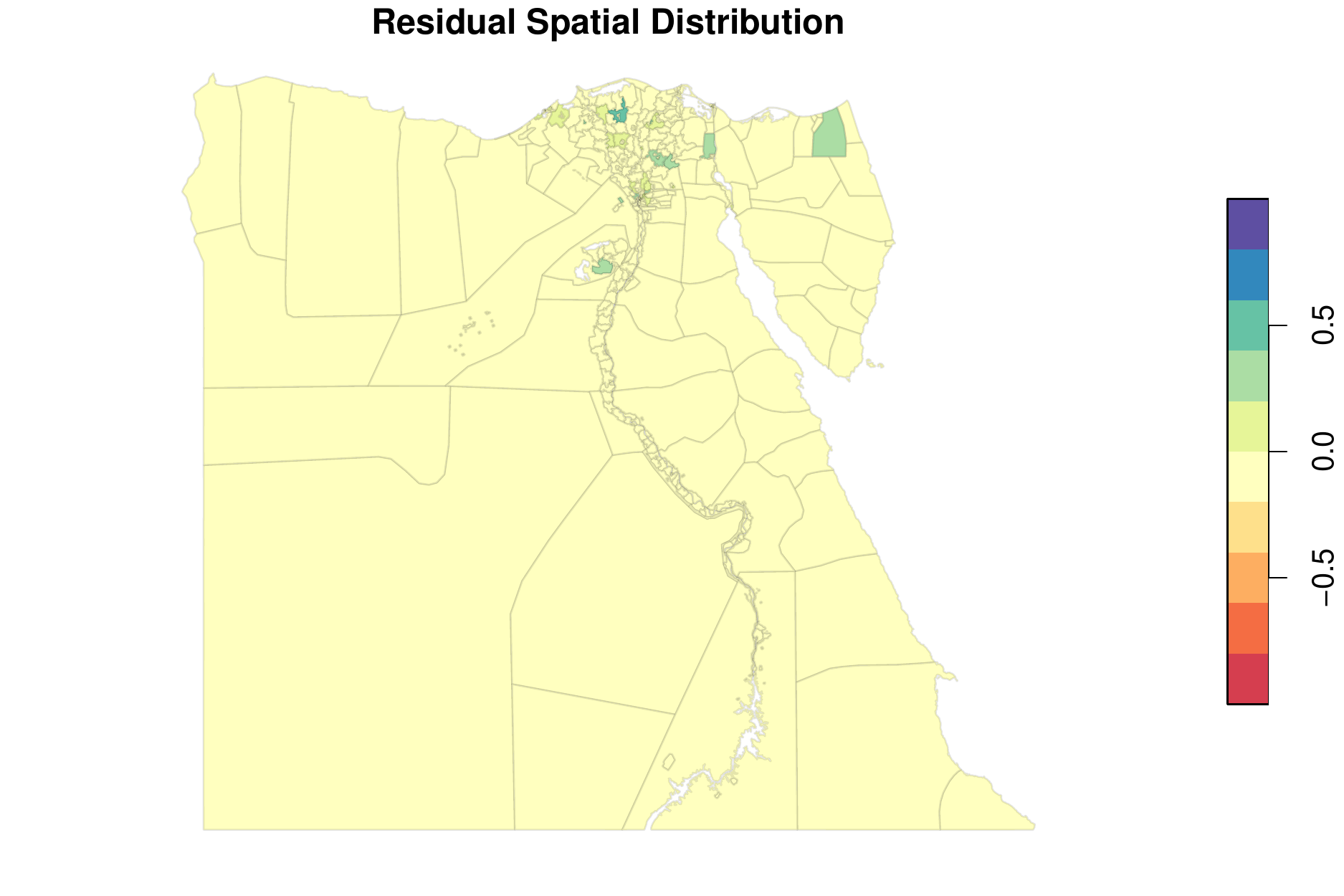}
\caption{} 
\end{subfigure}
\medskip
\begin{subfigure}{0.525\textwidth}
\includegraphics[width=\linewidth]{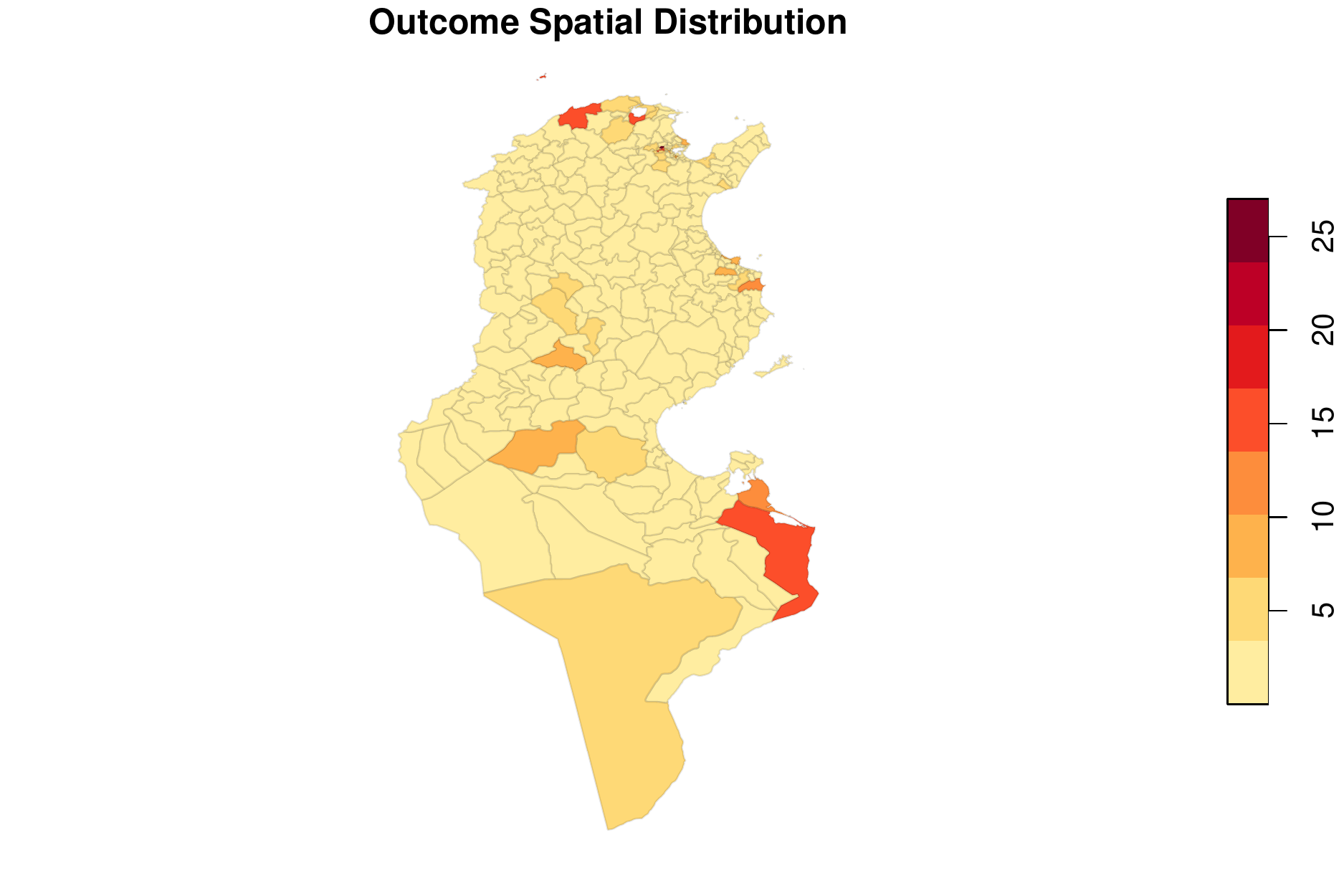}
\caption{} 
\end{subfigure}\hspace*{\fill}
\begin{subfigure}{0.525\textwidth}
\includegraphics[width=\linewidth]{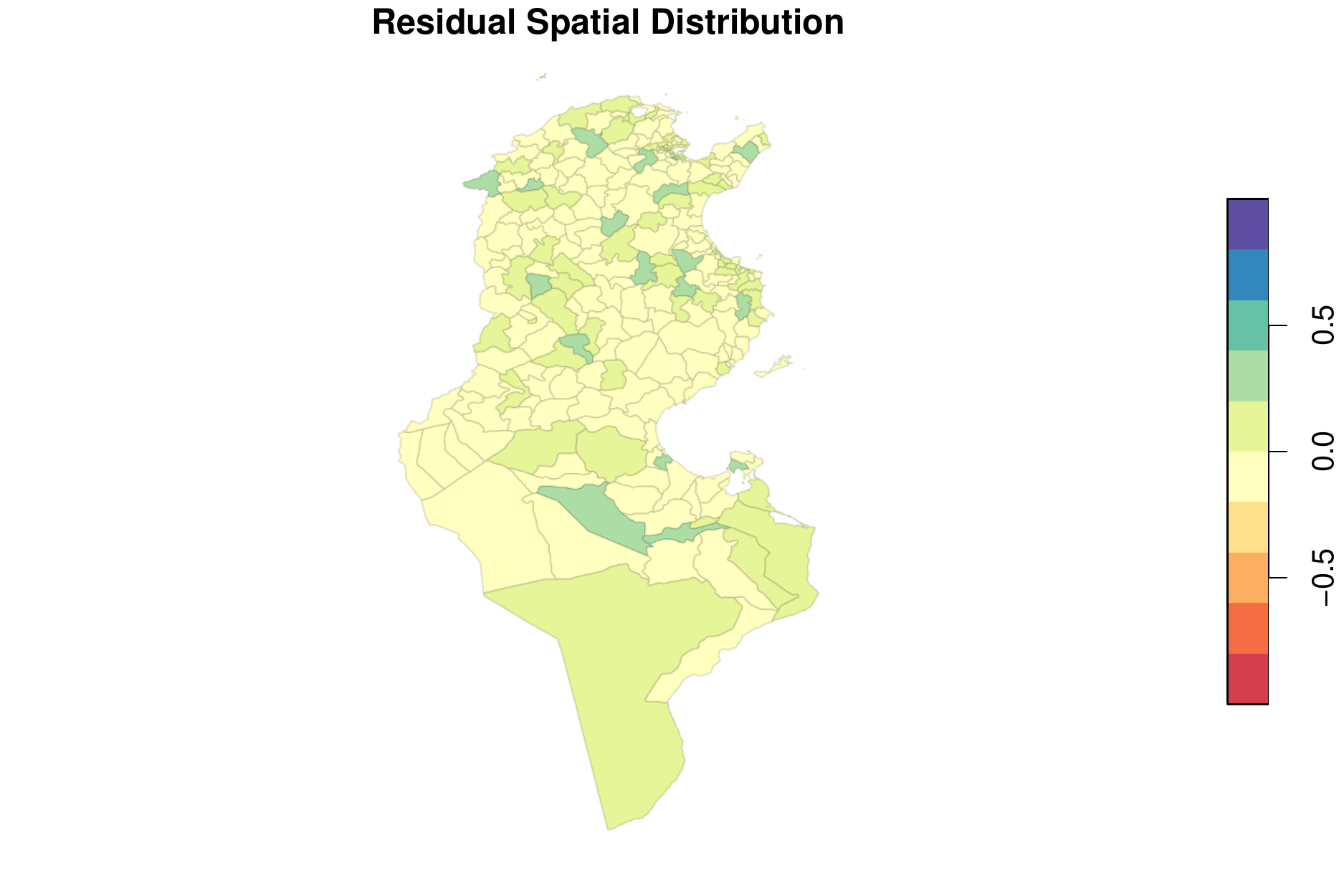}
\caption{} 
\end{subfigure}
\caption{\footnotesize{Spatial distribution of Egyptian observations (a) and residuals (b); Tunisian observations (c) and residuals (d) at the District level. (a) and (c) present the spatial distribution of recruits.}}
\label{fig::worm_spatial_outcome_residual}
\end{figure}

The residual plots in Figure \ref{fig::worm_spatial_outcome_residual}, along with the Moran's I presented in Figure \ref{fig::worm_moranI}, convincingly show we have extracted all spatial variance from the observations: the resulting Moran's Is are distributed around the null-value.

\begin{figure}[htp] 

\medskip
\begin{subfigure}{0.5\textwidth}
\includegraphics[width=\linewidth]{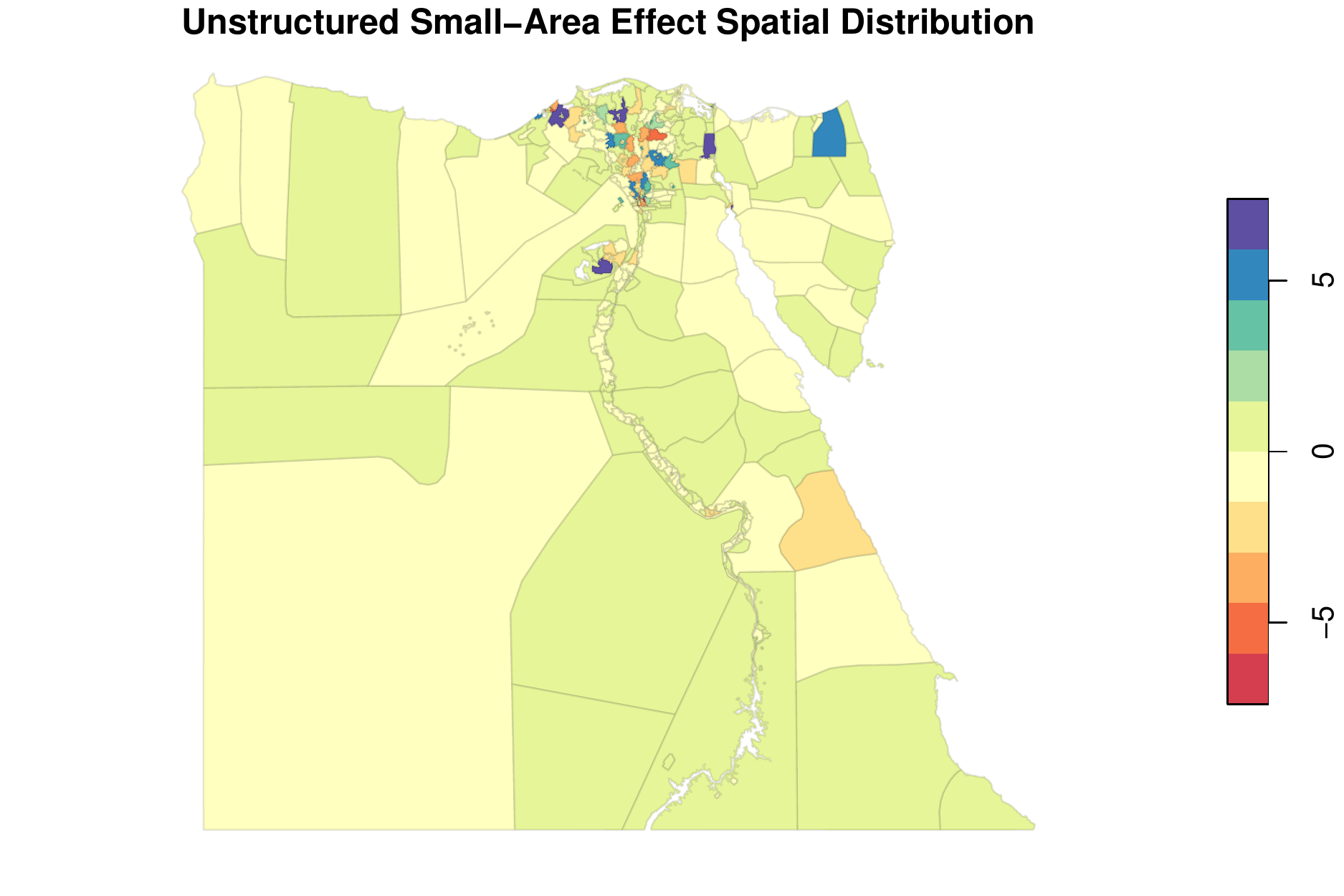}
\caption{} 
\end{subfigure}\hspace*{\fill}
\begin{subfigure}{0.5\textwidth}
\includegraphics[width=\linewidth]{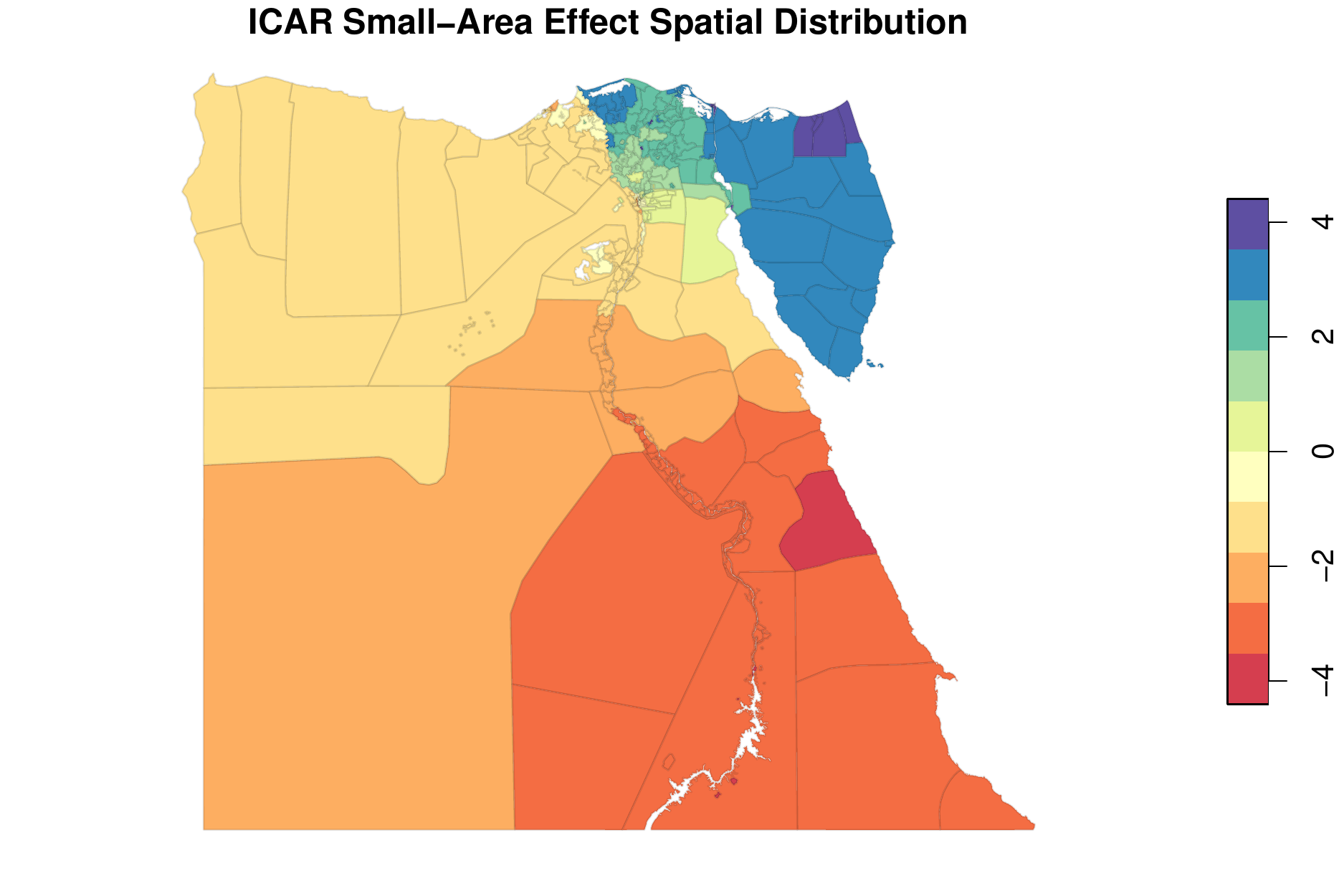}
\caption{} 
\end{subfigure}

\medskip
\begin{subfigure}{0.5\textwidth}
\includegraphics[width=\linewidth]{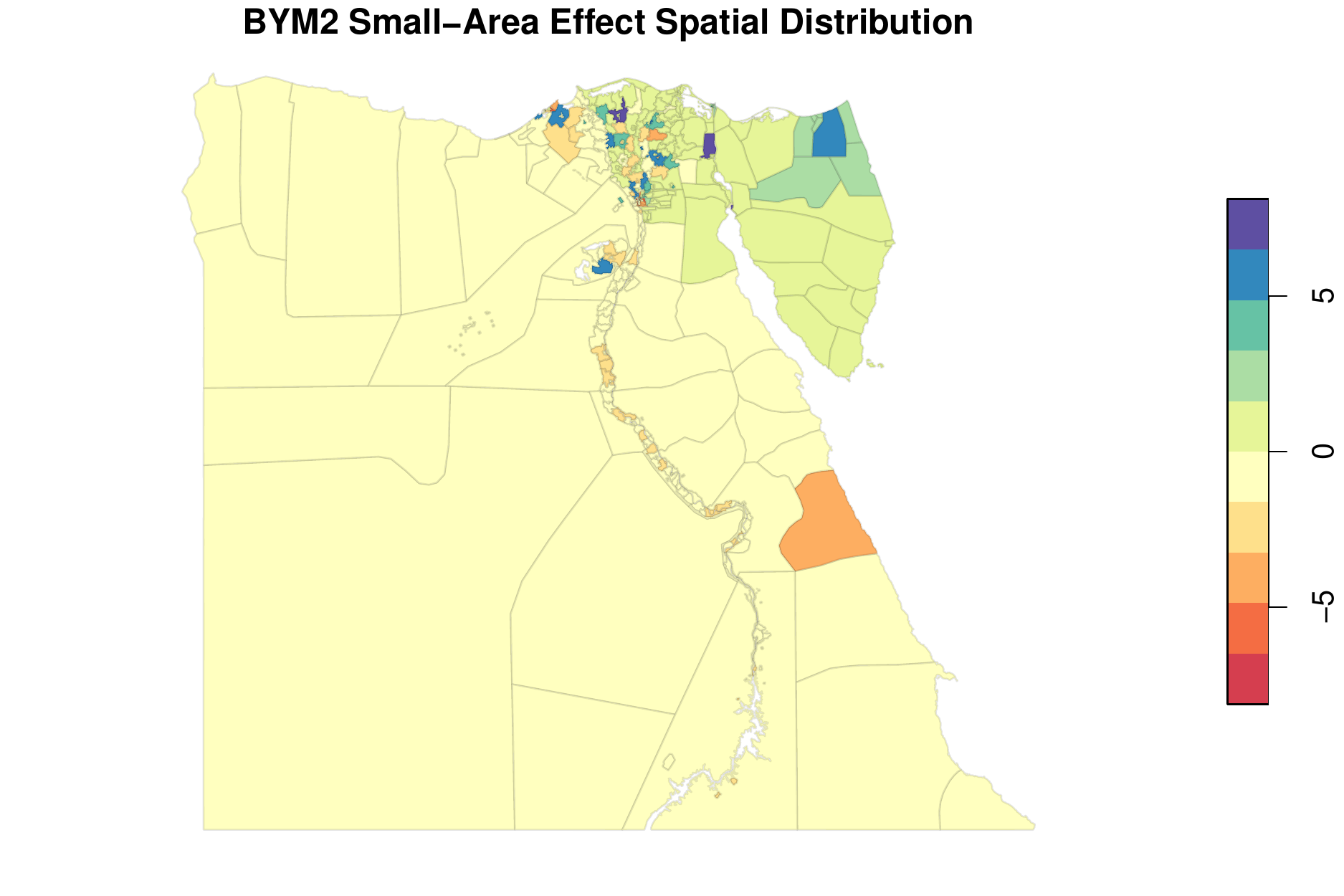}
\caption{} 
\end{subfigure}\hspace*{\fill}
\begin{subfigure}{0.5\textwidth}
\includegraphics[width=\linewidth]{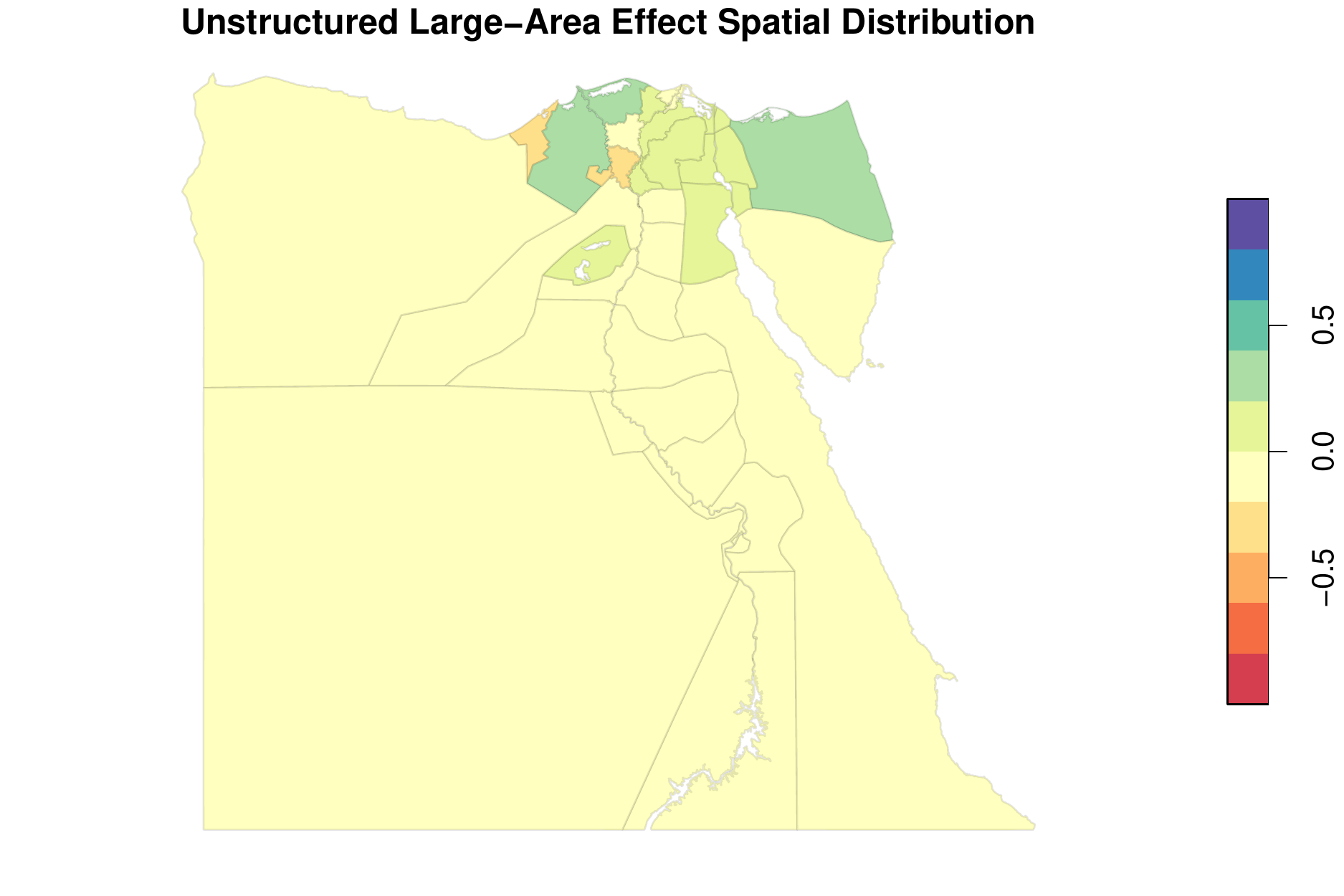}
\caption{} 
\end{subfigure}

\caption{\footnotesize{Egypt's Spatial distribution of: (a) the unstructured Governorate-level effect - $\phi$; (b) the spatial Governorate level effect - $\psi$; (c) the total Governorate effect - $\gamma = \sigma(\phi\sqrt{1-\lambda)} + \psi\sqrt{\lambda/s})$; (d) the unstructured Country effect - $\eta$. }} \label{fig:egypt_spatial_effects}
\end{figure}

\begin{figure}[htp] 

\medskip
\begin{subfigure}{0.5\textwidth}
\includegraphics[width=\linewidth]{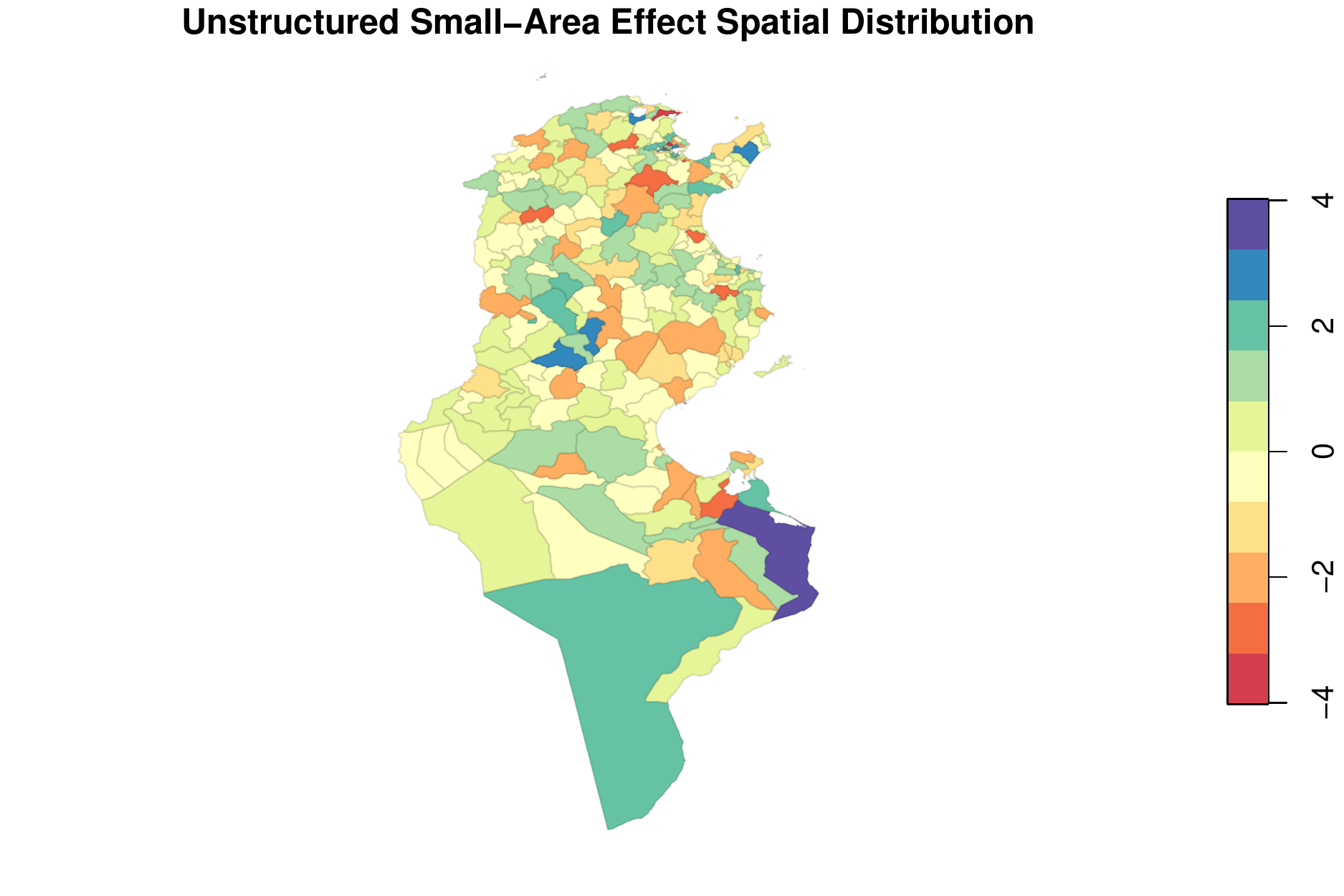}
\caption{} 
\end{subfigure}\hspace*{\fill}
\begin{subfigure}{0.5\textwidth}
\includegraphics[width=\linewidth]{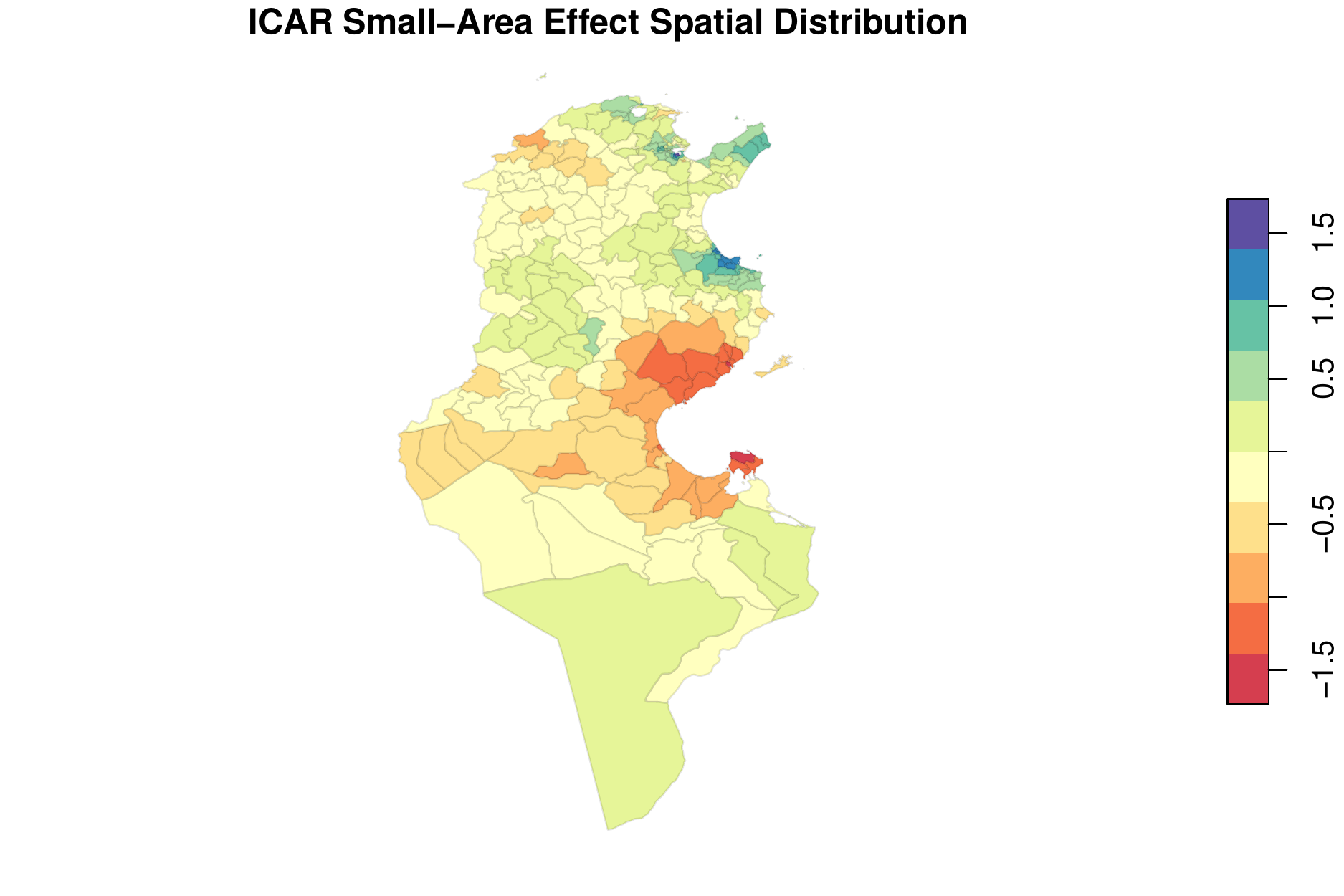}
\caption{} 
\end{subfigure}

\medskip
\begin{subfigure}{0.5\textwidth}
\includegraphics[width=\linewidth]{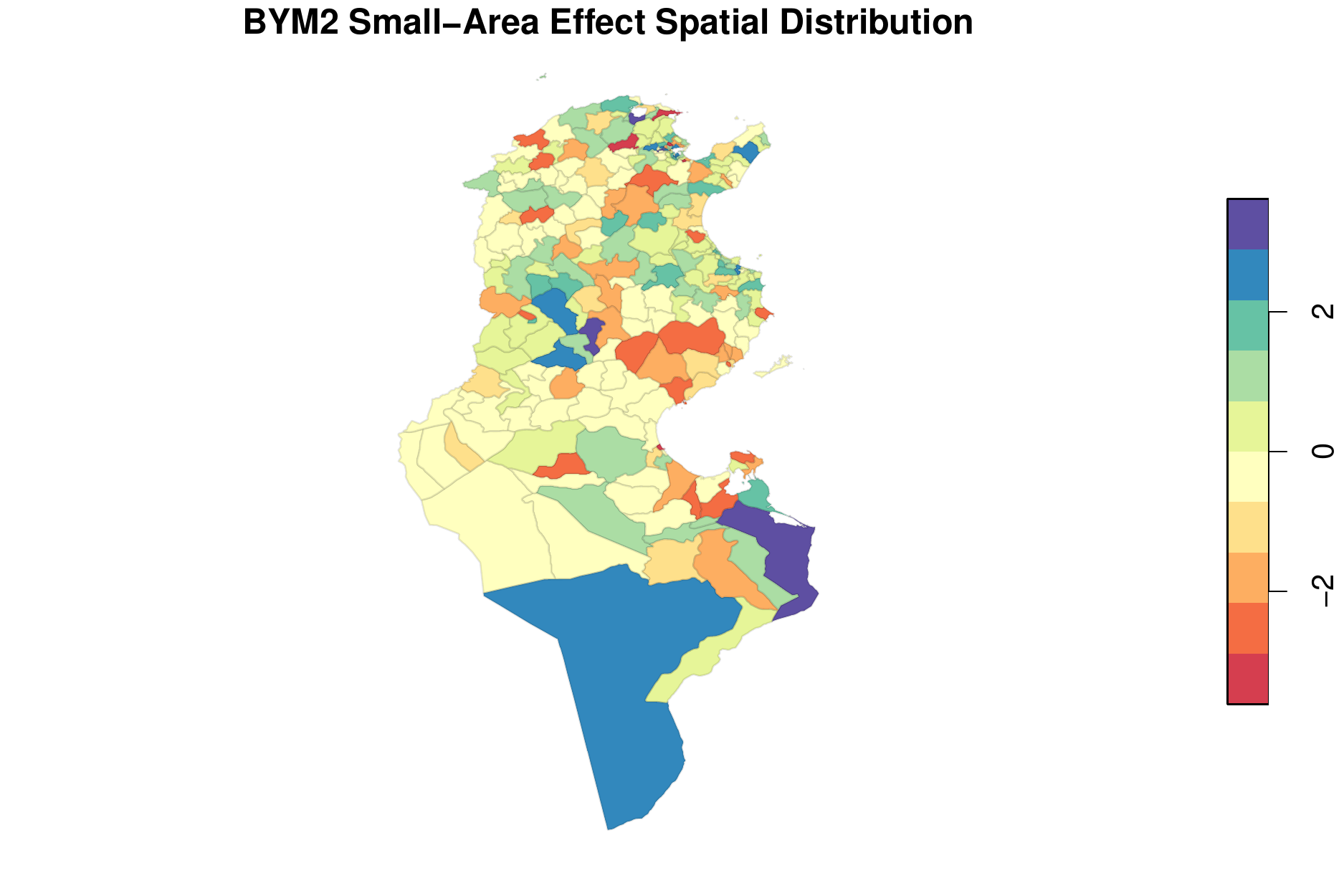}
\caption{} 
\end{subfigure}\hspace*{\fill}
\begin{subfigure}{0.5\textwidth}
\includegraphics[width=\linewidth]{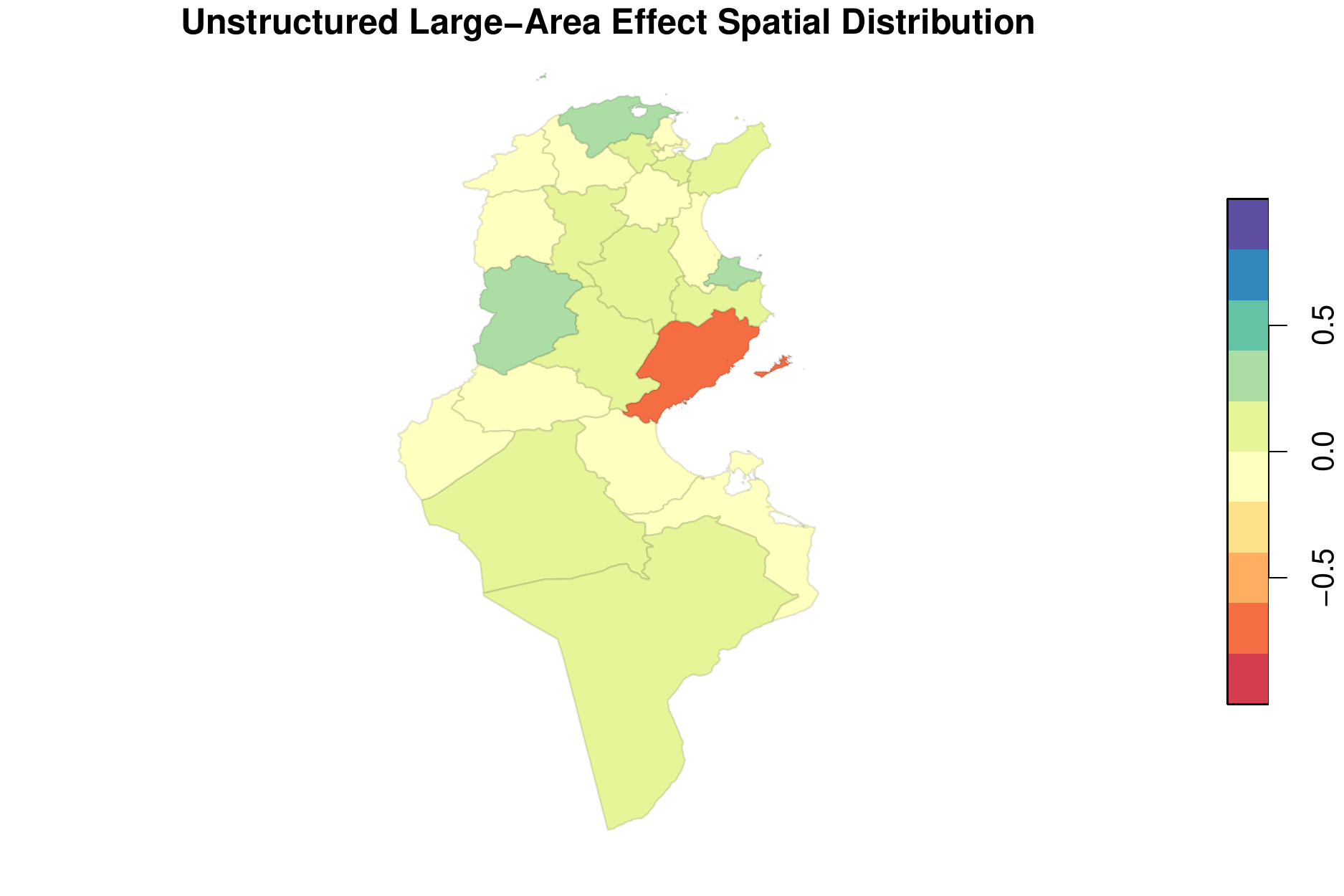}
\caption{} 
\end{subfigure}

\caption{\footnotesize{Tunisia's Spatial distribution of: (a) the unstructured Governorate-level effect - $\phi$; (b) the spatial Governorate level effect - $\psi$; (c) the total Governorate effect - $\gamma = \sigma(\phi\sqrt{1-\lambda)} + \psi\sqrt{\lambda/s})$; (d) the unstructured Country effect - $\eta$. }} \label{fig:tunisia_spatial_effects}
\end{figure}

Figures \ref{fig:egypt_spatial_effects} and \ref{fig:tunisia_spatial_effects} present the spatial distribution of point-estimates for the District and Governorate effects of Egypt and Tunisia respectively. The spatial distribution for Egypt indicates a substantially heightened propensity of recruitment in northeastern regions. No similar pattern is evident in Tunisia, though the mid-eastern costal areas do display systematically lower spatial recruitment effects than the rest of the country. Both countries estimate a number of highly significant district-level effects, which account for large portions of the variance in recruitment of both countries, with highly significant effects ranging from $-5$ to $+5$ log-odds points 
. In Tunisia, we also find evidence of a negative Sfax effect. Clearly, in order to be a recruit you must be subjected to unobserved area-level heterogeneity; individual-level covariates alone cannot counteract the underlying rarity of the event, as highlighted by the intercepts. 

\pagebreak

\pagebreak

\pagebreak

\begin{landscape}
\begin{figure}
\begin{subfigure}{0.7\textwidth}
\includegraphics[width=\linewidth]{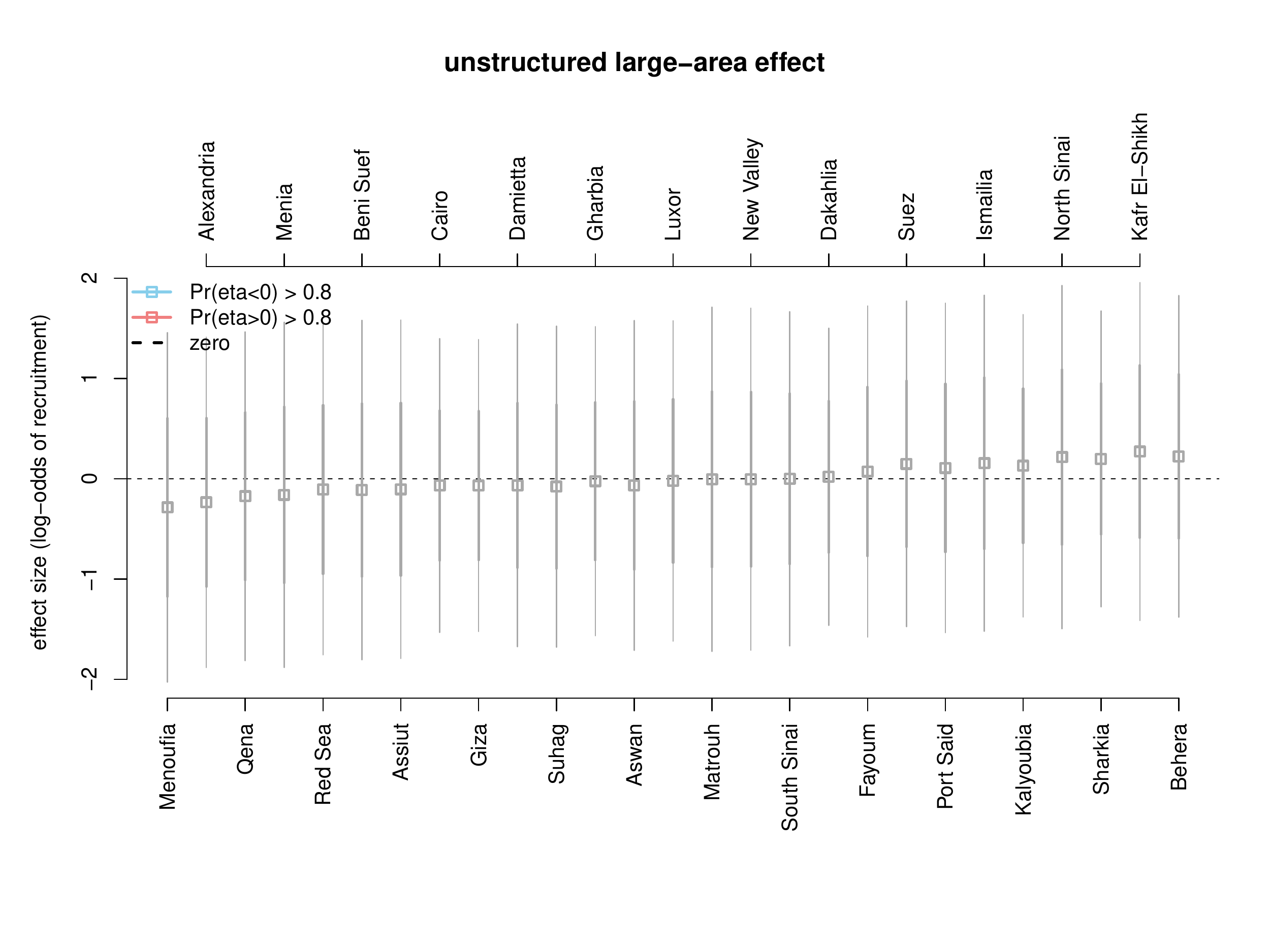}
\caption{Egypt} 
\end{subfigure}\hspace*{\fill}
\begin{subfigure}{0.7\textwidth}
\includegraphics[width=\linewidth]{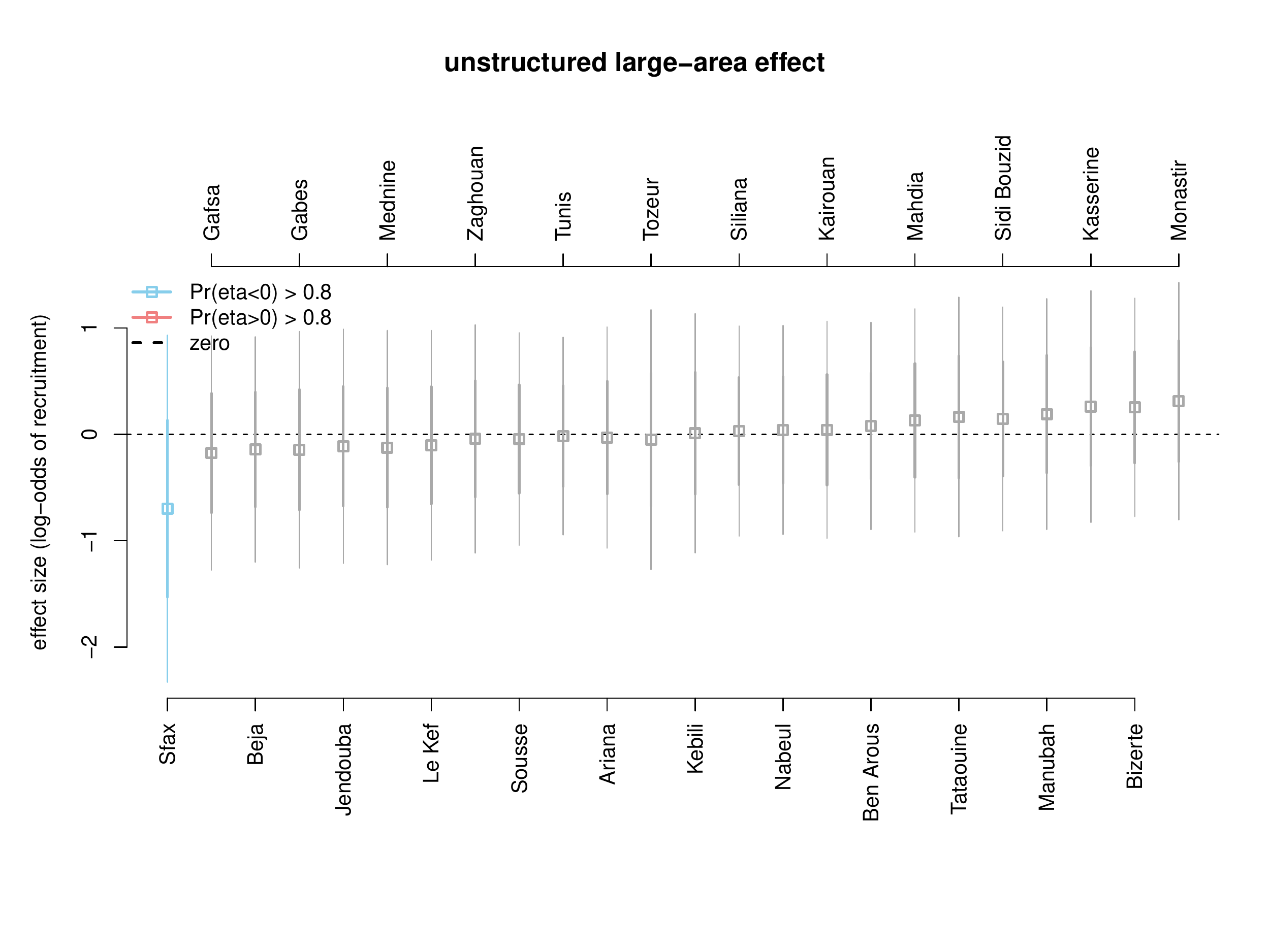}
\caption{Tunisia} 
\end{subfigure}
\caption{\footnotesize{Governorate effect ($\eta$) ordered by proportion of posterior simulations above zero.}}
\end{figure}
\end{landscape}
\pagebreak

\begin{landscape}
\begin{figure}[!htbp] 
\includegraphics[width=1.1\linewidth]{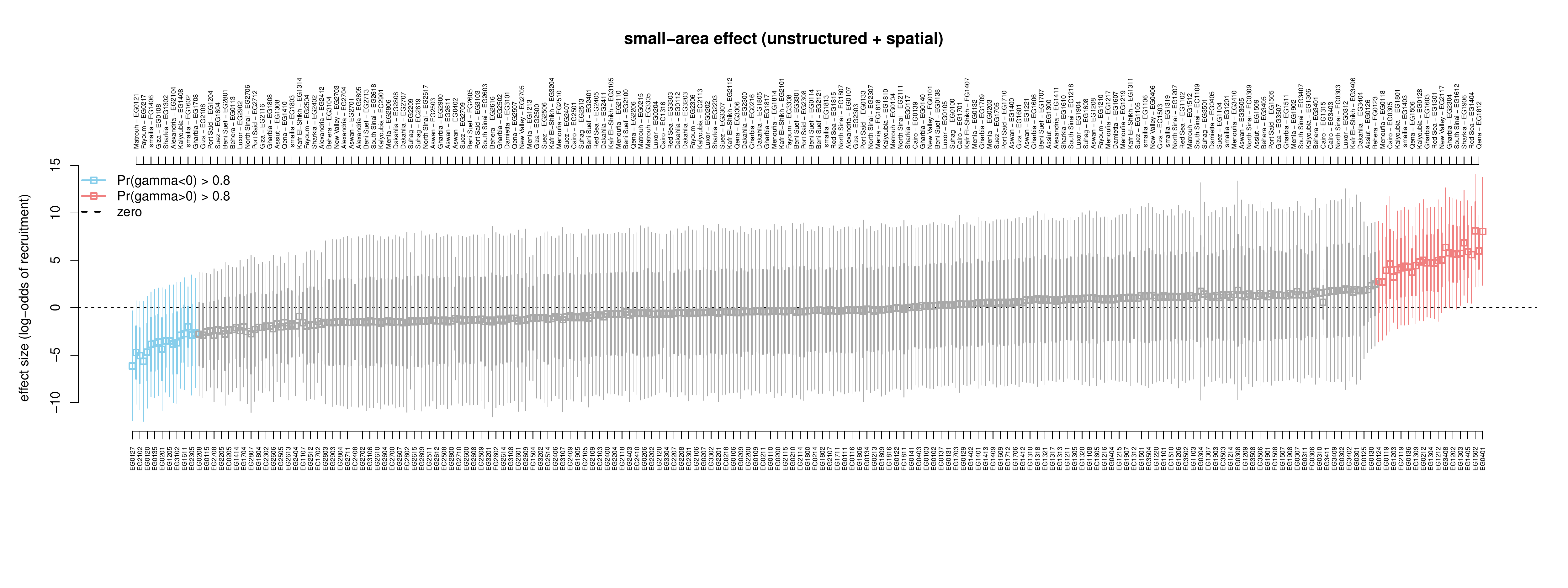}
\caption{\footnotesize{Total (structured + spatial) residual District effects in Egypt, ordered by proportion of posterior simulations above zero.}}
 \label{fig:egypt_dist} 
\end{figure}
\end{landscape}
\pagebreak

\begin{landscape}
\begin{figure}[!htbp] 
\includegraphics[width=1\linewidth]{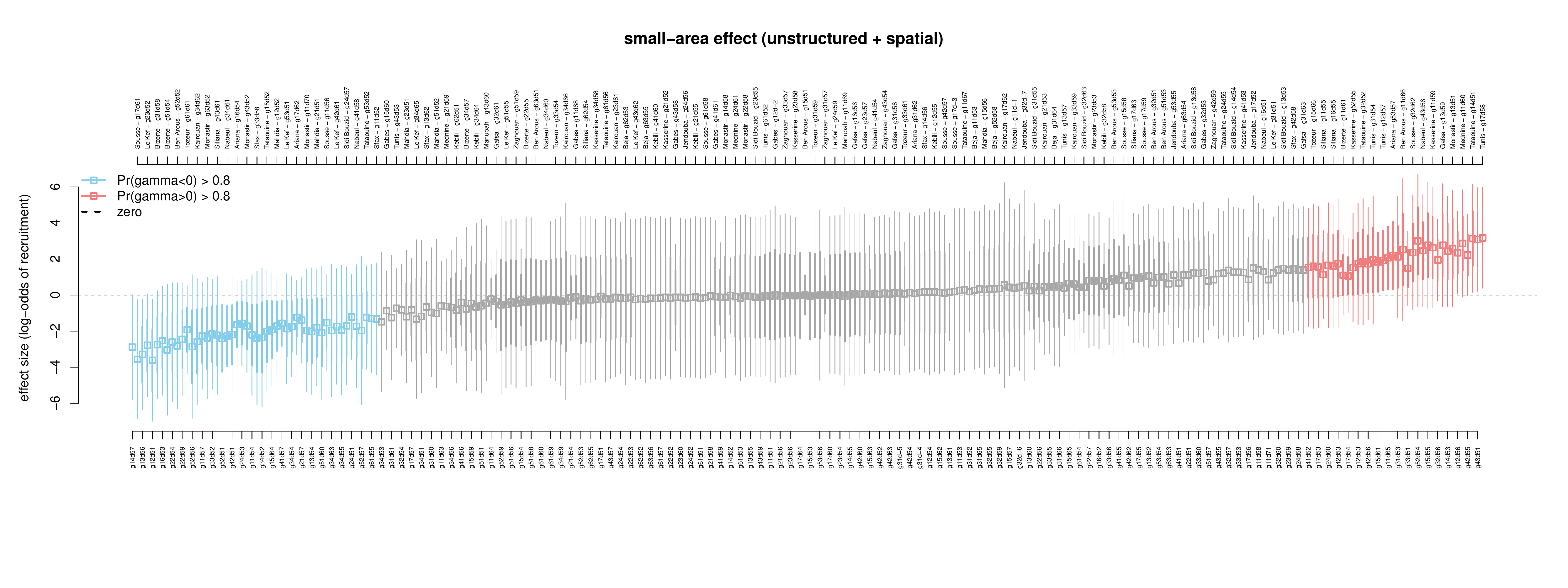}
\caption{\footnotesize{Total (structured + spatial) residual District effects in Tunisia, ordered by proportion of posterior simulations above zero.}}
 \label{fig:tunisia_dist} 
\end{figure}
\end{landscape}
\pagebreak

\begin{figure}[H]
    \centering
    \includegraphics[width=.95\textwidth]{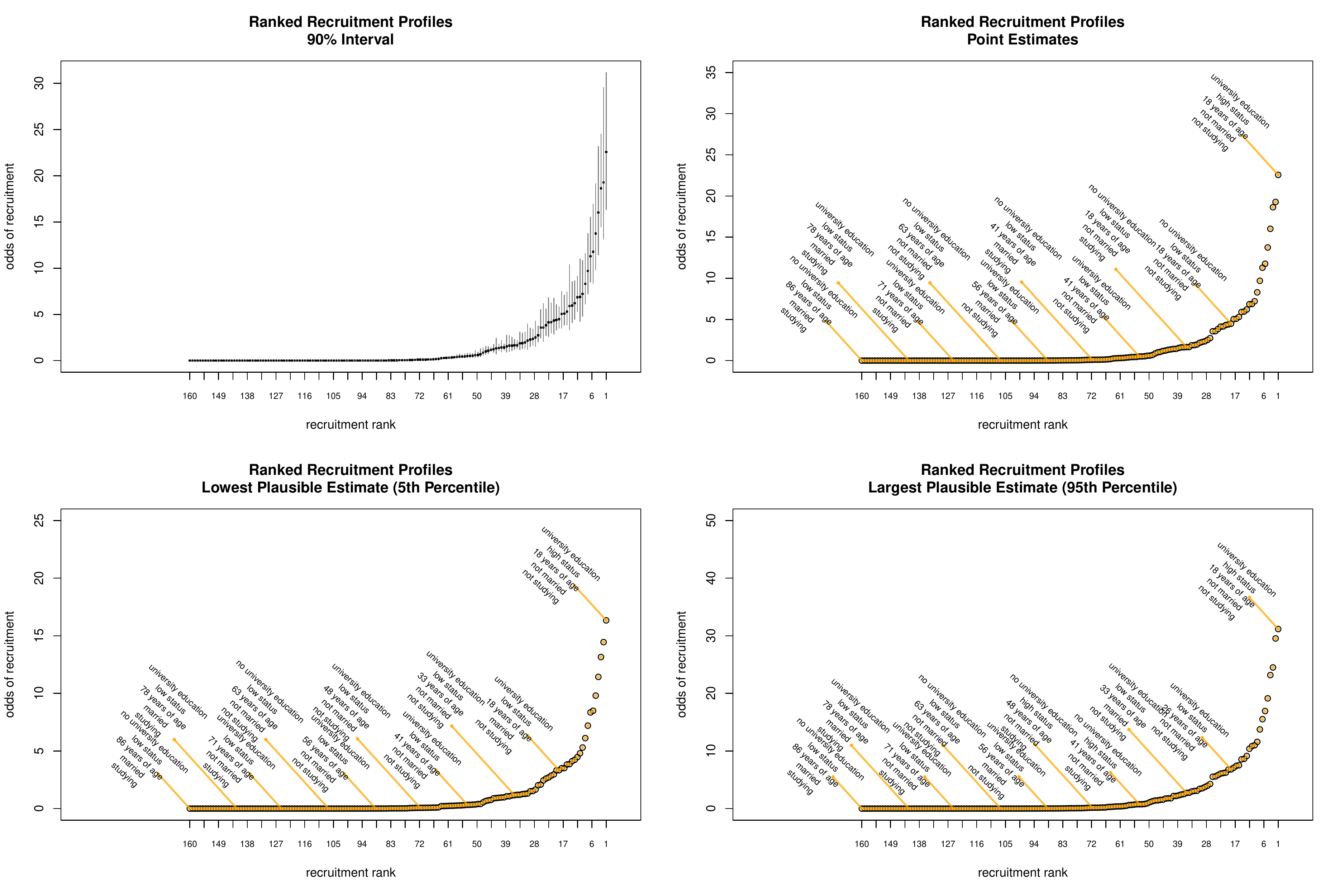}
    \caption{\scriptsize{Bird's Eye Distribution of predicted probabilities across a variety of hypothetical profiles. The distribution is presented on the odds relative to the average profile. To aid with interpretation, minimal and maximal estimates are presented separately. These plots help showcasing the sharp non-linearity across profile's recruitment propensities.}}
    \label{figure::predicted_probabilities_bird_odds}
    \includegraphics[width = .95\textwidth]{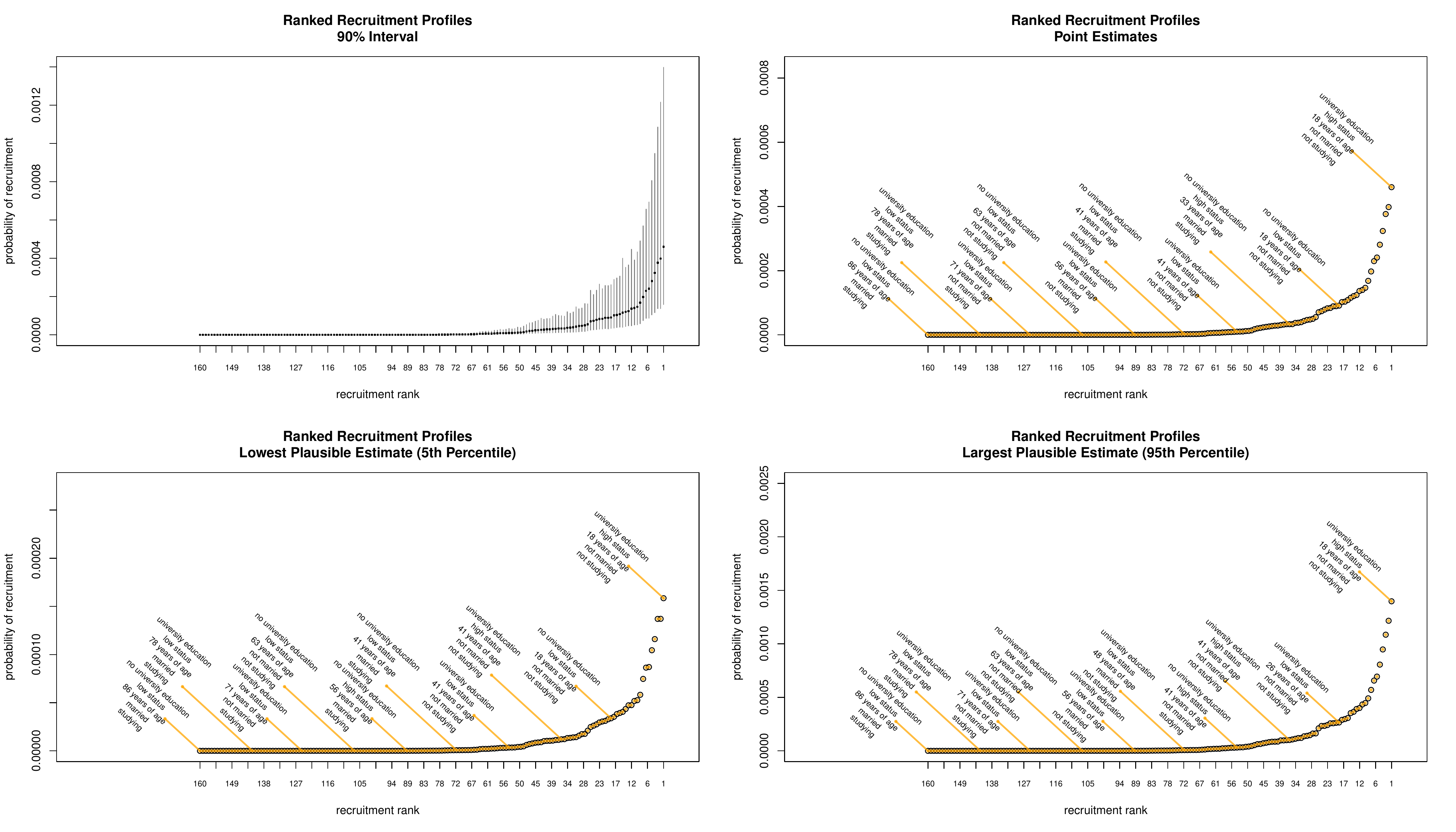}
    \caption{\scriptsize{Bird's Eye Distribution of predicted probabilities across a variety of hypothetical profiles. The distribution is presented on the probability scale. To aid with interpretation, minimal and maximal estimates are presented separately. These plots help showcasing the sharp non-linearity across profile's recruitment propensities.}}
    \label{figure::predicted_probabilities_bird_prob}
\end{figure}

\begin{figure}[H]
    \centering
    \includegraphics[width = .95\textwidth]{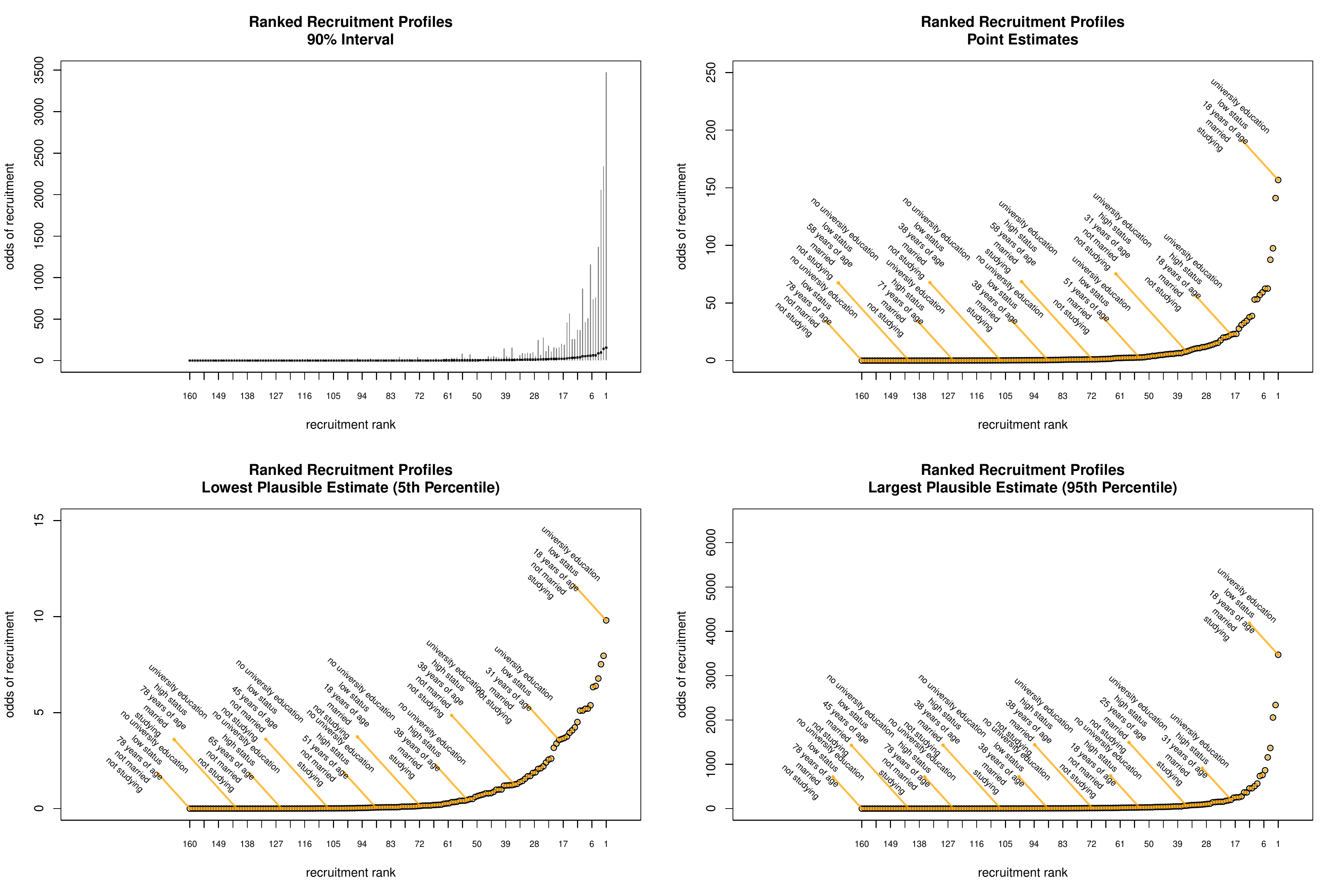}
    \caption{\scriptsize{Worm's Eye (Egypt) distribution of the predicted probabilities across a variety of hypothetical profiles. The distribution is presented on the odds relative to the average profile. To aid with interpretation, minimal and maximal estimates are presented separately. These plots help showcasing the sharp non-linearity across profile's recruitment propensities.}}
    \label{figure::predicted_probabilities_egypt_odds}
    \includegraphics[width = .95\textwidth]{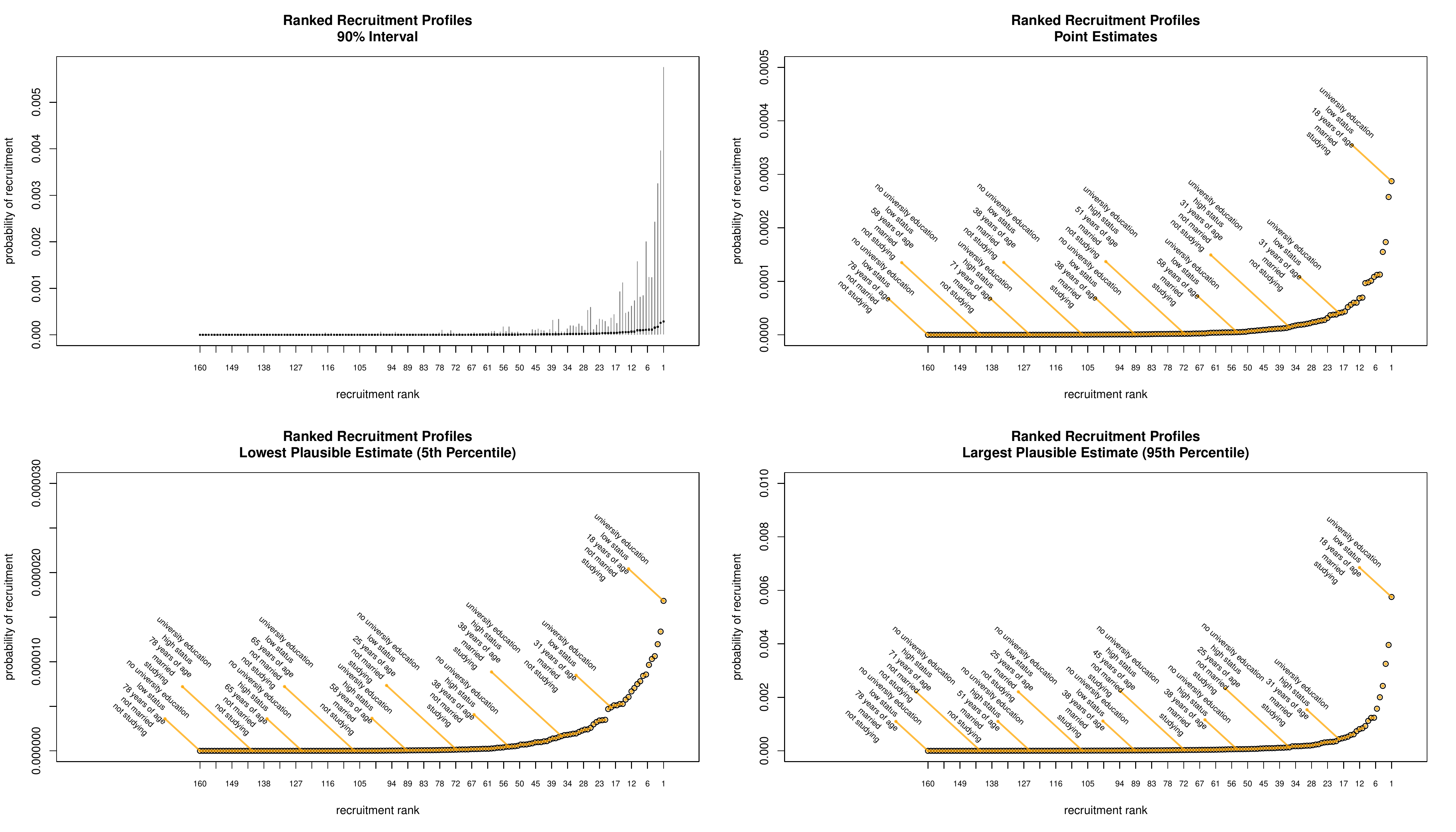}
    \caption{\scriptsize{Worm's Eye (Egypt) distribution of the predicted probabilities across a variety of hypothetical profiles. The distribution is presented on the probability scale. To aid with interpretation, minimal and maximal estimates are presented separately. These plots help showcasing the sharp non-linearity across profile's recruitment propensities.}}
    \label{figure::predicted_probabilities_egypt_prob}
\end{figure}

\begin{figure}[H]
    \centering
    \includegraphics[width = .95\textwidth]{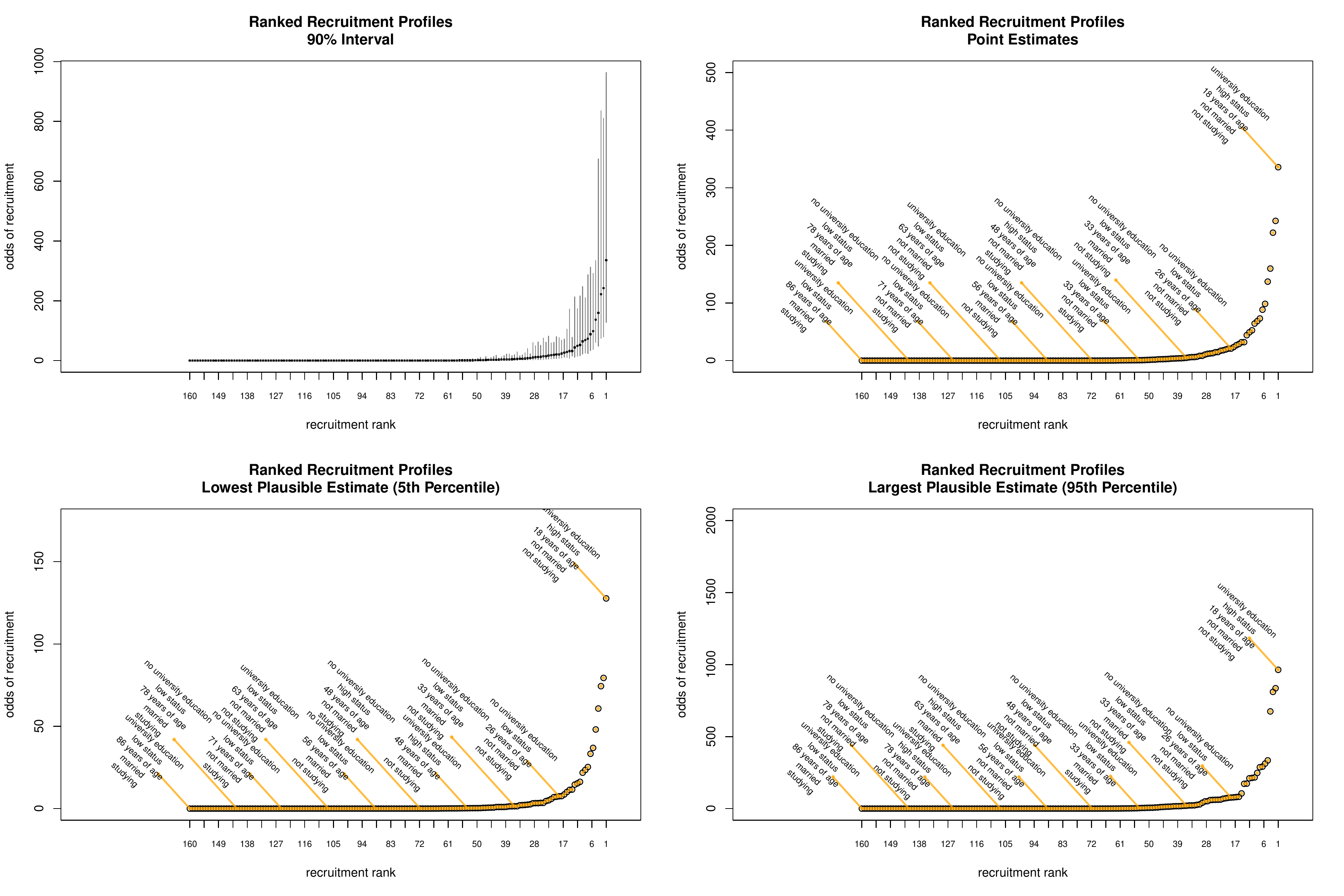}
    \caption{\scriptsize{Worm's Eye (Tunisia) distribution of the predicted probabilities across a variety of hypothetical profiles. The distribution is presented on the odds relative to the average profile. To aid with interpretation, minimal and maximal estimates are presented separately. These plots help showcasing the sharp non-linearity across profile's recruitment propensities.}}
    \label{figure::predicted_probabilities_tunisia_odds}
    \includegraphics[width = .95\textwidth]{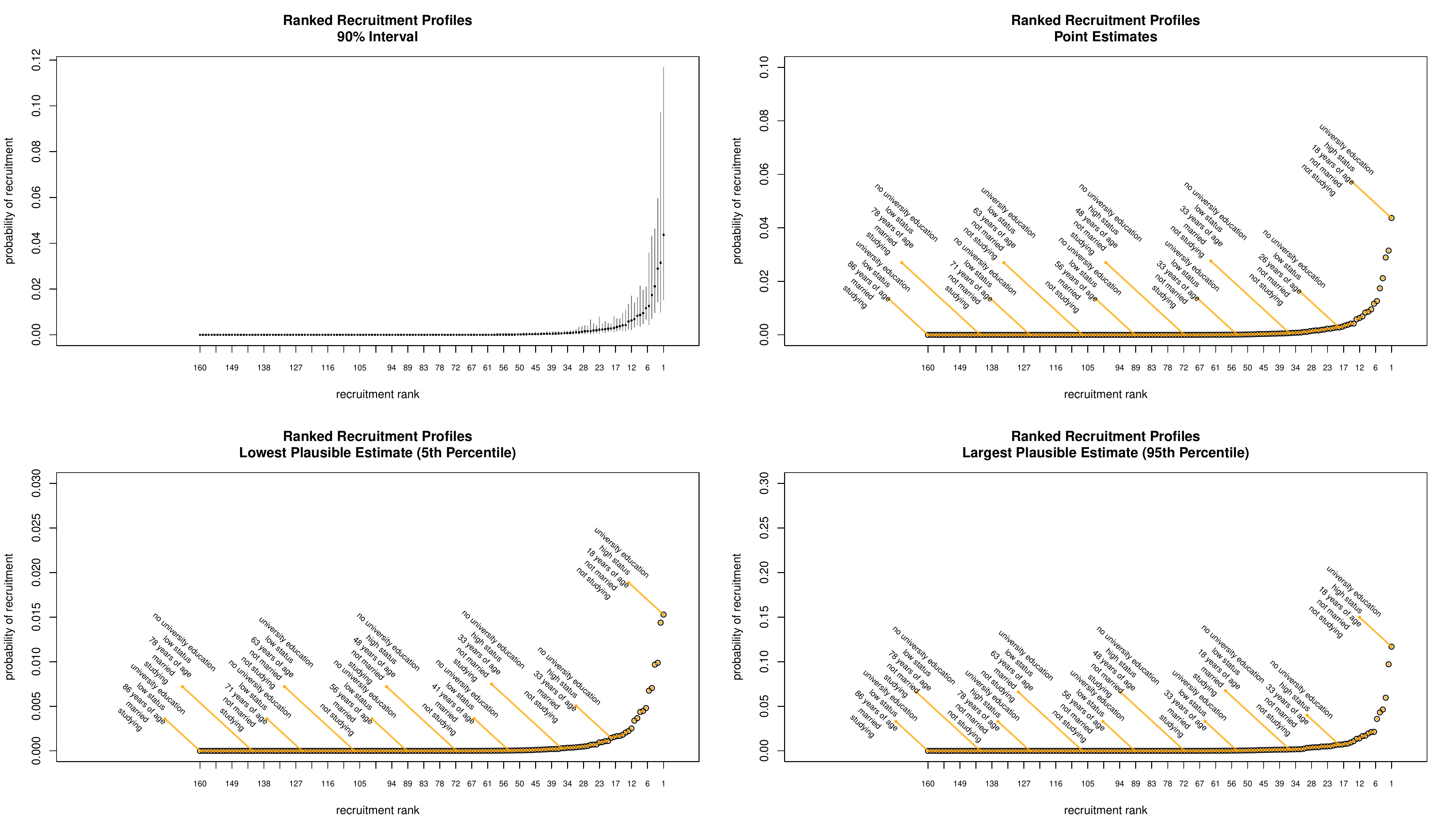}
    \caption{\scriptsize{Worm's Eye (Tunisia) distribution of the predicted probabilities across a variety of hypothetical profiles. The distribution is presented on the probability scale. To aid with interpretation, minimal and maximal estimates are presented separately. These plots help showcasing the sharp non-linearity across profile's recruitment propensities.}}
    \label{figure::predicted_probabilities_tunisia_prob}
\end{figure}
\end{spacing}{}

\end{document}